\documentclass[useAMS,usenatbib]{mn2e}
\usepackage[total={17.8cm,24.0cm},centering]{geometry}
\usepackage{graphicx}
\usepackage{color}
\usepackage{times}
\usepackage{amsmath}
\title[The ATLAS$^\mathrm{3D}$ Project - XXI. V$_\mathrm{esc}$ gradients]
  {The ATLAS$^\mathrm{3D}$ Project - XXI. Correlations between gradients of local escape velocity and stellar populations in early-type galaxies}
\author[N. Scott et al.]{Nicholas Scott$^{1,2}$\thanks{E-mail: nscott@astro.swin.edu.au}, Michele Cappellari$^2$, Roger L. Davies$^2$, Gijs Verdoes Kleijn$^3$,\newauthor Maxime Bois$^4$, Katherine Alatalo$^5$, Leo Blitz$^5$, Fr\'ed\'eric Bournaud$^{6}$, Martin Bureau$^2$, \newauthor Alison Crocker$^{7}$, Timothy A. Davis$^{2,8}$, P. T. de Zeeuw$^{8,9}$, Pierre-Alain Duc$^6$, \newauthor Eric Emsellem$^{8,10}$,  Sadegh Khochfar$^{11}$, Davor Krajnovi\'c$^8$, Harald Kuntschner$^{8}$,\newauthor Richard M. McDermid $^{12}$, Raffaella Morganti$^{3,13}$,Thorsten Naab$^{14}$, Tom Oosterloo$^{3,13}$, \newauthor  Marc Sarzi$^{15}$, Paolo Serra$^{13}$, Anne-Marie Weijmans$^{16}$\thanks{Dunlap Fellow} and Lisa M. Young$^{17}$\\
$^1$Centre for Astrophysics \& Supercomputing, Swinburne University of Technology, PO Box 218, Hawthorn, VIC 3122, Australia\\
$^2$Sub-Dept. of Astrophysics, Dept. of Physics, University of Oxford, Denys Wilkinson Building, Keble Road, Oxford, OX1 3RH, UK\\
$^{3}$Kapteyn Astronomical Institute, University of Groningen, Postbus 800, 9700 AV Groningen, The Netherlands\\
$^4$Observatoire de Paris, LERMA and CNRS, 61 Av. de lÔObservatoire, F-75014 Paris, France\\
$^5$Department of Astronomy, Campbell Hall, University of California, Berkeley, CA 94720, USA\\
$^6$Laboratoire AIM Paris-Saclay, CEA/IRFU/SAp Ð CNRS Ð Universit\'e Paris Diderot, 91191 Gif-sur-Yvette Cedex, France\\
$^7$Department of Astronomy, University of Massachusetts, Amherst, MA 01003, USA\\
$^8$European Southern Observatory, Karl-Schwarzschild-Str. 2, 85748 Garching, Germany\\
$^9$Sterrewacht Leiden, Leiden University, Postbus 9513, 2300 RA Leiden, the Netherlands\\
$^{10}$Universit\'e Lyon 1, Observatoire de Lyon, Centre de Recherche Astrophysique de Lyon and \\ \ \ \ Ecole Normale Sup\'erieure de Lyon, 9 avenue Charles Andr\'e, F-69230 Saint-Genis Laval, France\\
$^{11}$Max Planck Institut f\"ur extraterrestrische Physik, PO Box 1312, D-85478 Garching, Germany\\
$^{12}$Gemini Observatory, Northern Operations Centre, 670 N. A`ohoku Place, Hilo, HI 96720, USA\\
$^{13}$Netherlands Institute for Radio Astronomy (ASTRON), Postbus 2, 7990 AA Dwingeloo, The Netherlands\\
$^{14}$Max-Planck-Institut f\"ur Astrophysik, Karl-Schwarzschild-Str. 1, 85741 Garching, Germany\\
$^{15}$Centre for Astrophysics Research, University of Hertfordshire, Hatfield, Herts AL1 9AB, UK\\
$^{16}$Dunlap Institute for Astronomy \& Astrophysics, University of Toronto, 50 St. George Street, Toronto, ON M5S 3H4, Canada\\
$^{17}$Physics Department, New Mexico Institute of Mining and Technology, Socorro, NM 87801, USA\\}
\date{August 2012}
\pagerange{\pageref{firstpage}--\pageref{lastpage}} \pubyear{2012}

\def\LaTeX{L\kern-.36em\raise.3ex\hbox{a}\kern-.15em
    T\kern-.1667em\lower.7ex\hbox{E}\kern-.125emX}

\begin{document}

\label{firstpage}

\maketitle
\clearpage

\begin{abstract}
We explore the connection between the local escape velocity, V$_\mathrm{esc}$ , and the stellar population properties in the ATLAS$^\mathrm{3D}$ survey, a complete, volume-limited sample of nearby early-type galaxies. We make use of $ugriz$ photometry to construct Multi-Gaussian Expansion models of the surface brightnesses of our galaxies. We are able to fit the full range of surface brightness profiles found in our sample, and in addition we reproduce the results of state-of-the-art photometry in the literature with residuals of 0.04 mags. We utilise these photometric models and SAURON integral-field spectroscopy, combined with Jeans dynamical modelling, to determine the local V$_\mathrm{esc}$ derived from the surface brightness. We find that the local V$_\mathrm{esc}$ is tightly correlated with the Mg$\,b$ and Fe5015 linestrengths and optical colours, and anti-correlated with the H$\beta$ linestrength. In the case of the Mg$\,b$ and Colour - V$_\mathrm{esc}$ relations we find that the relation within individual galaxies follows the global relation between different galaxies. We intentionally ignored any uncertain contribution due to dark matter since we are seeking an empirical description of stellar population gradients in early-type galaxies that is ideal for quantitative comparison with model predictions. We also make use of single stellar population (SSP)  modelling to transform our linestrength index measurements into the SSP-equivalent parameters age (t), metallicity ([Z/H]) and $\alpha$-enhancement [$\alpha$/Fe]. The residuals from the relation are correlated with age, [$\alpha$/Fe], molecular gas mass and local environmental density. We identify a population of galaxies that occur only at low V$_\mathrm{esc}$ that exhibit negative gradients in the Mg$\,b$ - and Colour - V$_\mathrm{esc}$ relations. These galaxies typically have young central stellar populations and contain significant amounts of molecular gas and dust. Combining these results with N-body simulations of binary mergers we use the Mg$\,b$-V$_\mathrm{esc}$ relation to constrain the possible number of dry mergers experienced by the local early-type galaxy population -- a typical massive ETG can have experienced only $\sim 1.5$ major mergers before becoming a significant outlier in the Mg$\,b$-V$_\mathrm{esc}$ relation.
\end{abstract}

\begin{keywords}
 galaxies: elliptical and lenticular, cD -
 galaxies: abundances -
 galaxies: formation -
 galaxies: evolution.
\end{keywords}

\section{Introduction}
\label{sec:intro}
Early-type galaxies (ETGs including both ellipticals; E and lenticulars; S0) in the nearby universe represent an advanced stage of galaxy evolution. They likely result from the interplay of a wide range of physical processes. In the context of the ATLAS$^\mathrm{3D}$ survey \citep[][hereafter Paper I]{Cappellari:2011a} we found that ETGs are well separated in terms of their stellar angular momentum \citep[][hereafter Paper II and Paper III respectively]{Krajnovic:2011,Emsellem:2011}. Slow rotator ETGs can be explained as having accreted a major part of their mass due to hierarchical merging \citep[][hereafter Paper VIII]{Khochfar:2011}, while fast rotators likely experienced cold gas accretion \citep[][herafter Paper X and Paper XIII]{Davis:2011,Serra:2012} but then had their gas stripped by the cluster environment \citep[][hereafter Paper VII]{Cappellari:2011b} or AGN and supernova feedback. Their present-day properties -- their stellar populations, dynamics, gas content and environment -- contain information about their entire evolutionary history. Through a detailed study of these properties we can hope to disentangle some of this history and learn about the key physical mechanisms that have played a dominant role in shaping these objects.

Early-type galaxies have a few advantages when it comes to studies of this nature. They are massive and bright, allowing high quality data on both their stellar content and dynamical state to be obtained with relative ease. Their light is typically dominated by old, evolved stars \citep[see e.g.][]{Thomas:2005} -- the low levels of star formation allow us to make the assumption that their stars were formed in a single event. The morphology of their central regions is relatively featureless, allowing us to study how properties vary within these objects without the complication of substructure such as spiral arms or clumps of star formation.

Galactic archaeology is the study of the `fossil' record within local early-type galaxies. A detailed analysis of the stellar populations of these objects can reveal much about {\it when} the stars that now comprise these objects formed, but it reveals little of {\it where} they formed. In contrast, dynamical studies can reveal some of the {\it where} (in situ vs. accreted) but little of the {\it when}. By finding clear links between dynamical and stellar population quantities we can begin to answer all of these questions, forming a picture of when and under what conditions the stellar populations of early-type galaxies were formed and assembled.

There are several well-studied relations linking stellar population and dynamical quantities; the colour-magnitude relation \citep{Visvanathan:1977} and the Mg-$\sigma$ relation. Most early-type galaxies lie on the tight 'red sequence' of the colour-magnitude diagram \citep[e.g.][]{Bower:1998}, though some lie in the `green valley' transition region or even in the `blue cloud' \citep{Strateva:2001,Conselice:2006,van_den_Bergh:2007,Bernardi:2010}. The Mg-$\sigma$ relation has been studied by many authors \citep[examples include][]{Bender:1993,Jorgensen:1997,Colless:1999,Trager:2000b,Bernardi:2003b,Kuntschner:2006} with many hundreds of early-type galaxies and though the precise zero points and slopes vary between studies the small scatter compared to the range in velocity dispersion covered is universal.

The two above relations describe a connection between different galaxies, however we can also study the variation within individual galaxies. This was first done by \citet{Franx:1990} who studied the correlation of the {\it local} colour with the {\it local} $\sigma$ and escape velocity, V$_\mathrm{esc}$. They found that the global relation between different galaxies is the same as the local relation within individual galaxies. This was followed up by \citet{Davies:1993} and \citet{Carollo:1994}, who studied the connection between the Mg line strength and the local V$_\mathrm{esc}$. These works were all based on small samples of galaxies and relied upon long slit data, both in the measurement of line-strengths and as the input for the dynamical modelling necessary to derive V$_\mathrm{esc}$. Two dimensional spectroscopy was first used by \citet{Emsellem:1996}, who used IFU data to study the Mg$\,b$-V$_\mathrm{esc}$ relation in a single galaxy, M104. \citet[][hereafter S09]{Scott:2009} investigated the line strength - V$_\mathrm{esc}$ relations for the SAURON early-type galaxy sample \citep{deZeeuw:2002}, significantly increasing the sample size (to 48 objects) compared to previous studies and also fully utilising IFU spectroscopy, which allows us to study how kinematic and stellar population properties vary in two dimensions.

The SAURON survey was biased towards high mass galaxies by the survey selection criteria, which uniformly sampled galaxies in absolute magnitude. In this paper we extend our study of the role of the local V$_\mathrm{esc}$ to the complete, volume-limited ATLAS$^\mathrm{3D}$ survey of early-type galaxies (Paper I). In Section \ref{Sec:Sample} we present the sample selection and observations. In Section \ref{Sec:Method} we present our method of determining the local V$_\mathrm{esc}$ from the photometric and spectroscopic data. In Section \ref{Sec:Results} we present the results of our analysis, showing the correlation of various properties of early-type galaxies with the local V$_\mathrm{esc}$. In the same section we extend this analysis to the study of the stellar populations of our sample. Finally, in Section \ref{Sec:Discussion} we discuss our results in the context of cosmological and N-body simulations of early-type galaxy formation and state our conclusions in Section \ref{Sec:Conclusions}.

\begin{figure}
\includegraphics[width=3.25in]{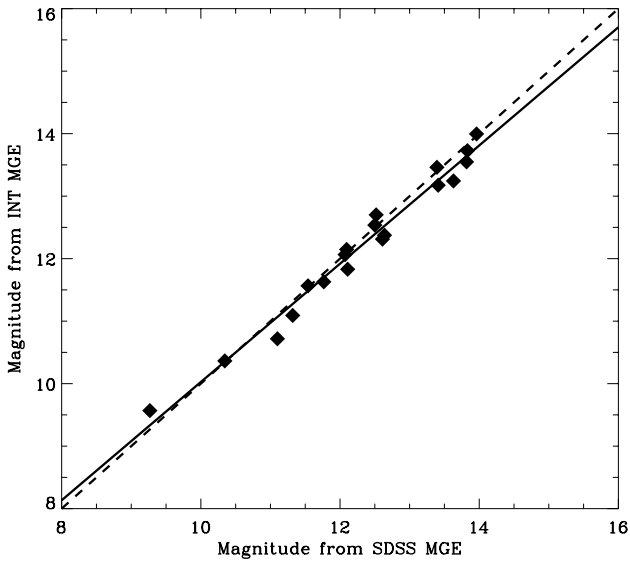}
\caption{Comparison between total {\it r}-band magnitudes derived from INT and SDSS DR8 images for the 18 ATLAS$^{\rm{3D}}$ galaxies included in the latest SDSS data release. The solid line shows a fit to the data and the dashed line shows a 1:1 relation. The data are consistent with this 1:1 relation, with a scatter of 0.17 magnitudes.}
\label{fig:INT_SDSS_comp}
\end{figure}

\begin{figure}
\includegraphics[width=3.25in]{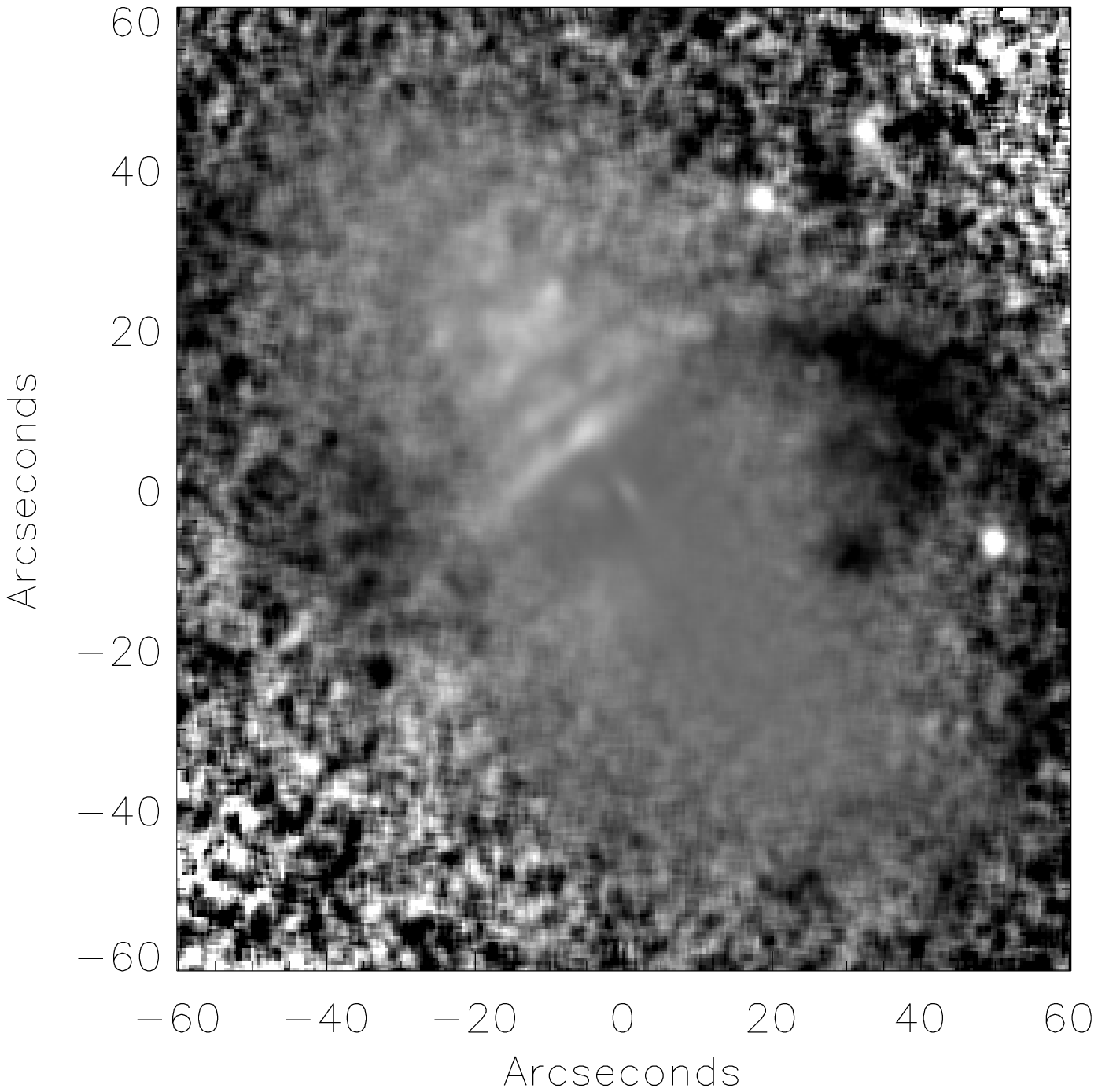}
\includegraphics[width=3.25in]{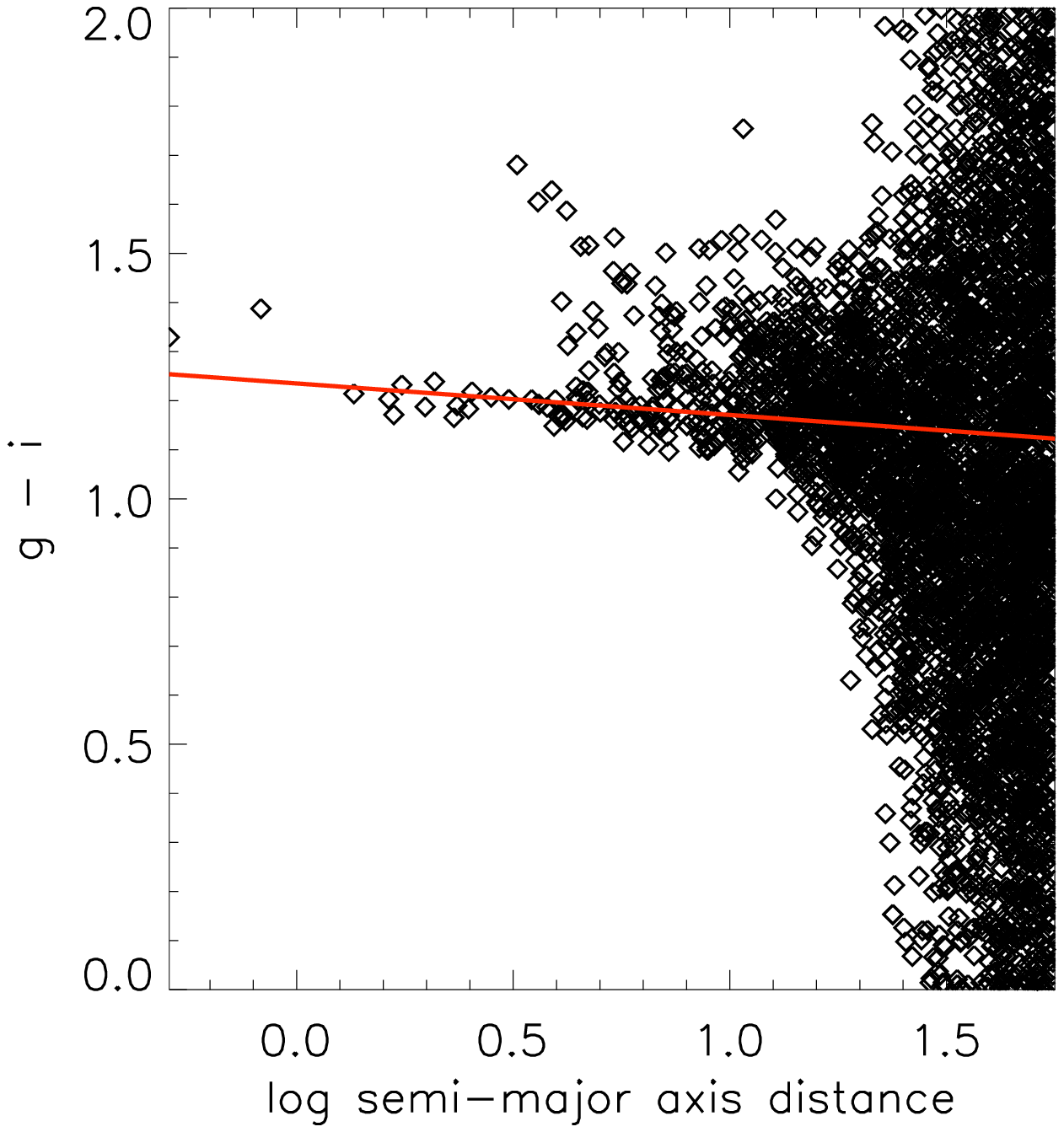}
\caption{Upper panel: {\it g-i} colour map of NGC4753. Dust filaments show prominently in the upper left quadrant of the figure. Lower panel: {\it g-i} vs. semi-major axis distance for individual pixels form the above colour map. The red line shows a fit to the underlying colour profile. Pixels lying significantly above this line are corrected for dust extinction as described in the text. At large radii the colour of individual pixels is dominated by noise and any uncertainty in the sky subtraction.}
\label{Fig:dust_cor}
\end{figure}

\section{Sample and Data}
\label{Sec:Sample}
The ATLAS$^\mathrm{3D}$ project (Paper I) is a volume limited complete survey of all nearby early-type galaxies (those not showing spiral structure in the visible photometry) brighter than  M$_K = -21.5$ (M$\ \sim 6 \times 10^9\ $ M$_\odot$) and within 42 Mpc. The sample consists of 260 galaxies spanning a broad range in central velocity dispersion, $\sigma_e$, environment and absolute magnitude. In this work we consider a sample of 256 objects, excluding 4 galaxies for which poor quality or missing data  (see below for details) prevented a full analysis.

All 260 galaxies in the sample were observed with the SAURON integral-field unit (IFU) on the William Herschel Telescope at the Roque de los Muchachos observatory in La Palma. The SAURON data consists of integral-field spectroscopy over the wavelength range $\sim 4800 - 5300$ \AA \ out to typically 1 effective radius (R$_e$). This wavelength range includes the three Lick stellar absorption line strength indices; Mg$\,b$, Fe5015 and H$\beta$. Maps of the stellar kinematics (parameterized by the mean velocity $v$ and velocity dispersion $\sigma$) and line strengths were produced as described in Paper I and McDermid et al., (in preparation) respectively. The velocity maps are presented in Paper II.

\begin{figure}
\includegraphics[width=3.25in]{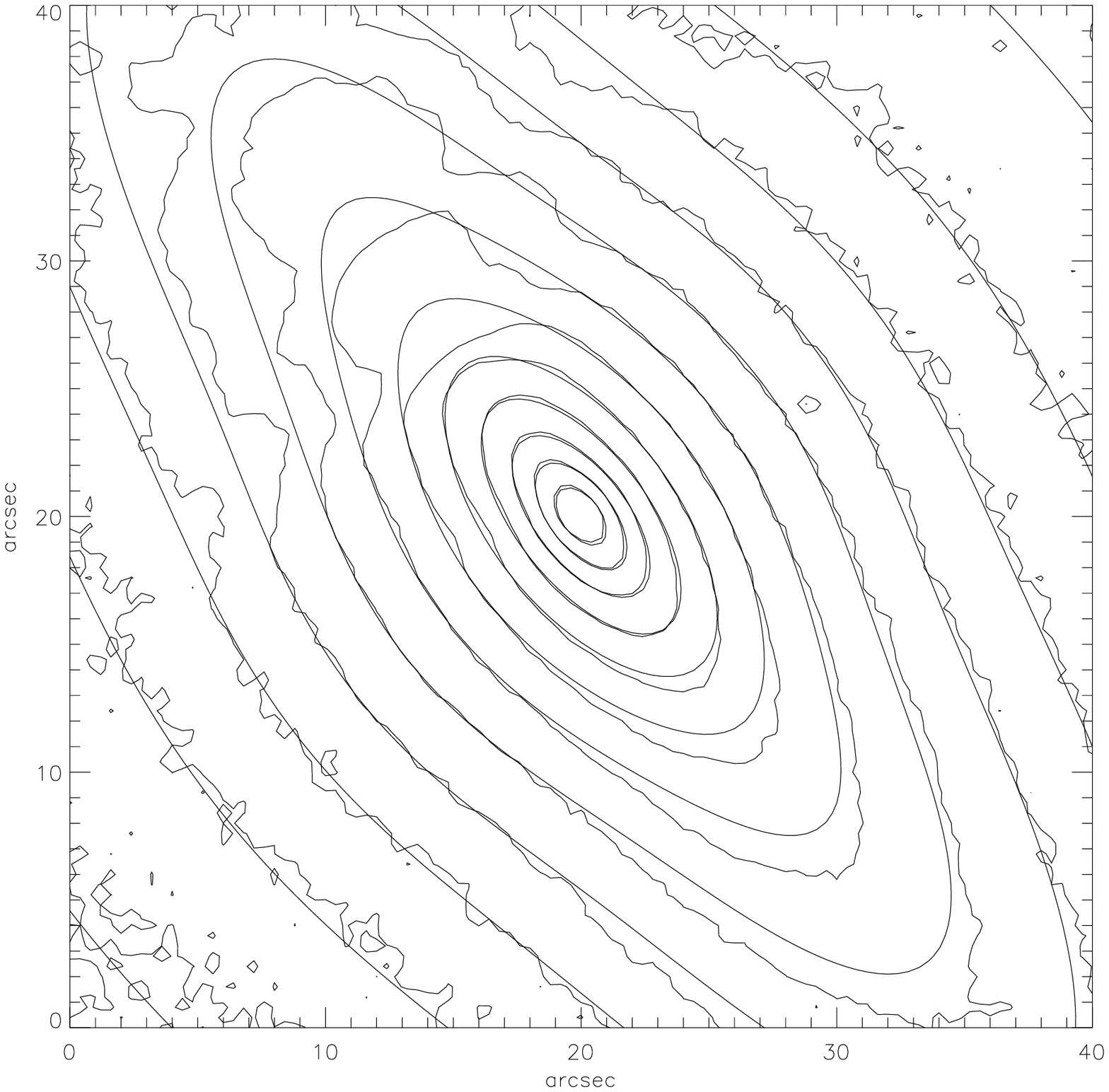}
\includegraphics[width=3.25in]{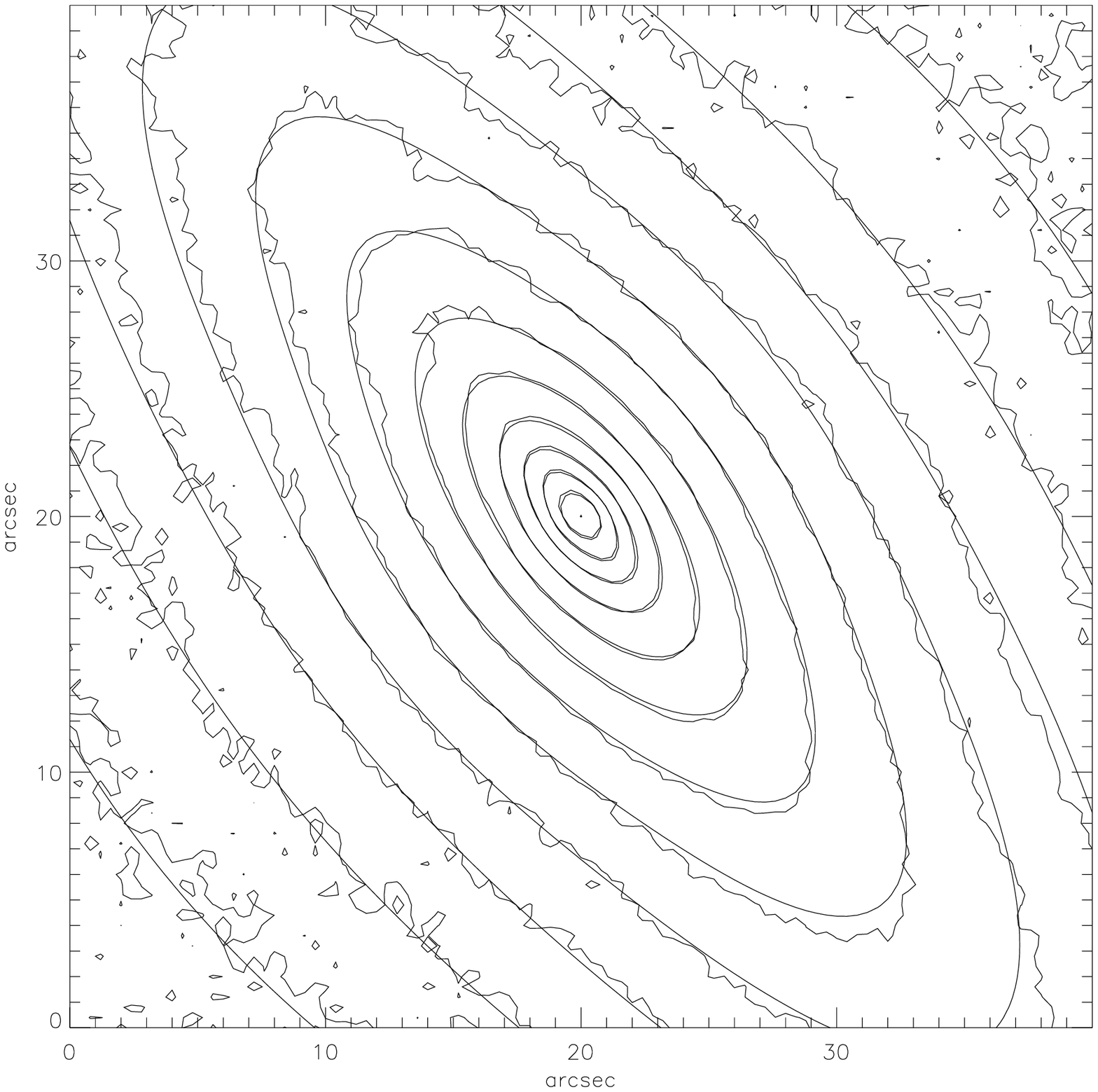}
\caption{Photometry of NGC2685, the Helix Galaxy, with MGE models overplotted before (upper panel) and after (lower panel) dust correction, as described in the text. The effect of dust is visible as the band-like structure in the upper left of the galaxy; this structure is absent in the dust-corrected image. }
\label{Fig:dust_example}
\end{figure}

\subsection{Photometry}

We have obtained {\it ugriz} photometry for 258 galaxies from the ATLAS$^\mathrm{3D}$ sample, taken from the Sloan Digital Sky Survey (SDSS) DR7 \citep{Abazajian:2009} where available, supplemented by our own observations with the Wide Field Camera (WFC) on the 2.5m Isaac Newton Telescope (INT), again at the Roque de los Muchachos observatory. Observing conditions prevented the acquisition of photometry for the remaining two galaxies. We also note that two other galaxies suffered from low signal-to-noise ratio in their SAURON observations (see McDermid et al, in preparation, for details) and we exclude these.

Observations with the INT WFC were carried out to obtain $ugri$-band imaging for galaxies not covered by SDSS DR7. The observations were obtained in 3 runs: 7-9 May 2007, 3 \& 4 November 2007 and 13-17 May 2008. Part of the observations were obtained under non-photometric conditions. Images were taken through the 5 filters for 55 galaxies from the ATLAS$^\mathrm{3D}$ sample. The filters that were used are listed in Table \ref{t:WFCfilters}. Integration times were typically 60 to 160 seconds, reaching sensitivities comparable to or deeper than the SDSS. Fifteen galaxies were already observed by SDSS DR5 and were used to perform cross-checks in general and to bring the INT imaging onto the same photometric system as SDSS in particular. A further eight galaxies were made available in SDSS DR7 - these were not used for the photometric calibration but replaced the corresponding INT observations. The images were reduced and calibrated using the Astro-WISE system \citep{Valentijn:2007}. \citet{McFarland:2011} give a detailed description of the Astro-WISE pipeline.

\begin{table}
\caption{WFC INT filters used for ATLAS$^\mathrm{3D}$ photometry}
\centering
\begin{tabular}{rccc}
Serial Number & Name & Central Wavelength & FWHM\\
& &  (\AA) &  (\AA) \\
\hline
204 & WFCRGOU    & 3581 & 638  \\
220 & WFCSloanG  & 4846 & 1285 \\
214 & WFCSloanR  & 6240 & 1347 \\
215 & WFCSloanI  & 7743 & 1519 \\
195 & WFCRGOZ    & 8763 & -    \\
\hline
\label{t:WFCfilters}
\end{tabular}
\end{table}

The observations were debiased and flatfielded using nightly biases and twilight flatfields. Hot and cold pixels are identified and indicated in the weight maps of the science images. Photometric calibration was derived in the Sloan {\it ugriz} system from nightly standard star field observations using SDSS DR5 \citep{Adelman-McCarthy:2007} photometry in those areas where the INT imaging overlapped the SDSS survey coverage. For the {\it g}- and {\it i}-bands we used a linear colour transformation \citep{Verdoes_Kleijn:2007} to convert from the WFC instrumental photometric system to the Sloan standard photometric system . They found no evidence for a transformation in other bands. We verified the photometric calibration by using galaxies with both INT and SDSS DR7 coverage by comparing magnitudes of stars found in both images. Astrometric calibration was performed using the USNO catalog as astrometric reference system \citep{Monet:2003}.

\begin{figure*}
\includegraphics[width=7in]{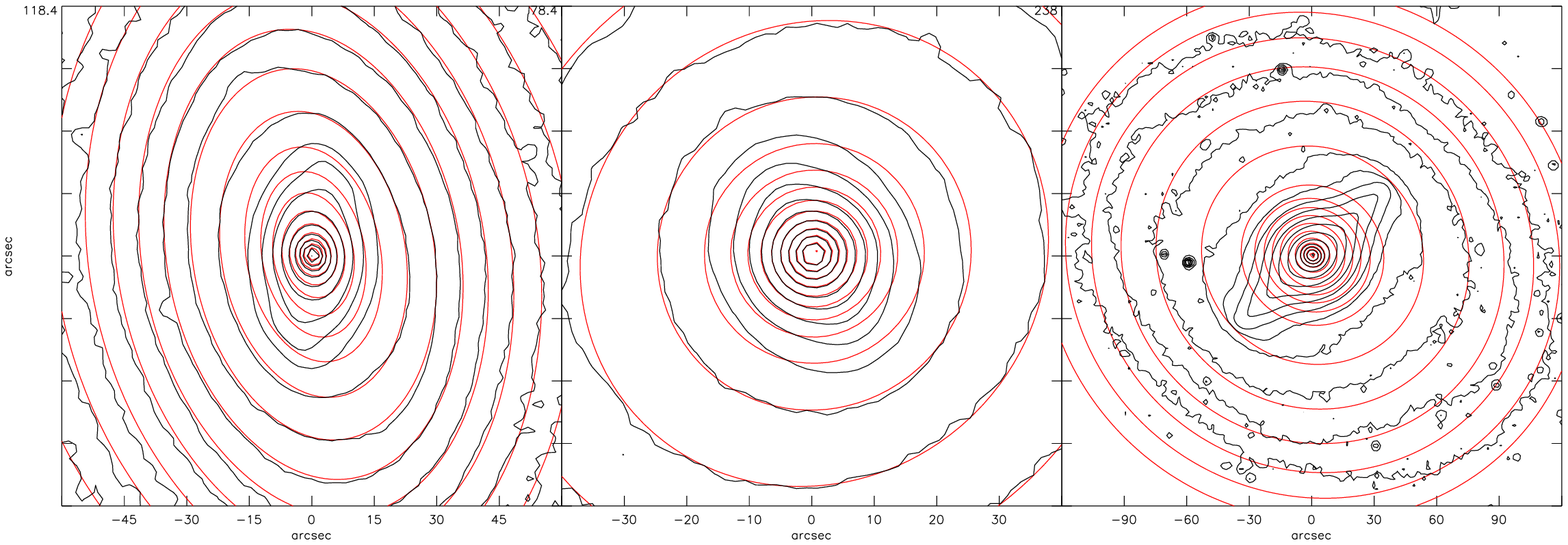}
\caption{Example MGE models of barred galaxies. Contours of the best-fitting MGE models (red) for NGC3941, NGC4267 and NGC4643 plotted over the photometric contours (black). The region shown was selected to approximately cover out to double the extent of the bar. The MGE matches the contours both inside and outside the region where the bar dominates. These models were produced automatically using the method described in the text to avoid fitting the bar. For galaxies with very strong bars (such as NGC4643) the fitting method is unable to match the photometry in detail, however the overall shape and surface brightness of the object are well reproduced.}
\label{fig:mge_bars}
\end{figure*}

In the SDSS DR8 data release \citep{Aihara:2011} imaging for 18 further ATLAS$^{\rm{3D}}$ galaxies was made available. We used this new SDSS photometry to assess the absolute photometric calibration of our INT photometry. We derived total {\it r}-band magnitudes from the two sets of images using the Multi Gaussian Expansion \citep[MGE][]{Emsellem:1994} models described in Section \ref{Sec:MGE_phot_testing}. The comparison is shown in Fig. \ref{fig:INT_SDSS_comp}. The data are consistent with a 1:1 relationship between the INT and SDSS derived magnitudes, with a scatter of 0.17 magnitudes, indicating our INT photometry is well-calibrated.

\section{Mass modelling}
\label{Sec:Method}
In order to determine the local V$_\mathrm{esc}$ we must first construct a dynamical model of each of our galaxies from our photometric and kinematic data. This involves several steps, summarised here and where necessary described in more detail below.
\newcounter{Lcount2}
\begin{list}{\roman{Lcount2})}
{\usecounter{Lcount2}}
\item The {\it r}-band image is corrected for internal extinction due to dust.
\item Nearby stars and galaxies are masked, as are regions where the dust correction failed.
\item A calibrated Multi Gaussian Expansion (MGE) model of the photometry is constructed from the masked and extinction corrected image using the method and software of \citet{Cappellari:2002}.
\item A series of Jeans Anisotropic MGE (JAM) models \citep{Cappellari:2008} are constructed using the photometric model with different values of inclination, $i$, mass to light ratio, M/L and anisotropy, $\beta$. The models do not explicitly include a dark matter halo and the measured M/L represents the total one, including both luminous matter and any possible contribution from dark matter.
\item The best fitting JAM model is selected by minimising the $\chi^2$ difference between the predicted second moments of the velocity field, $\sqrt{v^2 + \sigma^2}$ and the SAURON observations \citep[see][hereafter Paper XIX, for details of the fitting process]{Cappellari:2012b}.
\item This JAM model constrains the M/L normalisation and is used to transform the photometric model into a mass model. The JAM model also specifies the gravitational potential. In practise, we make use of the models presented in \citet{Cappellari:2012a}.
\item For each bin in the SAURON maps the local, line-of-sight, luminosity-weighted V$_\mathrm{esc}$ is calculated from the gravitational potential of the best fitting JAM model.
\end{list}

\subsection{Dust correction}
\label{Sec:Dust}
The presence of dust in a galaxy can significantly affect the accuracy of our MGE models, both through the dimming due to extinction and by altering the observed morphology of the galaxy. We attempt to remove this effect by correcting for this extinction and fitting the MGE model to the underlying surface brightness profile. This correction was applied to 25 of our galaxies, approximately 10 per cent of the sample.

We use the technique of \citet{Carollo:1997} (who assumed the dust is a screen in front of the stellar emission), further developed by \citet{Cappellari:2002} to correct for the extinction due to dust. We first construct a {\it g-i} colour map of  each galaxy. This is converted to a colour profile by plotting the {\it g-i} colour against the logarithm of semi-major axis distance along ellipses fixed to the global PA and $\epsilon$ (from Paper II), $\log m$, of each pixel in the image. We use a robust linear fit to this data, which minimises the influence of dusty regions, in order to determine the underlying colour gradient of the galaxy. We subtract this best fitting colour gradient from all pixels in the {\it g-i} colour map to determine the colour excess, E({\it g-i}). Using a cut in E({\it g-i}) (typically as a function of $\log m$) we select pixels which have been significantly affected by dust extinction and correct for this using the standard galactic extinction law of \citet{Schlegel:1998}. The r-band extinction, $A_r$, is related to the E({\it g-i}) colour excess by the equation: $A_r = 1.15\ \mathrm{E}(g-i)$. This process is illustrated in Fig. \ref{Fig:dust_cor}, where in the upper panel we show the {\it g-i} colour map for NGC 4753 (with the dust filaments prominently visible) and in the lower panel we show the robust linear fit to the colour profile.

While the extinction correction determined by this method is not perfect, it improves the MGE fit to the photometry in the majority of galaxies with significant amounts of dust. An example of this improvement is shown in Fig. \ref{Fig:dust_example}, for NGC2685. The upper panel shows the r-band image and MGE model fit without dust correction, the lower panel shows the MGE model fit to the dust-corrected image. In the two most extreme cases of the galaxies whose photometry was significantly affected by dust, NGC4710 and NGC5866, this procedure failed to provide a reasonable correction for the r-band photometry. For these galaxies we masked the dust-obscured areas from the MGE fitting procedure.

\subsection{Multi Gaussian Expansion modelling}

Using the MGE modelling technique of \citet{Emsellem:1994}, we constructed models of the surface brightness of each galaxy in the sample. We used the procedure of \citet{Cappellari:2002b} as the basis for our technique but embedded it into our photometric pipeline to facilitate dealing with a much larger sample of galaxies. This involved automating the process as much as possible. The position angle (PA) and centre of each galaxy was determined from the weighted second-moments of the surface brightness above a given level using the $find\_galaxy$ {\sc IDL} routine\footnote{Available as part of the MGE package of Cappellari (2002)}. The level was chosen to reflect the global behaviour of the galaxy and to avoid the effects of bars and other non-axisymmetric features in the inner regions. Foreground stars and nearby galaxies were masked by hand where necessary.

The MGE fits to the r-band photometry were performed by keeping the PA of the Gaussians constant in order to produce an axisymmetric MGE model to be used with the JAM modelling technique described below. Each MGE model was convolved with a Gaussian point spread function with $\sigma = 1.0$ (consistent to the typical seeing of the SDSS observations), before comparison to the observations and selection of the best-fitting MGE parameters. The resulting MGE models are all corrected for galactic extinction following \citet{Schlegel:1998}, as given by the NASA/IPAC Extragalactic Database (NED). The MGE models were converted to a surface density in solar units in the SDSS magnitude system using an $r$-band solar magnitude, M$_{r,\odot} = 4.64$ \citep{Blanton:2007}. 

\subsubsection*{Bars and other non-axisymmetric features}

The MGEs were fitted in such a way as to avoid following bars or other non-axisymmetric features in order to reflect the underlying mass distribution of each galaxy. In our fully automated procedure this was achieved by constraining the maximum ($q_{max}$) and minimum ($q_{min}$) axial ratios of the individual Gaussians contributing to the fit. Successive fits were carried out, using a narrower range of axial ratios in each each subsequent fit until the mean absolute deviation of the model at a given $q_{min}, q_{max}$ increased by $>10$ per cent over the previous step. This simple prescription proved effective in reproducing the surface brightness of the majority of galaxies at all radii while avoiding the effects of bars and other non-axisymmetric features. Only in a small number of cases of very strong bars was this prescription ineffective. In these cases the MGE fit was carried out by hand. Examples of the MGE models of several barred galaxies are shown in Fig. \ref{fig:mge_bars}. A detailed analysis of the recovery of M/L, $\beta_z$ and $i$ for barred objects is presented in \citet[hereafter Paper XII]{Lablanche:2012}, based on N-body simulations of axisymmetric and barred galaxies. There we find that the M/L recovered from the JAM modelling is typically accurate to within a few per cent, but for face-on objects can differ from the true value by up to 15 per cent.

\subsection{Jeans Anisotropic MGE (JAM) Modelling}
The dynamical models used in this paper were presented in \citet{Cappellari:2012a} and described in more detail in Paper XIX. The MGE models described in this work are used as input for the JAM modelling method, which calculates a prediction of the line-of-sight second velocity moments $\langle v^2_{los} \rangle$ for a given set of model parameters and fits this to the observed $V_{rms}$ \citep{Cappellari:2008}. We make use of the simplest set of models from \citet{Cappellari:2012a}, model {\bf (A)}. These are self-consistent axisymmetric JAM models, in which the dark matter is assumed to be proportional to the stellar mass.

The potential $\Phi$ is calculated as in \citet{Emsellem:1994} and V$_\mathrm{esc}$ is simply related to this by V$_\mathrm{esc} =  \sqrt{2|\Phi(R,z)|}$. In order to deproject our 3-dimensional potential to compare with our SAURON index maps we assume that V$_\mathrm{esc}$ is related to the indices by a power law relation of the form: Index $\propto \mathrm{V}_\mathrm{esc}^{\gamma}$. With this assumption we can extract the luminosity-weighted average V$_\mathrm{esc,p}$ of the local V$_\mathrm{esc}$ along the line-of-sight and produce V$_\mathrm{esc}$ maps across the entire SAURON field (see S09 for details). We emphasise that the quantity we derive here and describe as the local V$_\mathrm{esc}$ is essentially derived from the stellar surface brightness with a global scaling determined by the dynamical modelling. Our V$_\mathrm{esc}$ is likely close to the true V$_\mathrm{esc}$ in our galaxies, given that stars likely dominate the centre of ETGs (Paper XIX). However the quantity we compute does not intend to represent the true V$_\mathrm{esc}$, which is empirically uncertain due to the possible contribution of dark matter. Our aim is to find a purely empirical relation between quantities that can be robustly measured on real galaxies from photometry and integral-field spectroscopy. What matters is that our procedure can be replicated on simulated galaxies, to test whether they behave like real galaxies. When comparing our V$_\mathrm{esc}$ to simulated galaxies one should determine the local V$_\mathrm{esc}$ from the stellar particles alone.

\subsection{Tests of MGE Photometry}
\label{Sec:MGE_phot_testing}
While MGE models of galaxy photometry have generally only been used as input to dynamical models, we would like to emphasise that these models provide a robust and flexible  description of galaxy surface brightness profiles. This is true across a wide range of morphological types and is not based on any assumption about galaxy structure. In this section we compare photometric quantities (both total magnitudes and surface brightness profiles) derived from the MGE models described in Section \ref{Sec:Method} to studies in the literature using standard techniques to derive their photometric quantities. Total magnitudes were derived from the MGE models using the following expression:
\begin{equation}
L_{Tot} = \sum_{j=1}^N 2\pi I^\prime_j \sigma_j^2 q^\prime_j.
\end{equation}
where the sum is over the $N$ component Gaussians and $I_j^\prime$, $\sigma_j$ and $q_j^\prime$ are the peak intensity, dispersion and axial ratio of the $j$th Gaussian component. The primes denote quantities as measured on the sky plane \citep[see][for details]{Cappellari:2002}. The surface brightness at a given position on the sky is given by:
\begin{equation}
\Sigma (x^\prime,y^\prime) =  \sum_{j=1}^N I^\prime_j \exp \left [ { - \frac{1}{2\sigma^2_j} } \left ( x^{\prime 2} + \frac{y^{\prime 2}}{q_j^{\prime 2}}\right ) \right ].
\end{equation}

\begin{figure}
\includegraphics[width=3.25in]{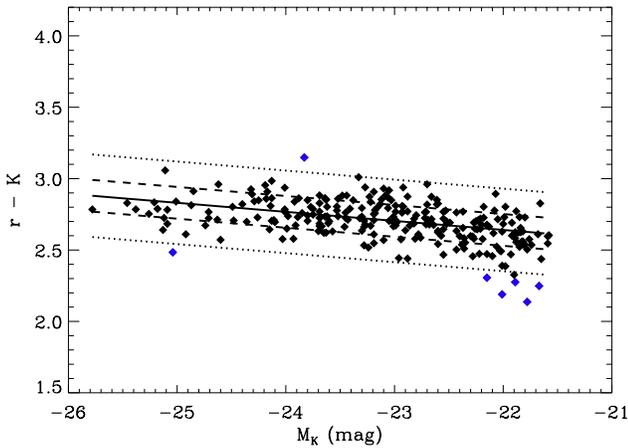}
\caption{2MASS absolute {\it K}-band magnitude vs. r-K colour derived from our {\it r}-band MGE models. The solid black line shows the best-fitting relation, with the dashed and dotted lines indicating the 1 and 2.6 $\sigma$ bounds. The blue points indicate $>2.6 \sigma$ outliers from the relation which were not included in the fit.}
\label{fig:2MASS_MGE_comp}
\end{figure}

The only previous data set in the literature that contains photometry for the entire ATLAS$^\mathrm{3D}$ sample is the 2MASS Extended Source Catalog \citep{Jarrett:2000}. This allows us to test the accuracy and detect possible outliers over the full sample. In Fig. \ref{fig:2MASS_MGE_comp} we compare the absolute {\it r}-band magnitude derived from our MGE models (and distance-corrected using the distances in Paper I) to the absolute {\it K}-band magnitude from 2MASS. Overplotted is the best-fitting relation derived from an linear fit to the datapoints that minimises the absolute residuals after rejecting $2.6 \sigma$ outliers (enclosing 99\% of values for a Gaussian distribution). We measure an observed rms scatter of 0.11 magnitudes. This is a firm upper limit to the true error and intrinsic colour differences must be significant.

\begin{figure}
\includegraphics[width=3.25in]{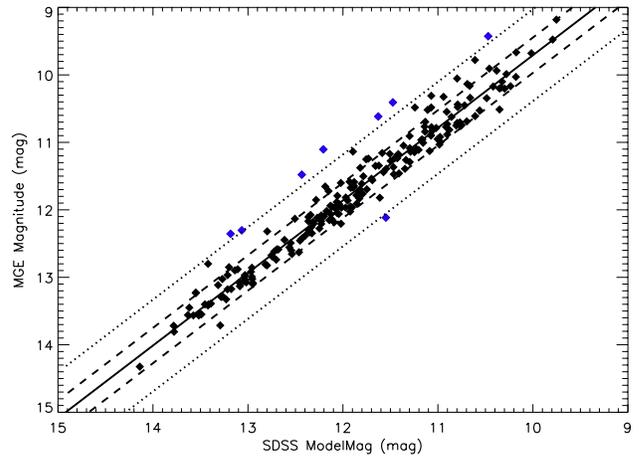}
\caption{Comparison between total {\it r}-band magnitudes derived from our MGE models and from the SDSS DR8 photometric object catalogue. Both magnitudes were derived from the same imaging data. The solid line shows the best-fitting relation, with the dashed and dotted lines indicating the 1 and 2.6 $\sigma$ levels. Blue points were excluded from the fit.}
\label{fig:SDSS_MGE_mag_comp}
\end{figure}

The comparison with 2MASS magnitudes is between two datasets utilising different imaging data in very different bands. We also compared our derived apparent {\it r}-band magnitudes to apparent {\it r}-band {\it ModelMags} taken from the SDSS DR8 photometric object catalogue \citep{Aihara:2011}. We restrict the comparison to galaxies which have no warning flags in their SDSS photometry, yielding a comparison of 214 galaxies. The comparison is shown in Fig. \ref{fig:SDSS_MGE_mag_comp}. For total magnitudes derived from the same images the scatter is significant, after rejecting 2.6 $\sigma$ outliers the rms scatter is 0.17 magnitudes. There is a systematic trend for the magnitude of the largest galaxies (R$_e > 30"$) to be underestimated by the SDSS {\it ModelMags}, with respect to our MGE based {\it r}-band magnitudes. The offset is $\sim 0.5$ magnitudes. This trend was partially (but not completely) addressed in the SDSS DR8 photometric pipeline \citep{Aihara:2011}. \citet{Blanton:2011} introduce an improved sky subtraction method that solves this issue, however it has only been applied to SDSS mosaic images and not the photometric catalogue. A detailed inspection of the SDSS DR8 photometric catalogue {\it ModelMags} for ATLAS$^\mathrm{3D}$ galaxies reveals several further problems. Some galaxies are significantly affected by confusion, where multiple PhotoObjects are identified for a single galaxy (e.g. NGC4710). Other ATLAS$^\mathrm{3D}$ galaxies are entirely missing from the photometric catalogue, despite lying within the survey area (e.g. NGC4486). Finally, a number of magnitudes in individual bands show significant errors (for example the ({\it g-i}) colour of -5.8 for NGC3384). These issues appear most significant for the {\it ModelMag} values, however the {\it PetrosianMag} values also exhibit similar problems.

To place the observed rms scatters of 0.11 and 0.17 magnitudes (from the comparison to 2MASS $K-$band and SDSS $r-$band magnitudes respectively) in context, we compare it to the set of photometric comparisons made by \citet{Chen:2010}. Among various comparisons they present from the literature, the most accurate is the one between their determination of total magnitudes $g_ T$ and the one by \citet{Janz:2008}, based on the same $g$-band SDSS images and basically the same curve-of-growth method. They find an rms scatter in the two determinations of 0.09 mag (0.036 dex). This scatter is likely to represent the best of what can be achieved, when only minimal differences in the methods play a role. When comparing their $g_T$ values to those from the state-of-the-art ACSVCS survey \citep{Ferrarese:2006} in the same band, they find that the scatter increases to 0.19 mag (0.076 dex), and the same applies for their comparison against the VCC $g$-band magnitudes of \citet{Binggeli:1984}. Contrary to those comparison, we are using two very different photometric bands ($r$ and $K_s$), where intrinsic colour differences must be significant. This is confirmed by the fact that the $g-r$ colour-magnitude relation, for 43 objects in common, between our r-band MGE and \citet{Chen:2010} $g$-band curve-of-growth determinations (their $g_A$ values) has an rms scatter of 0.062 dex, which would indicate an error of 11\% in each of the two determinations. We conclude that our MGE photometry is as accurate as that of other state-of-the-art surveys. Moreover we only detect two possible outliers, while there is no evidence for possible calibration issues for the entire sample. We therefore assign an error of 10\% in our M$_r$.

\begin{figure*}
\includegraphics[height=2.95in]{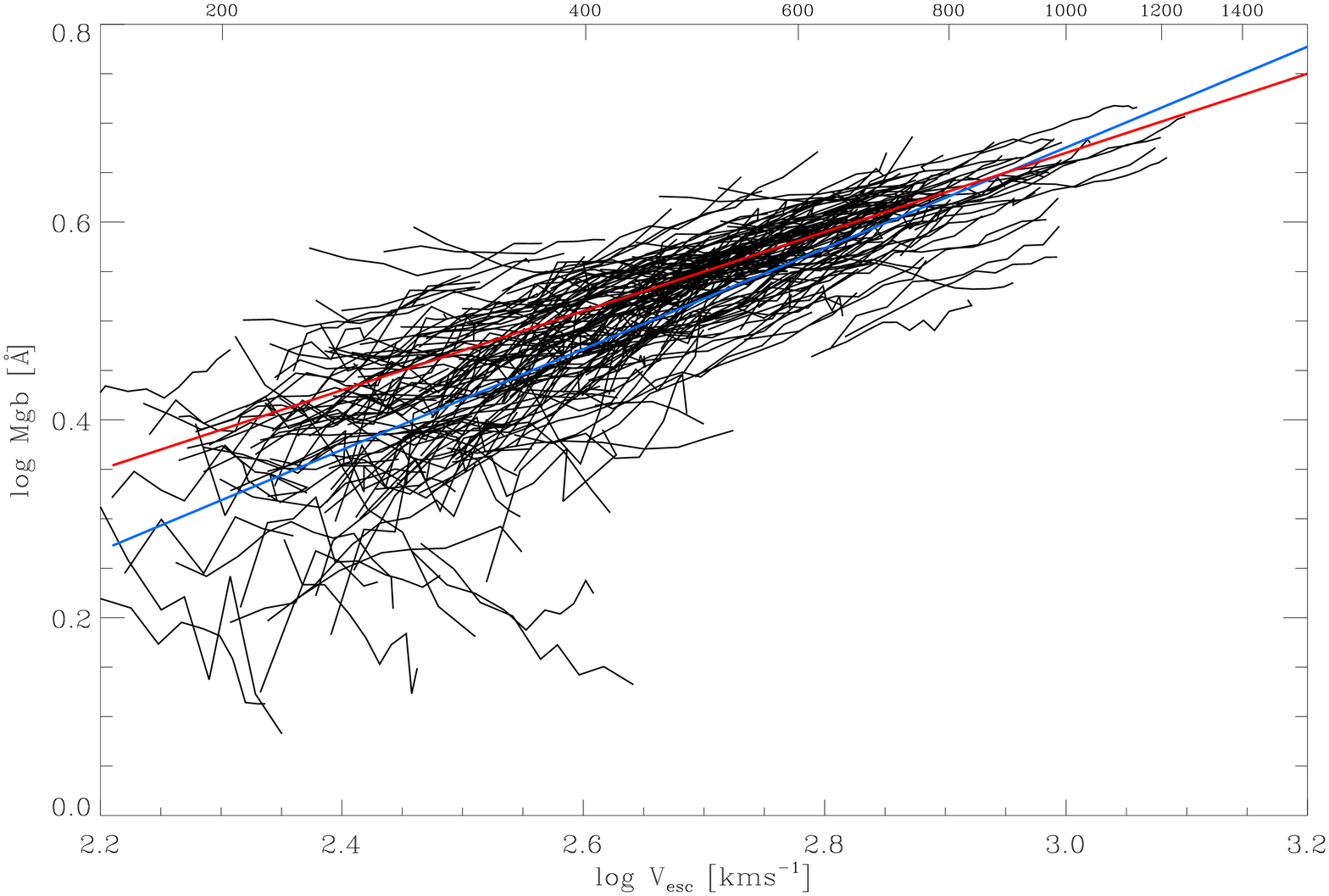}
\includegraphics[height=2.95in]{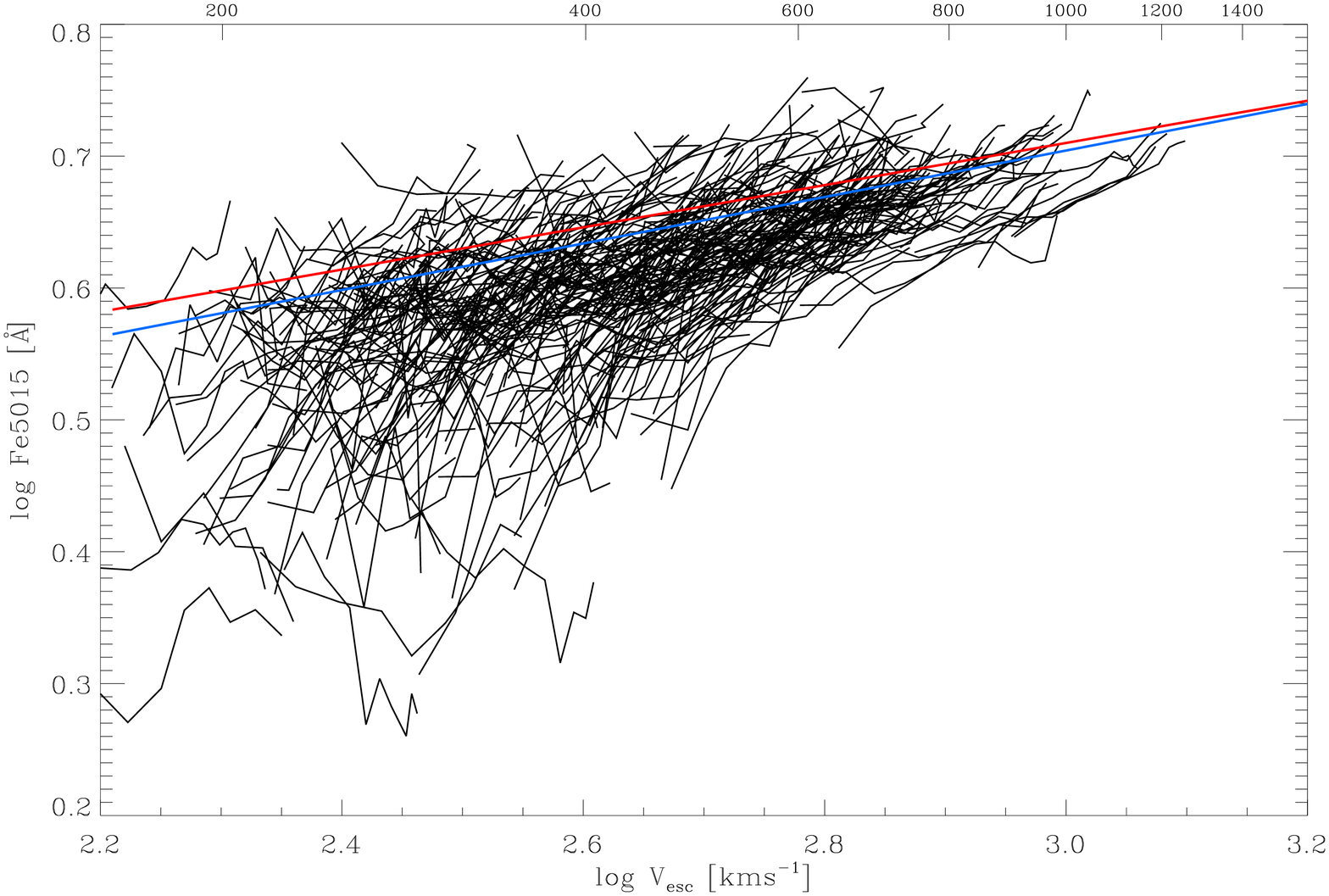}
\includegraphics[height=2.95in]{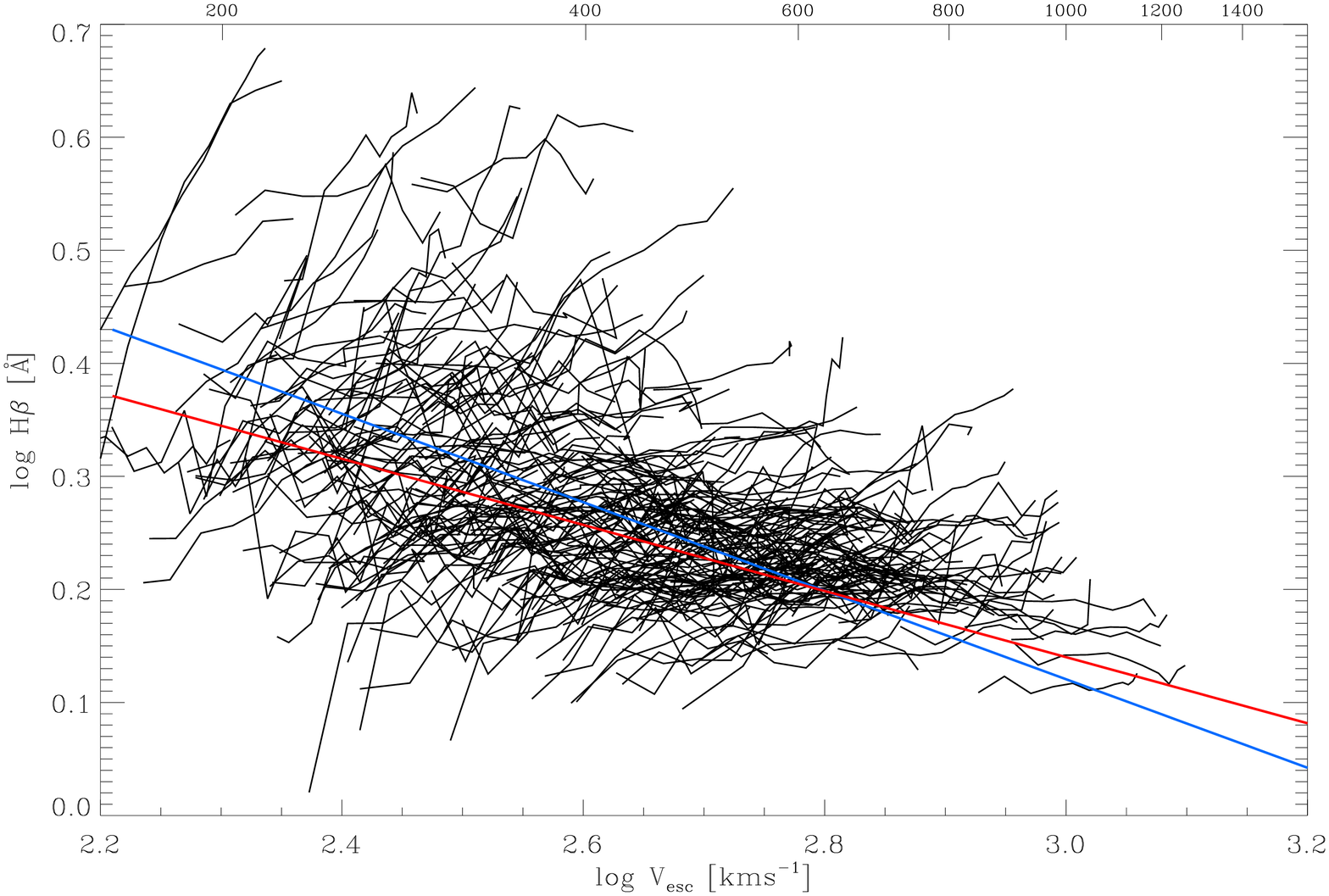}
\caption{The Index-V$_\mathrm{esc}$ relations for the whole sample. Blue line is a fit to R$_\mathrm{e}$ circular aperture values for all galaxies. Red line is a fit to the same aperture but for only those galaxies with V$_\mathrm{esc,Re} > 400\ \mathrm{kms}^{-1}$. With all three indices the relation is much tighter above V$_\mathrm{esc} > 400$ kms$^{-1}$ (log V$_\mathrm{esc} > 2.6$). Below this value the scatter is significantly increased, largely due to a population of galaxies with unusually low Mg$\,b$ values in their central regions.}
\label{Fig:Ind-Vesc}
\end{figure*}

Finally, to demonstrate that we are able to not only derive total magnitudes from our MGE models but full surface brightness profiles we compare the major-axis surface brightness profiles derived from our MGE models to the major-axis surface photometry of \citet{Kormendy:2009}. There are 23 galaxies in common between their morphologically selected sample of Virgo cluster galaxies and the ATLAS$^\mathrm{3D}$ sample. Surface brightness profiles for all 23 galaxies are shown in Appendix B. We restrict the comparison to the region over which our MGE models are reliable; 1.5" $< r < 2 \mathrm{max}(\sigma_\mathrm{MGE})$, where max($\sigma_\mathrm{MGE}$) is the $\sigma$ (in arcseconds) of the largest Gaussian in the MGE model for each galaxy. To account for the difference between the bands we shift our surface brightness profiles by a constant value such that the mean surface brightness over the region of comparison matches that of the \citet{Kormendy:2009} profiles. The agreement between the profiles is excellent, with mean residuals (after applying the constant shift) of $\pm 0.04$ magnitudes. We accurately reproduce the surface brightness profiles for the full range of ellipticities in the \citet{Kormendy:2009} sample. In summary, our photometric quantities, both total magnitudes and surface brightness profiles, derived from the MGE modelling are of excellent quality. The full set of calibrated MGE models will be made available from our website\footnote{http://purl.org/atlas3d} at the conclusion of our project. An example MGE model is given in Table \ref{tab:mge_example} -- the columns (described in the table caption) are the same as those available from our website. The MGE models for the full sample (overplotted as contours on the $r-$band imaging) are shown in Appendix A.

\begin{table}
\caption{MGE parameters for the deconvolved {\it r}-band surface brightness for NGC4570, as an example of the ascii MGE parameter files available to download from our website.}
\label{tab:mge_example}
\begin{center}
\begin{tabular}{c c c}
\hline
$\log I^\prime_j$ & $\log \sigma_j$ & q$^\prime_j$ \\
(L$_{\sun}$ pc$^{-2}$) & (arcsec) & \\
\hline
\hline
45198.69 & 0.318 & 0.72 \\
9747.41 & 1.142 & 0.80 \\
3741.94 & 2.743 & 0.65 \\
1947.36 & 3.796 & 0.72 \\
935.15 & 9.019 & 0.54 \\
45.97 & 13.037 & 0.80 \\
495.96 & 19.671 & 0.20 \\
255.63 & 36.549 & 0.23 \\
37.80 & 60.620 & 0.28 \\
6.07 & 60.620 & 0.80 \\
\hline
\end{tabular}
\end{center}
Notes: Column (1): Total intensity of the $j$th Gaussian component. Column (2): Dispersion of the $j$th Gaussian component. Column (3): Axial ratio of the $j$th Gaussian component.
\end{table}

\section{Results}
\label{Sec:Results}
Using the SAURON line-strength maps and the V$_\mathrm{esc}$ maps described above we constructed profiles by taking the average within elliptical annuli. The ellipticity chosen was a global ellipticity measured using the weighted second moments of the surface brightness and was taken from Paper II. S09 found that this choice of ellipticity minimised the errors on each point within a galaxy's profile, and that their results did not depend on the choice of aperture (circular, fixed or varying $\epsilon$ or major axis). We find a stronger trend of increased errors for non-optimal choices of aperture (likely due to the lower signal-to-noise (S/N) of the ATLAS$^\mathrm{3D}$ IFU data compared to the SAURON survey), however in general we confirm this lack of dependence on aperture. We also find that the relations presented in this work are robust against the choice of aperture. The error on each point for the line-strengths is the rms sum of the measurement errors and the rms scatter within each annulus. Foreground stars and bad bins due to low S/N were masked in the line-strength maps before extracting the profiles.

\begin{figure}
\includegraphics[width=3in]{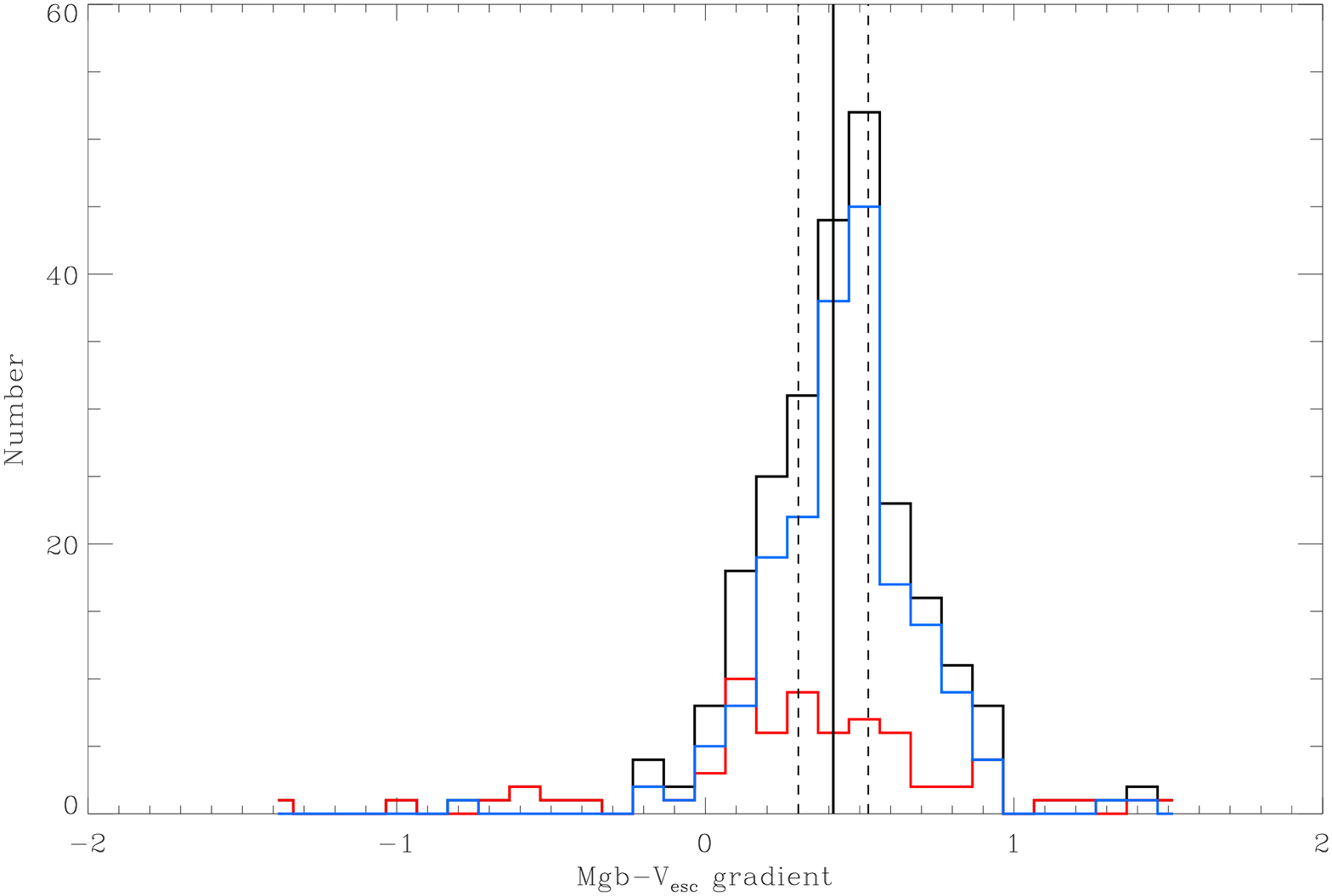}
\includegraphics[width=3in]{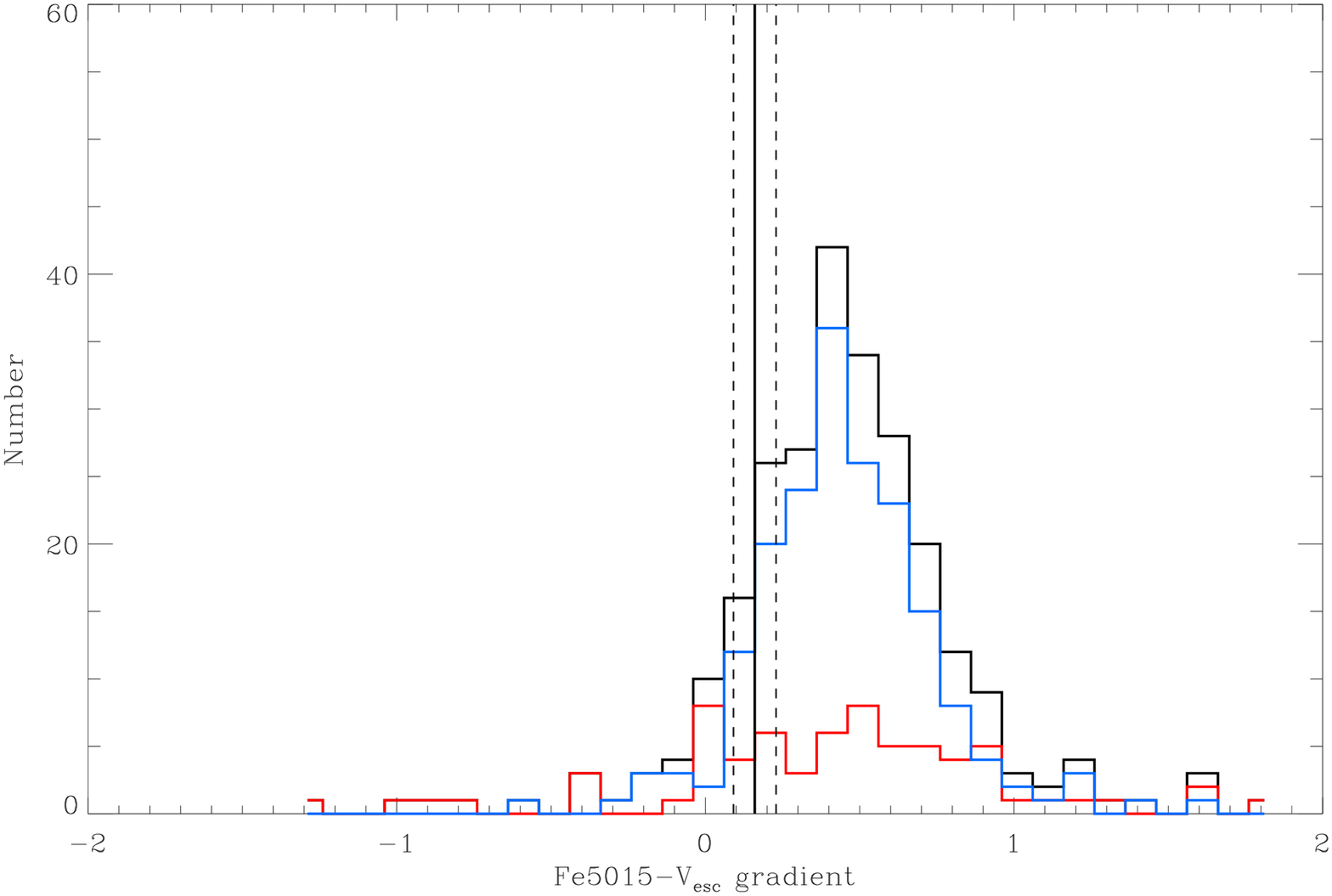}
\includegraphics[width=3in]{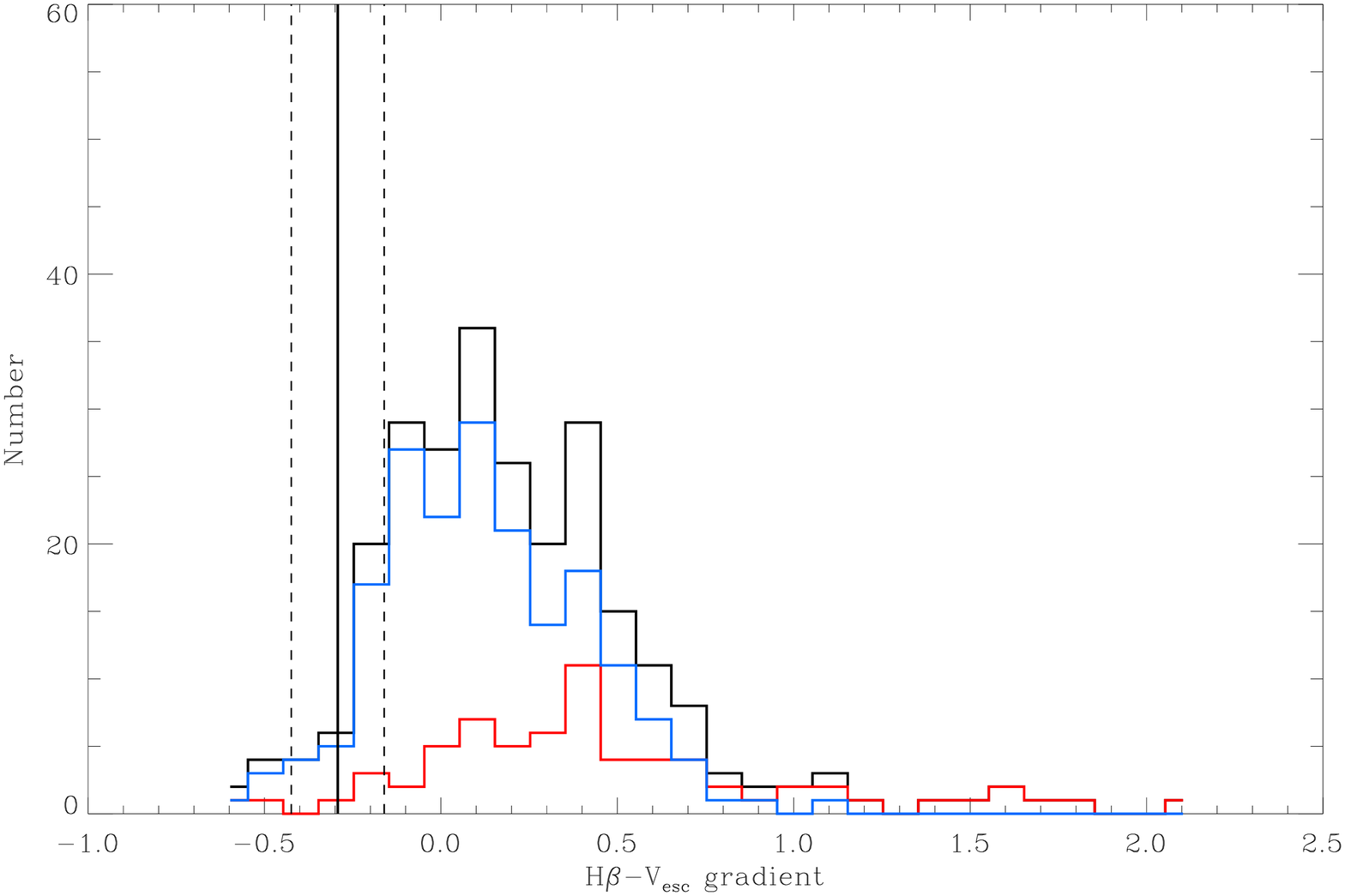}
\caption{Each histogram shows the Index-V$_\mathrm{esc}$ gradients of the individual galaxies determined from a robust linear fit to each galaxy profile (black histogram). The red and blue histograms show the same data for galaxies with V$_\mathrm{esc,Re}$ greater and less than 400 kms$^{-1}$ respectively. The solid vertical line shows the global gradient determined from a linear fit to the central R$_e$ aperture values for the whole sample, with the dashed lines showing the $3\sigma$ errors on this global gradient. For the Mg$\,b$ gradient histogram, the peak of the distribution for individual galaxy gradients coincides exactly with the global gradient, though there is significant scatter in the individual galaxy gradients. For the Fe5015 and H$\beta$ gradient histograms there is a significant offset between the peak of the distribution of individual galaxy gradients and the global gradient.}
\label{Fig:mgb-hist}
\end{figure}

\subsection{The Index-V$_\mathrm{esc}$ relations}
\label{Sec:Index_Vesc}
The Index-V$_\mathrm{esc}$ relations are shown in Fig. \ref{Fig:Ind-Vesc}, with the profile for each galaxy shown as a black line. The blue line shows a fit to the central R$_\mathrm{e}$ circular aperture values for the whole sample. The red line shows a fit only to the galaxies with V$_\mathrm{esc,Re} > 400\ \mathrm{kms}^{-1}$ ($\log$ V$_\mathrm{esc} > 2.6$). All three indices show a tight correlation with V$_\mathrm{esc}$, though the scatter increases at lower V$_\mathrm{esc}$. This is particularly evident in the Mg$\,b$-V$_\mathrm{esc}$ relation, where a clear break occurs at V$_\mathrm{esc} \sim 400\ \mathrm{kms}^{-1}$, though the effect is present with all three indices. This increase in scatter is primarily due to a population of galaxies having negative Mg$\,b$- and Fe5015-V$_\mathrm{esc}$ gradients. In the H$\beta$-V$_\mathrm{esc}$ diagram these galaxies are seen to have steeply positive gradients, compared to the typically flat profiles of the bulk of galaxies in the sample.

The behaviour of the individual galaxy gradients can be better seen in Fig. \ref{Fig:mgb-hist}. The histogram shows the individual galaxy gradients determined from robust linear fits to each galaxy profile. The solid vertical line shows the global gradient, determined from a robust linear fit to the line-strengths and V$_\mathrm{esc}$ as measured within a circular R$_e$ aperture. In the case of Mg$\,b$ the individual galaxy gradients are consistent with the global gradient, though there is significant scatter in the distribution about the global value. This is not true for Fe5015 and H$\beta$, where the peak of the distribution of individual galaxy gradients is significantly offset from the global gradient. When we divide the sample into high (V$_\mathrm{esc} > 400\ \mathrm{kms}^{-1}$) and low V$_\mathrm{esc}$ samples we find that the low V$_\mathrm{esc}$ sample is biased to negative gradients. In the high V$_\mathrm{esc}$ sample we find only 2.7 per cent of galaxies have negative gradients, however this rises to 15.9 per cent in the low V$_\mathrm{esc}$ sample. If we remove these negative gradient galaxies from the sample we no longer observe an increase in scatter at low V$_\mathrm{esc}$. For the remainder of this section we exclude the negative gradient galaxies from our analysis, however they will be discussed further in Section \ref{sec:RSF}. The Index - V$_\mathrm{esc}$ relations determined from galaxies with V$_\mathrm{esc} (\mathrm{R}_e) > 400\ \mathrm{kms}^{-1}$ (red lines in Fig. \ref{Fig:Ind-Vesc}) are given below:
\begin{eqnarray*}
\log \mathrm{Mg\,b} = (0.40 \pm 0.02) \log \mathrm{V}_\mathrm{esc}+(-0.53 \pm 0.04); \sigma = 0.04\\
\log \mathrm{Fe5015} = (0.16 \pm 0.02) \log \mathrm{V}_\mathrm{esc}+(0.23 \pm 0.03); \sigma = 0.03\\
\log \mathrm{H}\beta = (-0.29 \pm 0.04) \log \mathrm{V}_\mathrm{esc}+(1.02 \pm 0.06); \sigma = 0.06
\end{eqnarray*}
Here, $\sigma$ indicates the robust rms deviation orthogonal to the fitted linear relation. The Mg$\,b$ and Fe5015 relations are $\sim 3 \sigma$ steeper than those found in S09, with improved formal uncertainties. The relation for H$\beta$ is consistent with the S09 result within the uncertainties. This is because the ATLAS$^\mathrm{3D}$ sample contains a large number of lower mass ETGs, a population not well sampled by the SAURON survey.

\begin{figure*}
\includegraphics[width=6.5in]{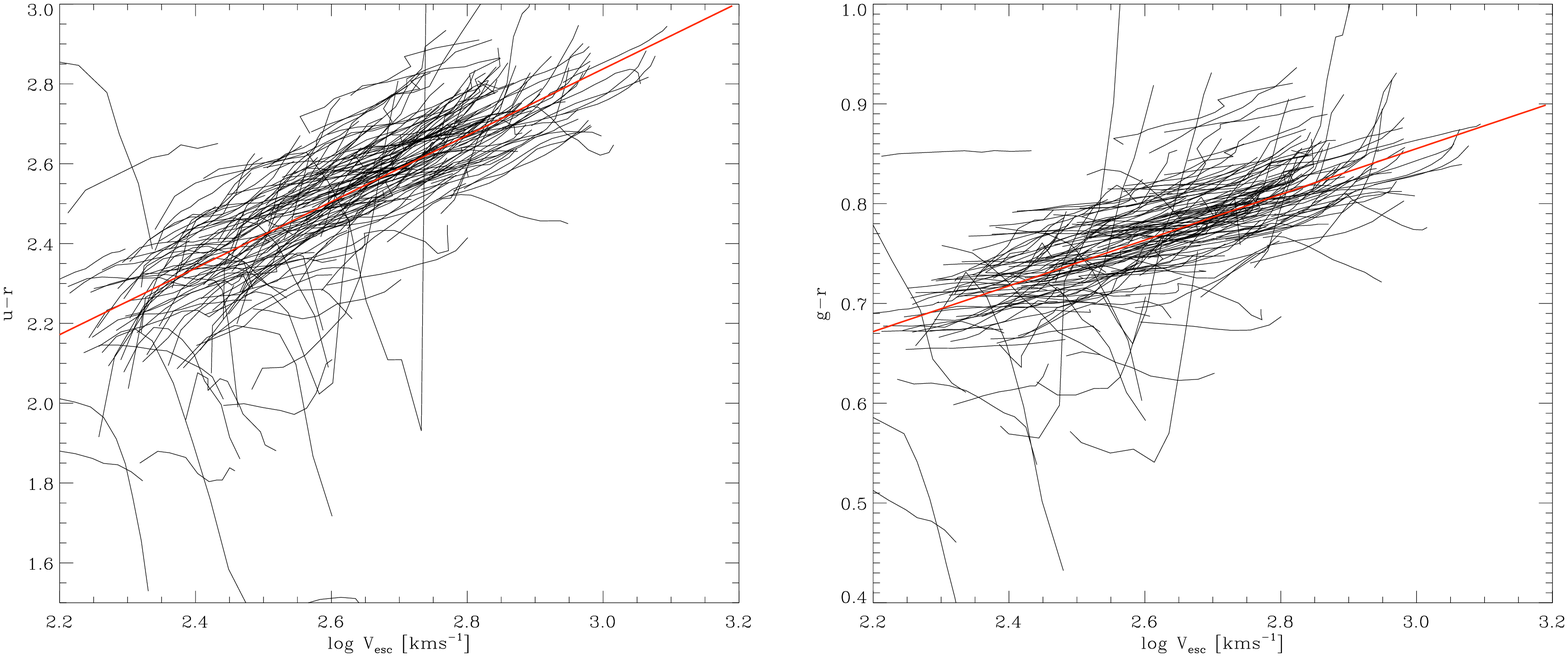}
\includegraphics[width=3.25in]{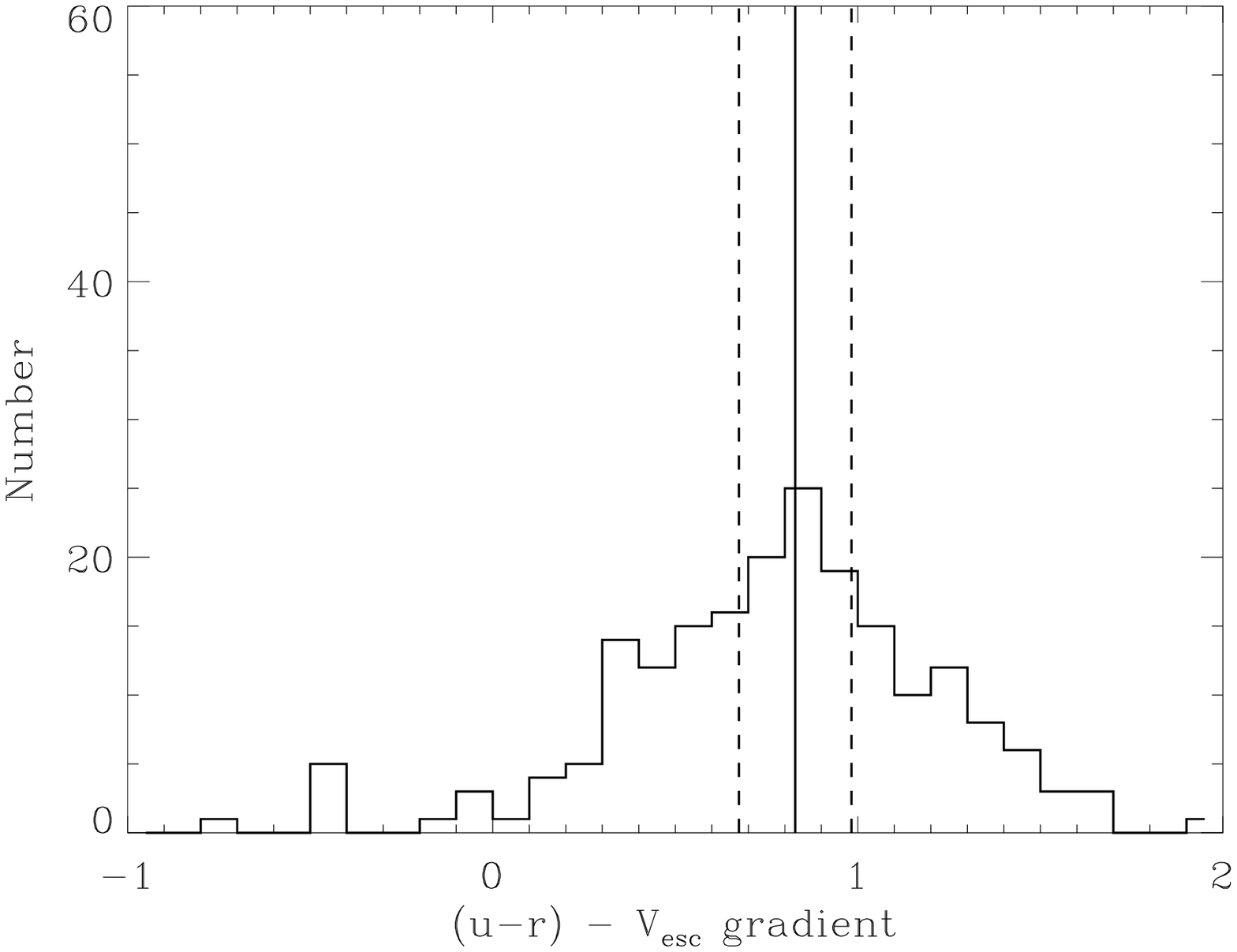}
\includegraphics[width=3.25in]{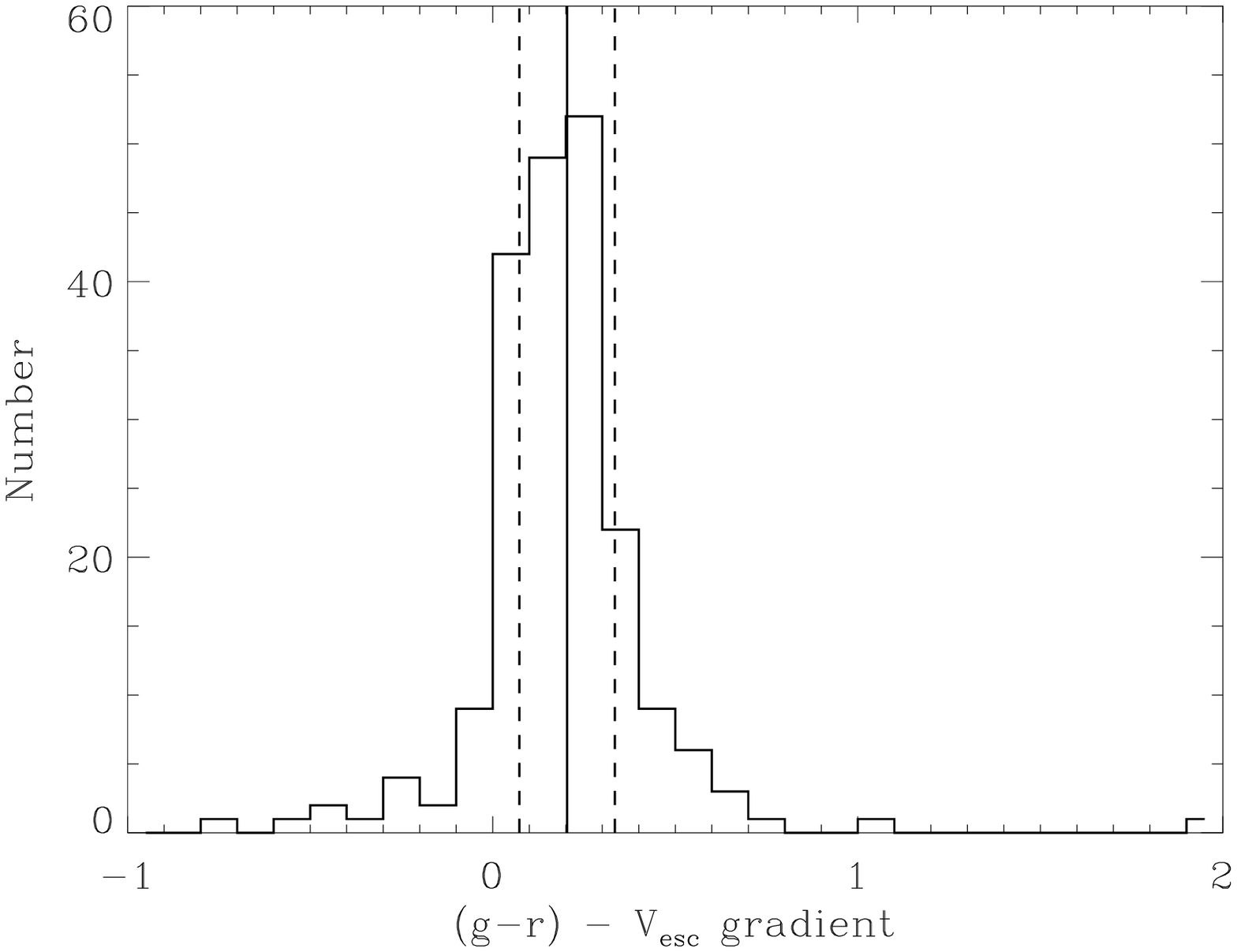}
\caption{Upper panels: The {\it u-r} and {\it g-r} - V$_\mathrm{esc}$ relations. As before individual galaxy profiles are shown as the black solid lines with the global relation, again determined from R$_e$ aperture values, plotted in red. Both relations show the same local and global behaviour as the Mg$\,b$-V$_\mathrm{esc}$ relation. We again see an increase in scatter below V$_\mathrm{esc} \sim 400 \mathrm{kms}^{-1}$ due to a population of negative gradient, `blue core' galaxies. Lower panels: Histogram of the individual galaxy gradients for the {\it g-r} - and {\it u-r} - V$_\mathrm{esc}$ relation, determined by robust linear fits to each galaxy profile. The global relation is shown as the solid vertical line with the 3 $\sigma$ errors shown as vertical dashed lines. The peak of the distribution of individual galaxy gradients coincides with the global gradient, and the width of the distribution is consistent with the error on the global gradient. The {\it u-r}-V$_\mathrm{esc}$ relation also exhibits the same local and global connection.}
\label{Fig:Col-Vesc}
\end{figure*}

\subsection{The Colour-V$_\mathrm{esc}$ relations}
\label{Sec:Colour}
\citet{Franx:1990} originally studied the Colour-V$_\mathrm{esc}$ relations, but recent work has focused exclusively on the correlation with line-strengths. With the ATLAS$^\mathrm{3D}$ sample we have complete {\it ugriz} coverage of the sample, allowing us to re-examine their original relation. Colour profiles were extracted in the same fashion as the line-strength and V$_\mathrm{esc}$ profiles; median colours were measured within elliptical annuli on the sky-subtracted, dust-corrected and masked SDSS and INT images. In Fig. \ref{Fig:Col-Vesc} we show the {\it g-r} and {\it u-r} - V$_\mathrm{esc}$ relations. We selected these colours as the {\it g-} and {\it r-} bands have the highest S/N and hence show the tightest relation, while the {\it u-}band is more sensitive to young populations. We find the same behaviour as in the Mg$\,b$ - V$_\mathrm{esc}$ relation; a tight correlation with increasing scatter appearing below V$_\mathrm{esc} \sim\ 400 \mathrm{kms}^{-1}$. This increase in scatter is more evident with the {\it u-r} colour than with {\it g-r}. We also find the same local and global behaviour described above (see lower panel of Fig. \ref{Fig:Col-Vesc}), in that individual galaxy gradients are consistent with the global relation determined from central values. Fitting to the R$_e$ aperture values for those galaxies with V$_\mathrm{esc,Re} > 400\ \mathrm{kms}^{-1}$ we find that the colours are given by:
\begin{eqnarray}
g-r = (0.26 \pm 0.03) \log \mathrm{V}_\mathrm{esc} + (0.08 \pm 0.05); \sigma = 0.04\\
u-r = (0.80 \pm 0.08) \log \mathrm{V}_\mathrm{esc} + (0.47 \pm 0.14); \sigma = 0.12
\end{eqnarray}
Below V$_\mathrm{esc} \sim 400\ \mathrm{kms}^{-1}$ we again find a significant increase in the fraction of negative gradient galaxies; 3.9 per cent at high V$_\mathrm{esc}$ increasing to 22.0 per cent below $\sim 400\ \mathrm{kms}^{-1}$.

\subsection{V$_\mathrm{esc}$ and the globally averaged SSP parameters}
\label{Sec:SSP}
Line strengths and colours can be indicative of trends in the stellar populations of our target galaxies but the information we are really interested in is the star formation histories of these galaxies. These histories will typically be quite complex, involving multiple periods of star formation over the course of a galaxy's lifetime with different durations, contributions to the total stellar mass and chemical compositions. As an initial test we can simplify this scenario by treating all our galaxies as single stellar populations (SSPs), i.e. all the stars were formed in a single, instantaneous burst, and we can characterise these SSPs by three parameters, the age, metallicity and enhancement of alpha elements. While the SSP scenario is a simple one it can provide a useful starting point for studying early-type galaxies, where a significant fraction of their stars were typically formed at high redshift \citep[see e.g.][]{Thomas:2005}. In these cases the approximation of a single star-formation episode is a reasonable one. In particular it is important not to place too much emphasis on the precise values of the SSP parameters of individual objects, but the SSP models can be used to study the differences between objects and any trends arising from this. 

We adopt the single stellar population models of \citet{Schiavon:2007} to interpret our global line-strength measurements of Mg$\,b$, Fe5015 and H$\beta$ in terms of the SSP parameters, age (t), metallicity ([Z/H]) and alpha enhancement ([$\alpha$/Fe]). We follow the approach previously described by \citet{Kuntschner:2010}, where the models of \citet{Schiavon:2007} were found to provide physically realistic SSP-equivalent ages, metallicities and alpha enhancements for the full range of line strength index measurements found in that work. In particular, the \citet{Schiavon:2007} models were found not to saturate in age at high values of the H$\beta$ index. \citet{Kuntschner:2010} conducted a detailed comparison between the predictions of the \citet{Schiavon:2007} models and other competing SSP models, finding their derived parameters did not depend significantly on the choice of SSP model. Further details of our global stellar population modelling are given in McDermid et al. (in preparation). While the determination of each of the three parameters depends on all three indices, some parameters depend more heavily on one index over another. H$\beta$ can be broadly thought of as an age indicator whereas Fe5015 and Mg$\,b$ are more strongly influenced by [Z/H] and [$\alpha$/Fe]. 

\begin{table}
\caption{Principle components analysis}
\label{Tab:PCA}
\begin{center}
\begin{tabular}{c c c c c c c}
\hline
& V$_{esc}$' & Age' & [Z/H]' & [$\alpha$/Fe]' & Eigen- &\% of \\
& & & & & value &variance \\
\hline
PC1 & 0.680 & 0.187 & -0.137 & -0.696 & 1.87 & 47\\
PC2 & 0.350 & -0.609 & -0.644 & 0.304 & 1.38 & 34\\
PC3 & 0.522 & 0.554 & 0.065 & 0.646 & 0.59 & 15\\
PC4 & 0.379 & -0.537 & 0.750 & 0.078 & 0.16 & 4\\
\hline
\end{tabular}
\end{center}
Notes: The primed variables are standardised versions of the corresponding variables with zero mean and unit variance. The coefficients of the principal components are scaled to the variance and sensitive to the range of each variable, in the sense that variables that only vary by a small amount tend to have a large coefficient.
\end{table}

Rather than examining the individual SSP-V$_\mathrm{esc}$ relations (for a discussion of the SSP-V$_\mathrm{esc}$ relations see S09) we consider the four-dimensional parameter space defined by V$_\mathrm{esc}$, age, [Z/H] and [$\alpha$/Fe]. To analyse these data we use the technique of Principal Component Analysis (PCA). A description of the PCA technique and the interpretation of its output can be found in \citet{Faber:1973} and \citet{Francis:1999}.
\begin{figure}
\includegraphics[width=3.25in]{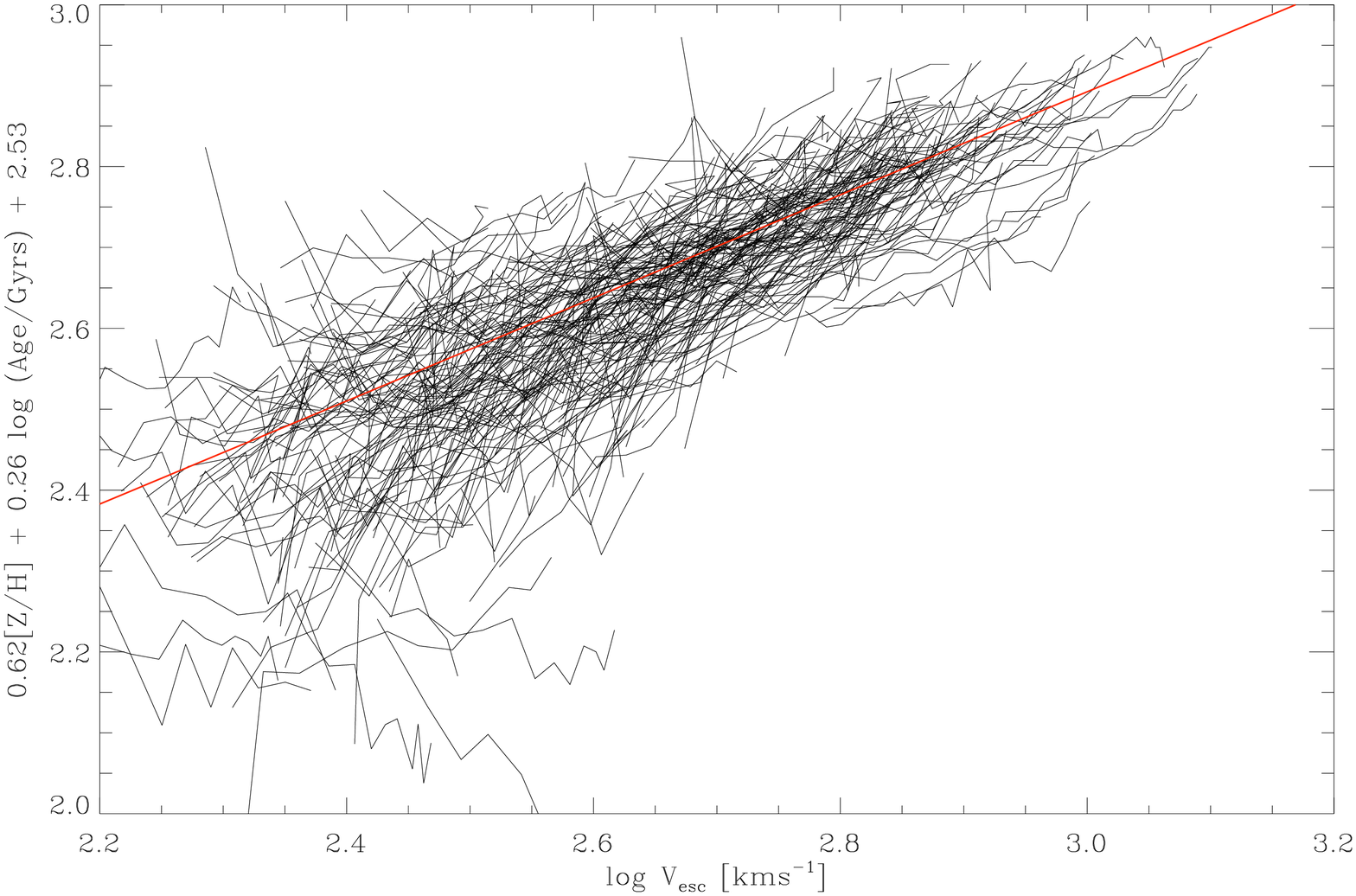}
\includegraphics[width=3.25in]{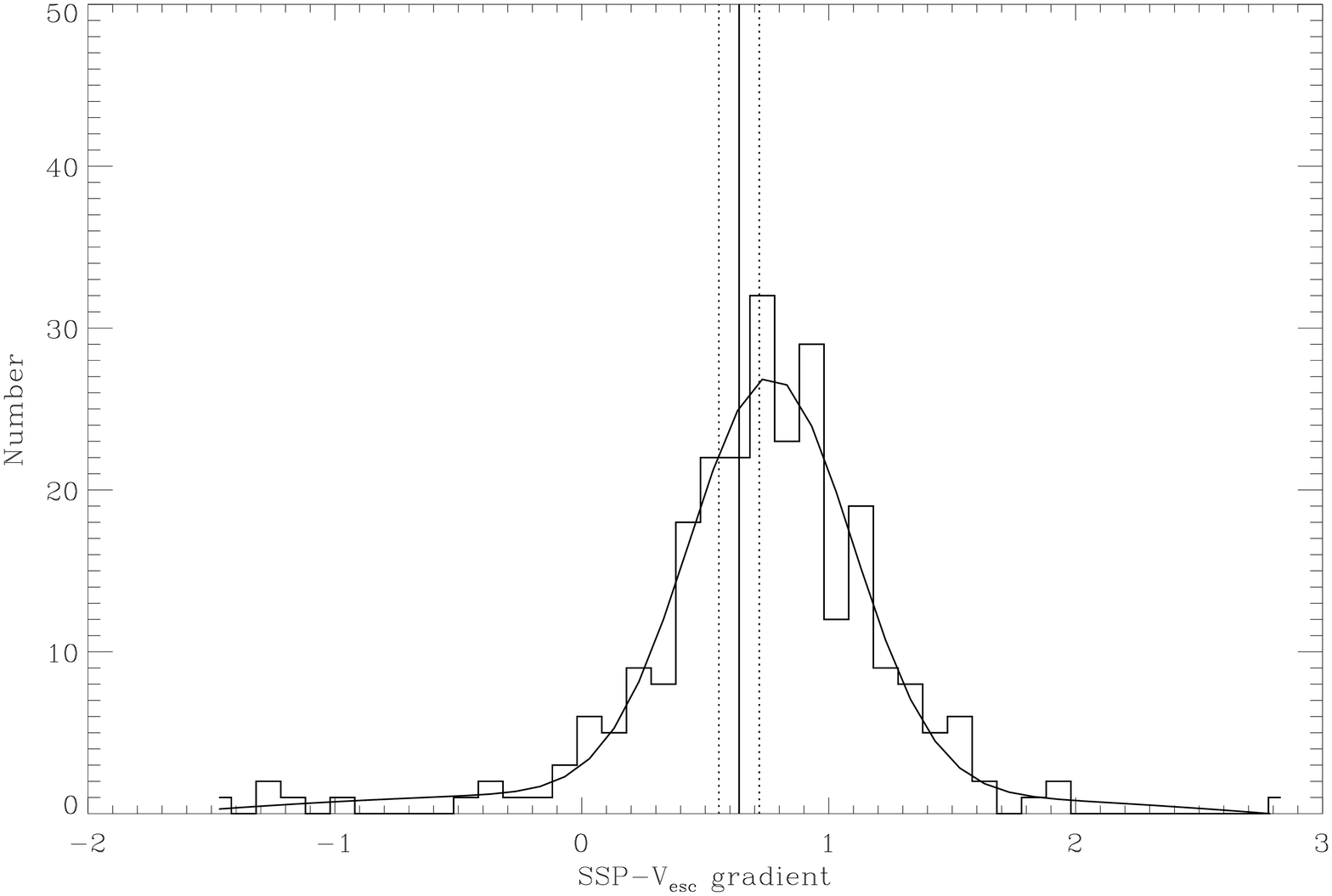}
\caption{Upper panel: The best-fitting SSP - V$_\mathrm{esc}$ relation determined by robustly fitting a plane to the local V$_\mathrm{esc}$, age and [Z/H] of our galaxies. As before, the solid black lines show individual galaxy profiles and the red line is a robust linear fit to the R$_e$ aperture values for the whole sample. Lower panel: The distribution of the individual galaxy gradients. The solid vertical line shows the global gradient determined from the R$_e$ apertures and the dotted lines show the 3-sigma errors on this gradient.}
\label{Fig:SSP}
\end{figure}

Applying PCA to our data we find that $\sim 81$ per cent of the scatter is confined to the first two principal components (Table \ref{Tab:PCA}), indicating that in this four-dimensional space local early-type galaxies are effectively confined to a plane. While this is consistent with the findings in S09 and the weights associated with each physical variable in the corresponding PCs are very similar, we do find a significantly increased scatter associated with the less significant third and fourth principal components -- we find a `thicker' plane than in S09. Part of this effect can likely be attributed to the somewhat poorer quality of the IFU data in the ATLAS$^\mathrm{3D}$ survey compared to the SAURON survey due to i) 1 hour instead of 2 hour exposures and ii) a higher proportion of faint galaxies. Large values of PC3 and PC4 are typically associated with extreme values of age or [$\alpha$/Fe], lying on the boundaries of our SSP model grid. Interestingly, the negative gradient galaxies do not occupy any special part of the parameter space and, in this respect, are indistinguishable from typical early-types.

We take this idea further by attempting to fit a single plane to the local V$_\mathrm{esc}$, age and [Z/H] data points for our full sample. Using a robust, linear plane fit to our data we find that local early-type galaxies can be described by the relation:
\begin{equation}
	\log\left(\frac{\mathrm{V}_{\mathrm{esc}}}{500\mathrm{kms}^{-1}}\right)= 0.62 \left[\frac{\mathrm{Z}}{\mathrm{H}}\right] + 0.26 \log \left(\frac{\mathrm{t}}{\mathrm{Gyrs}}\right) - 0.17
	 \label{Eq:SSP}
\end{equation} 
with $\sigma = 0.07$. This relation is shown in Fig. \ref{Fig:SSP}. This SSP - V$_\mathrm{esc}$ relation is much tighter than any of the individual age, [Z/H] or [$\alpha$/Fe] - V$_\mathrm{esc}$ relations. However, this relation shows an offset between the global gradient ($0.63 \pm 0.03$) and the distribution of the local gradients ($0.75 \pm 0.51$, see the lower panel of Fig. \ref{Fig:SSP}). If we use the relation previously found by S09 we find a somewhat increased scatter of $\sigma = 0.11$ (though still significantly tighter than the individual SSP-V$_\mathrm{esc}$ relations) but we find a better match between the local ($0.97 \pm 0.04$ ) and global ($0.95 \pm 0.71$) gradients. We also find that this SSP - V$_\mathrm{esc}$ relation displays the same local and global behaviour as the Mg$\,b$- and Colour-V$_\mathrm{esc}$ relations. 

\subsection{Deviations from the Mg$\,b$-V$_\mathrm{esc}$ relation}
\label{sec:residuals}
Because of the tightness as well as the global and local behaviour of the Mg$\,b$-V$_\mathrm{esc}$ relation we can look for correlations of the residuals from the relation with other galaxy parameters as a further tool to unraveling the processes involved in galaxy evolution. We find that there is no correlation of the individual galaxy gradients with any other dynamical, morphological or stellar population property discussed in this study (excluding the previously mentioned negative gradient galaxies), therefore we concern ourselves with examining only the offsets from the mean relation. 

\begin{figure}
\includegraphics[width=3.25in]{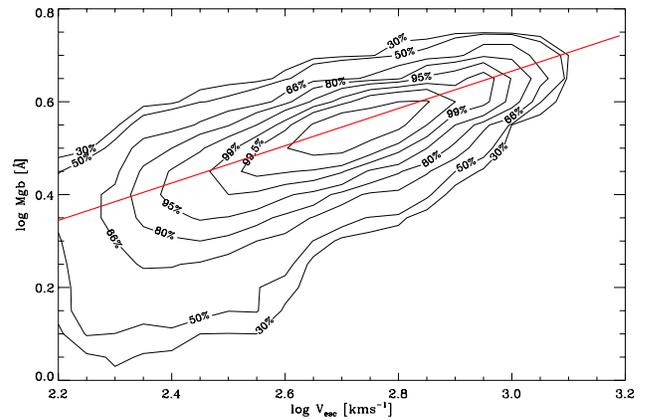}
\caption{Mg$\,b$-V$_\mathrm{esc}$ relation determined from individual bins as opposed to averaging over elliptical annuli. The solid red line in shows the best-fitting relation to the central aperture measurements. The contours of the individual bins follow the central relation, with the expected deviation at low V$_\mathrm{esc}$ due to negative gradient galaxies.}
\label{fig:mgb_vesc_bins}
\end{figure}

\begin{figure*}
\begin{minipage}{5.5in}
\includegraphics[width=5.5in]{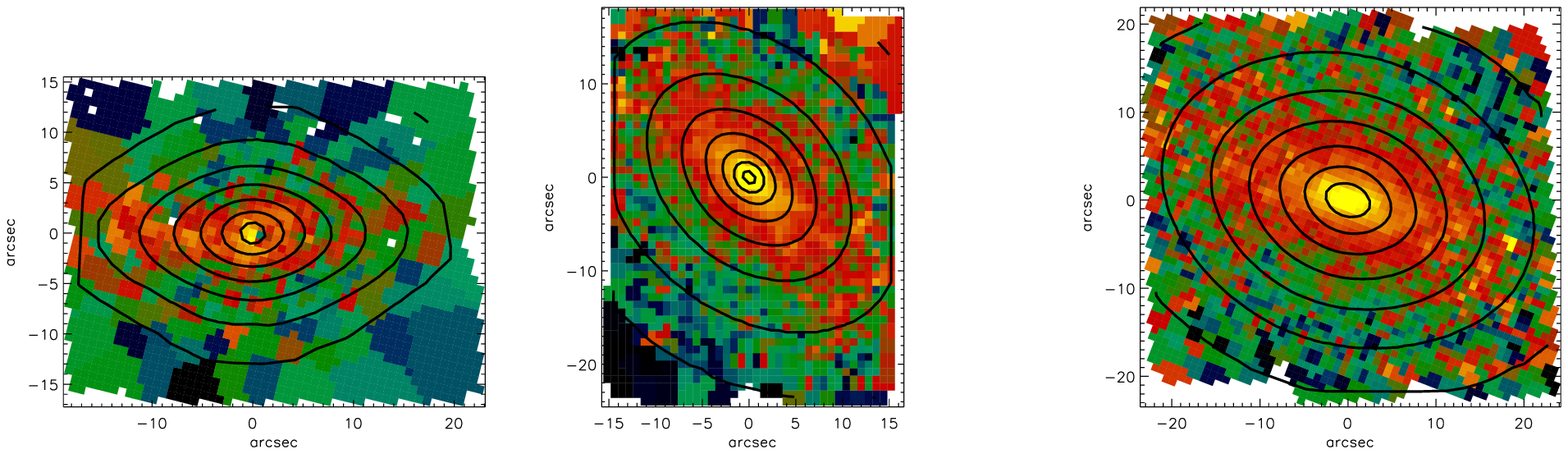}
\includegraphics[width=5.5in]{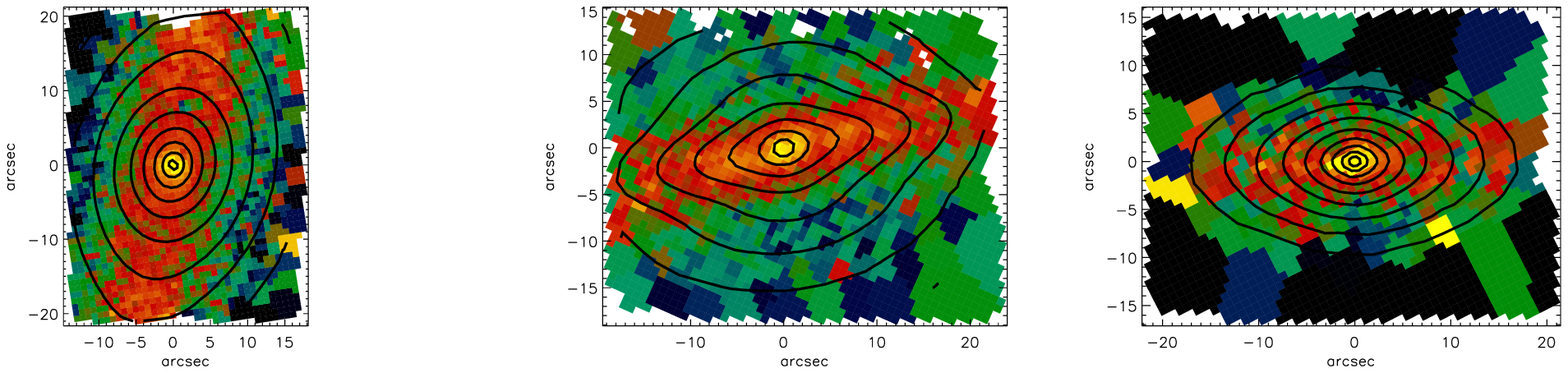}
\end{minipage}
\begin{minipage}{0.5in}
\includegraphics[width=0.5in]{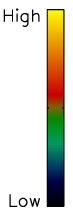}
\end{minipage}
\caption{Examples of Mg$\,b$ disk structures in the IFU maps. From top left the galaxies shown are: NGC2577, NGC3377, NGC4473, NGC4570, NGC4638 and NGC5611. While the colour scaling is different for each galaxies, the colour bar indicates the general trend in Mg$\,b$ for all panels The flux contours from the collapsed SAURON cubes are overplotted. The Mg$\,b$ isocontours are flatter than the isophotes in each case.}
\label{fig:mgb_disks_examples}
\end{figure*}

While the global residual is an interesting quantity it does not take full advantage of the real power of integral-field spectroscopy: it's spatially resolved nature. The key result to take from the Mg$\,b$-V$_\mathrm{esc}$ relation is that galaxies show the same behaviour both locally and globally. The Mg$\,b$-V$_\mathrm{esc}$ relation holds whether it is measured in a central aperture or in annuli. This scale-independent behaviour suggests that the Mg$\,b$-V$_\mathrm{esc}$ relation should hold at all points within the central regions of ETGs. It is easy to test this with IFU data by plotting the Mg$\,b$-V$_\mathrm{esc}$ relation for individual bins, completely ignoring any information as to which galaxies any given bin belonged to. The result of this analysis is shown in the Fig. \ref{fig:mgb_vesc_bins}. Because of the sheer number of individual bins in the ATLAS$^\mathrm{3D}$ data it is necessary to display this in a different fashion to the elliptical annuli data, with the contours reflecting the number of bins at any given point in the Mg$\,b$-V$_\mathrm{esc}$ plane. As usual the best-fitting global relation is overplotted as a solid red line, which follows the contours precisely. The bin data are fully consistent with that determined from elliptical annuli. The overall behaviour of the relation is also the same, with significantly increasing scatter below V$_\mathrm{esc} = 400 \mathrm{kms}^{-1}$. This tight relation found even when considering individual bins allows us to go further than the globally-averaged mean residual described above and consider the residuals of individual bins from the Mg$\,b$-V$_\mathrm{esc}$ relation. The individual bin residuals are defined following as above but, instead of using global values, using values from individual IFU bins. We will make use of both the global residuals and individual IFU bin residuals in the following analysis.

\subsubsection{Deviations from axisymmetry}
As a sanity check on the way our MGE modelling deals with bars and other non-axisymmetric features we separate the sample into a `clean' sample of axisymmetric, fast-rotating galaxies that are well described by the MGE models and a `dirty' sample of all other galaxies where there are significant deviations between the observed photometry and the MGE model. We also include all galaxies classified as slow rotators in Paper III as these are likely to be weakly triaxial, as well as galaxies with bars, rings, isophotal twists and shells or other merger signatures. When we examine the Index-V$_\mathrm{esc}$ relations of these two subsamples separately we find that they are consistent with following the same relation, though there is larger scatter in the dirty sample. We find $\sigma = 0.037$ for the clean sample, whereas the dirty sample has $\sigma=0.042$. The increased scatter we find in the dirty sample can account for 15 per cent of the scatter in the relation for the complete sample.

\begin{figure}
\includegraphics[width=3.25in]{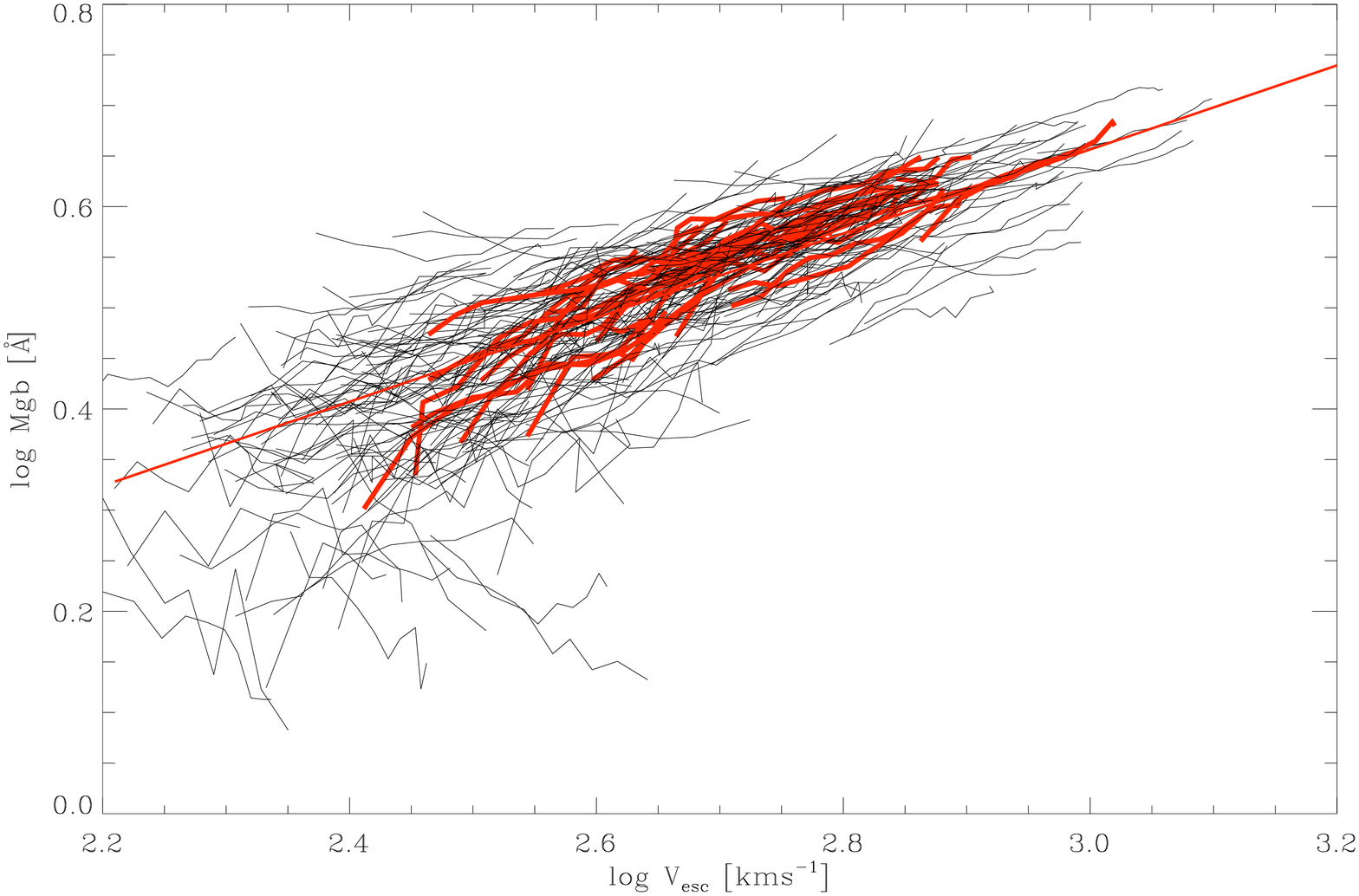}
\includegraphics[width=3.25in]{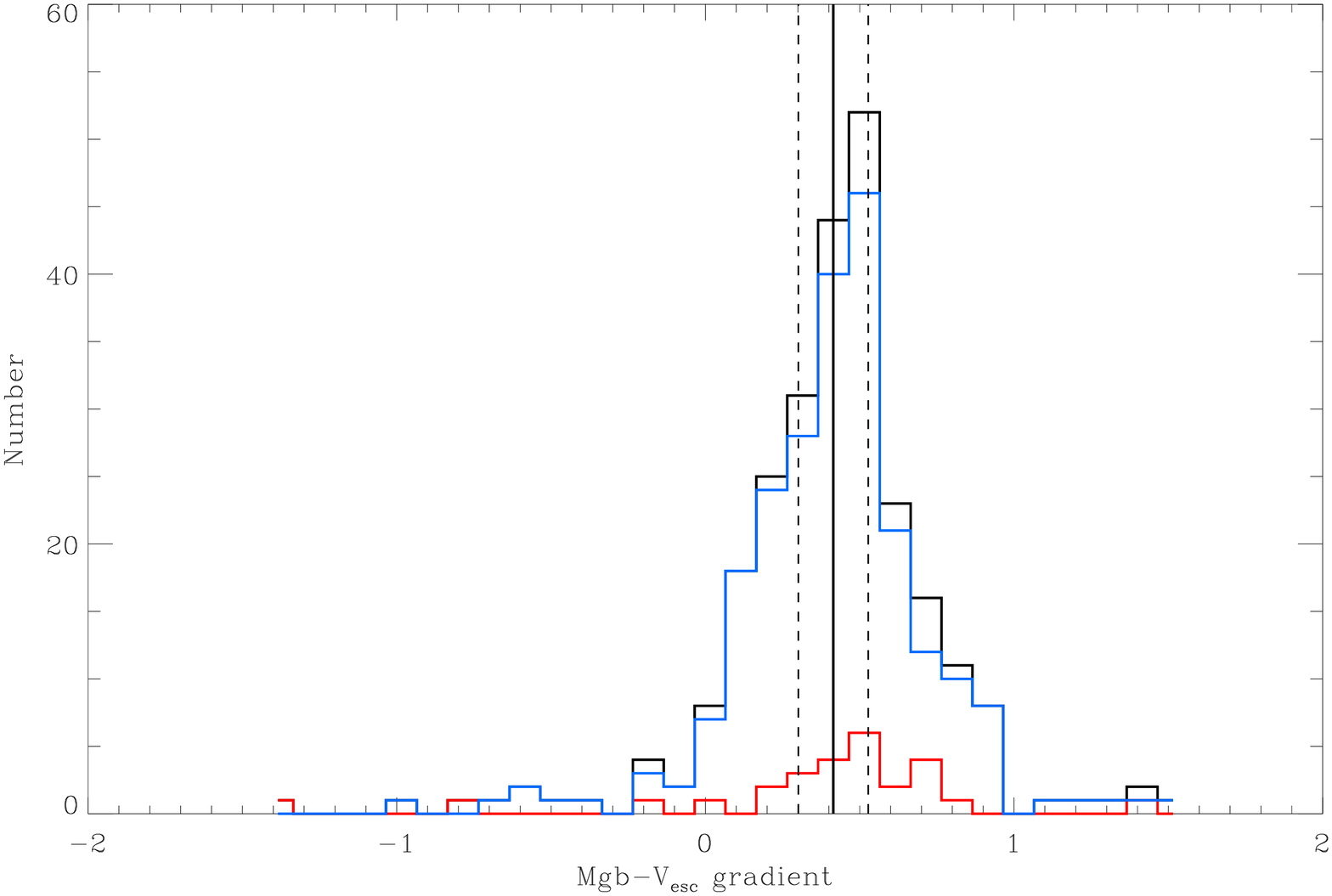}
\caption{Upper panel: The location of Mg$\,b$-disk galaxies in the Mg$\,b$-V$_\mathrm{esc}$ relation. The Mg$\,b$-disk galaxies, indicated by the red profiles, have the same distribution as the normal galaxies. Lower panel: Distribution of individual galaxy gradients for Mg$\,b$-disk (red) and non-disk galaxies (blue). The Mg$\,b$-disk galaxies are offset to steeper gradients than non-disk galaxies, however the offset is only marginally significant in a KS-test.}
\label{Fig:mgb_disks}
\end{figure}

\subsubsection{Mg$\,b$ structures within galaxies}
While we use the same elliptical annuli to measure both the index and V$_\mathrm{esc}$ profiles, the index and V$_\mathrm{esc}$ maps often have somewhat different isocontour shapes. While some of this variation can be accounted for by measurement errors there are a number of cases where the isocontours of the Mg$\,b$ and Fe5015 maps show a prominent disk-like structure. This effect is most pronounced in the Mg$\,b$ maps. This issue was discussed briefly in \citet{Kuntschner:2006} and \citet{Krajnovic:2008}, and more extensively in \citet[][see their fig. 9 and accompanying discussion]{Kuntschner:2010}. Here we explore how these structures in the index maps affect the Mg$\,b$-V$_\mathrm{esc}$ relation. 

We selected by eye all galaxies showing a prominent flattened Mg$\,b$ component in the Mg$\,b$ maps, yielding a sample of 27 objects. A few examples of galaxies with flattened Mg$\,b$ structures are shown in Fig. \ref{fig:mgb_disks_examples}. These galaxies do not occupy any particular region of the Index-V$_\mathrm{esc}$ diagrams, being indistinguishable from early-types that do not exhibit a flattened Mg$\,b$ component (upper panel of Fig. \ref{Fig:mgb_disks}). However, these galaxies have steeper gradients than the typical early-type (as well as being steeper than the global relation) as shown in the lower panel of Fig. \ref{Fig:mgb_disks}. The shift in slope is small, only 1.5$\sigma$ above the mean local gradient for the whole sample. When we exclude the negative gradient galaxies (as these are clearly a different population) a Kolmogorov-Smirnov test (KS-test) returns the probability that the two sets of gradients are drawn from the same distribution is 0.17, which is inconclusive. However, these galaxies have residuals from the Mg$\,b$-V$_\mathrm{esc}$ relation consistent with the full sample of galaxies. This suggests that our choice of elliptical apertures is robust against structure in the line strength maps.

\begin{figure*}
\includegraphics[width=3.25in]{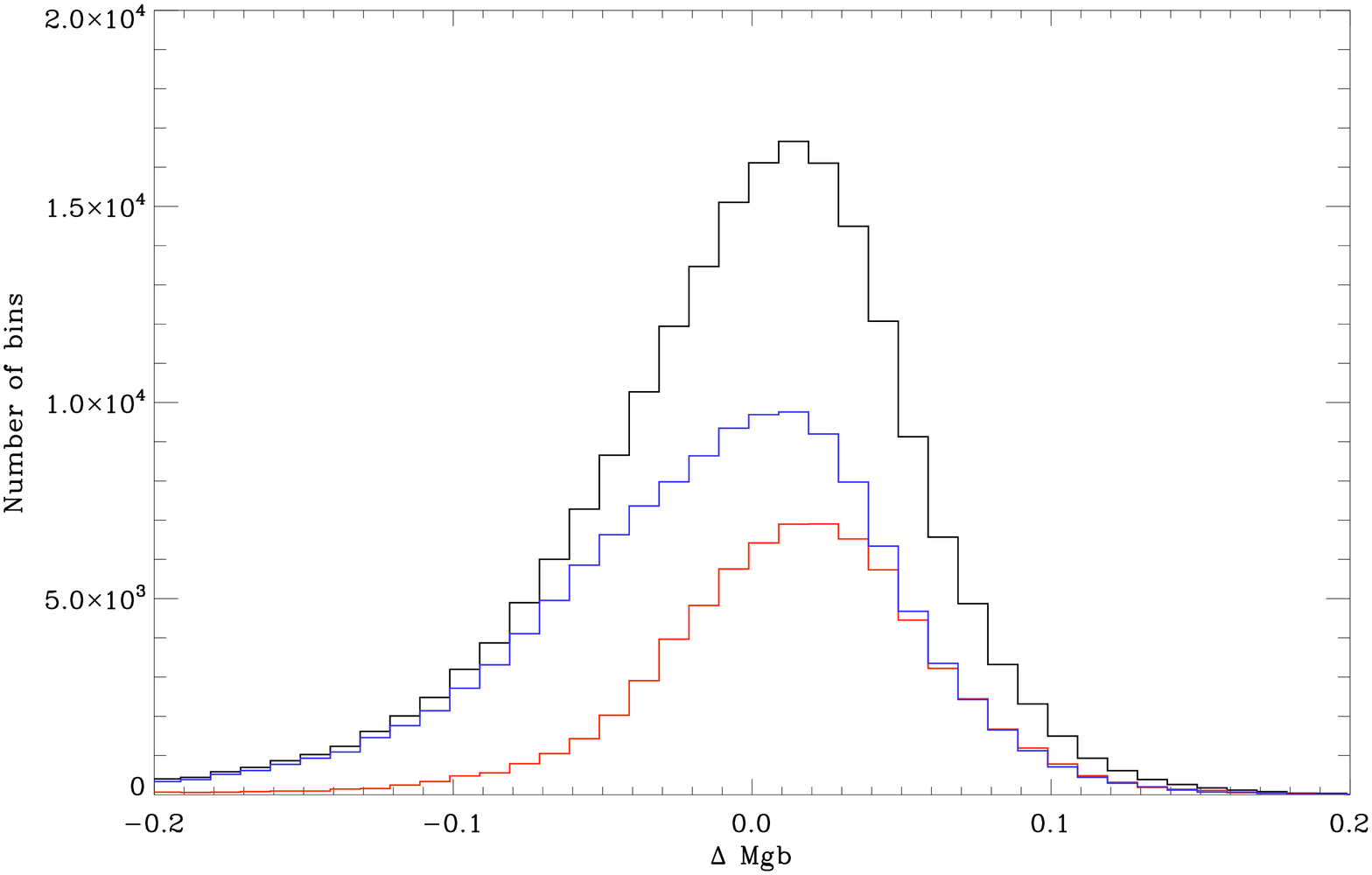}
\includegraphics[width=3.25in]{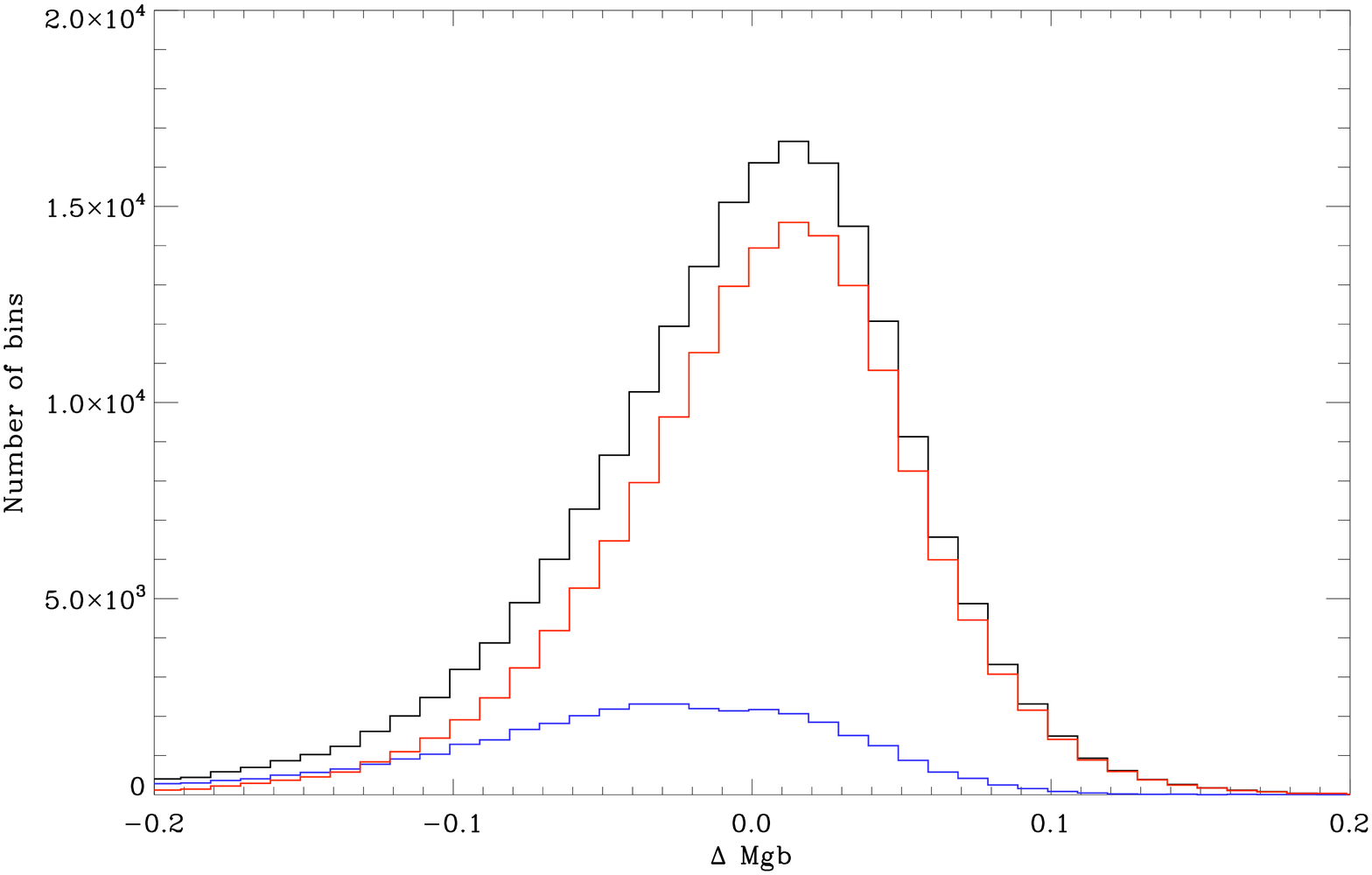}
\caption{Left panel: Histogram of $\Delta$ Mg$\,b$ for individual bins, divided into all (black), Virgo (red) and non-Virgo (blue) samples. Bins belonging to galaxies in the Virgo cluster have higher Mg$\,b$ at fixed V$_\mathrm{esc}$. The Virgo sample does not exhibit the prominent tail of points due to negative gradient galaxies found in the full and non-Virgo samples.
Right panel: Histogram of $\Delta$ Mg$\,b$ for individual bins, divided according to the presence of dust (blue) or not (red) in the associated galaxy (not in that specific bin). Dusty galaxies tend to have low $\Delta$ Mg$\,b$ at a given V$_\mathrm{esc}$ and also frequently exhibit negative gradients.}
\label{Fig:density}
\end{figure*}

\subsubsection{Other influences}
In McDermid et al., in preparation we will show that the environment and gas content of an early-type galaxy have a small yet significant influence on the global properties of its stellar population. These effects persist independent of the size of aperture used to determine the SSP properties. We briefly revisit these results in this paper, using our bin-by-bin analysis and examining the residuals from the Mg$\,b$-V$_\mathrm{esc}$ relation (rather than from the SSP-$\sigma$ relations) in the context of the local and global behaviour described above.

In the left panel of Fig. \ref{Fig:density} we show the histogram of $\Delta$ Mg$\,b$ for individual bins, divided into bins belonging to all (black), Virgo (red) and non-Virgo (blue) galaxies. Bins belonging to Virgo cluster galaxies have more positive residuals, in the sense that bins belonging to Virgo cluster galaxies have higher Mg$\,b$ at fixed V$_\mathrm{esc}$ compared to bins belonging to galaxies in lower density environments. In the right panel of Fig. \ref{Fig:density} we show the histogram of the residuals from the Mg$\,b$-V$_\mathrm{esc}$ of individual bins associated with galaxies from the full (black), dusty (blue) and non-dusty (red) samples (as classified in Paper II). As can be seen these galaxies are strongly biased towards having negative residuals - at a given V$_\mathrm{esc}$ a bin from a dusty galaxy will typically have a weaker Mg$\,b$ line-strength than a dust-free galaxy. The same result is true of CO and HI detected galaxies -- bins belonging to galaxies with detectable cold gas components are generally found below the relation. 

These results are consistent with our findings in McDermid et al., in preparation. However the use of spatially resolved quantities allows us to go one step further in our analysis. As previously stated, the Mg$\,b$-V$_\mathrm{esc}$ gradient does not depend on any galaxy property, including environment and gas and dust content (with the exception of the negative gradient galaxies). Here we have shown that individual bins within galaxies in high-density environments or with high dust or gas contents are offset from the main relation (in a positive and negative sense respectively). These two observations show that the offset is a systematic one -- all bins within the central $\sim 1$ R$_e$ are offset from the mean relation by the same amount for a given galaxy. This does not apply to the negative gradient galaxies which show a broad range of behaviours and no systematic offset.

\begin{figure*}
\includegraphics[width=4.3in]{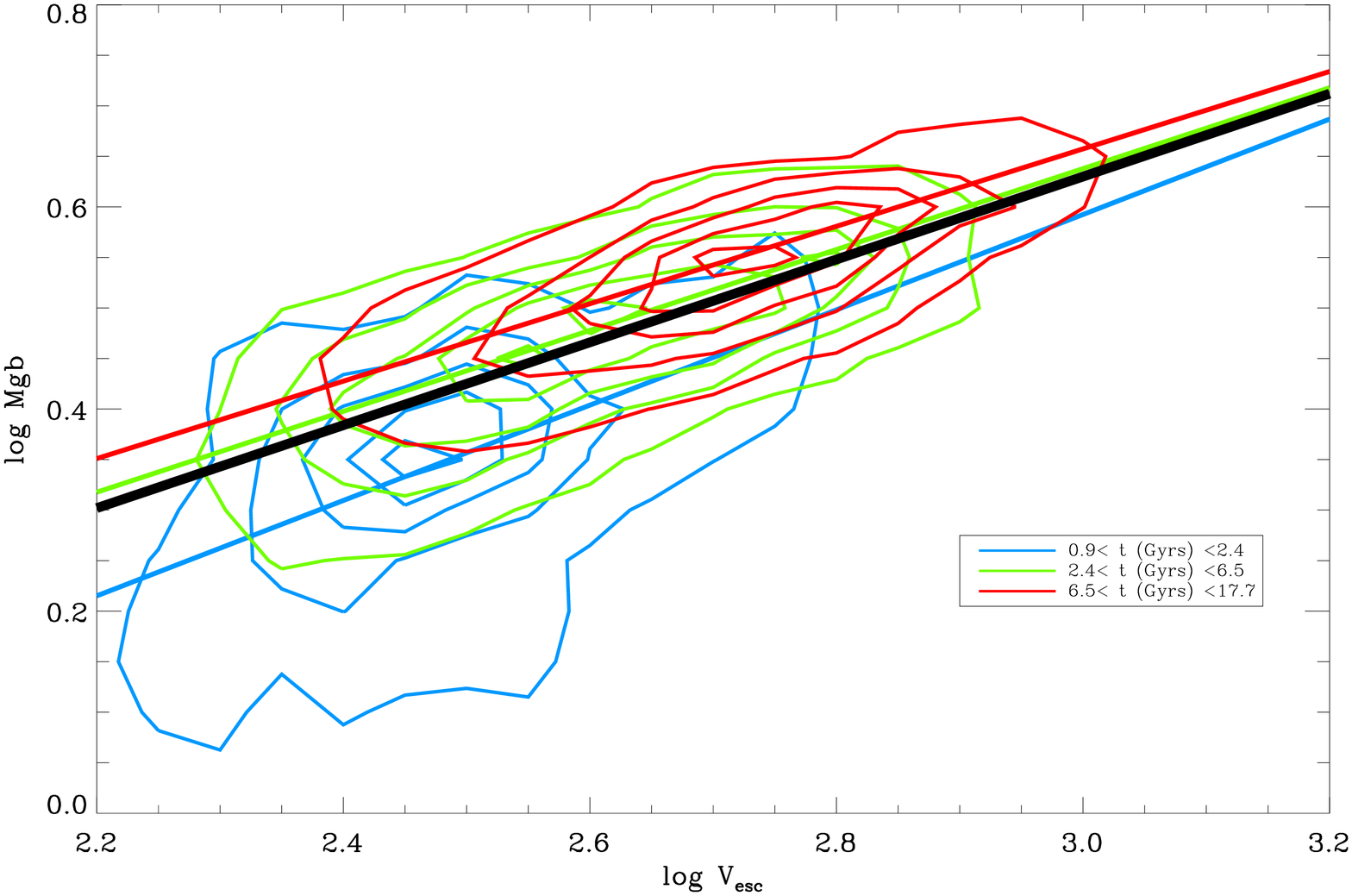}
\includegraphics[width=4.3in]{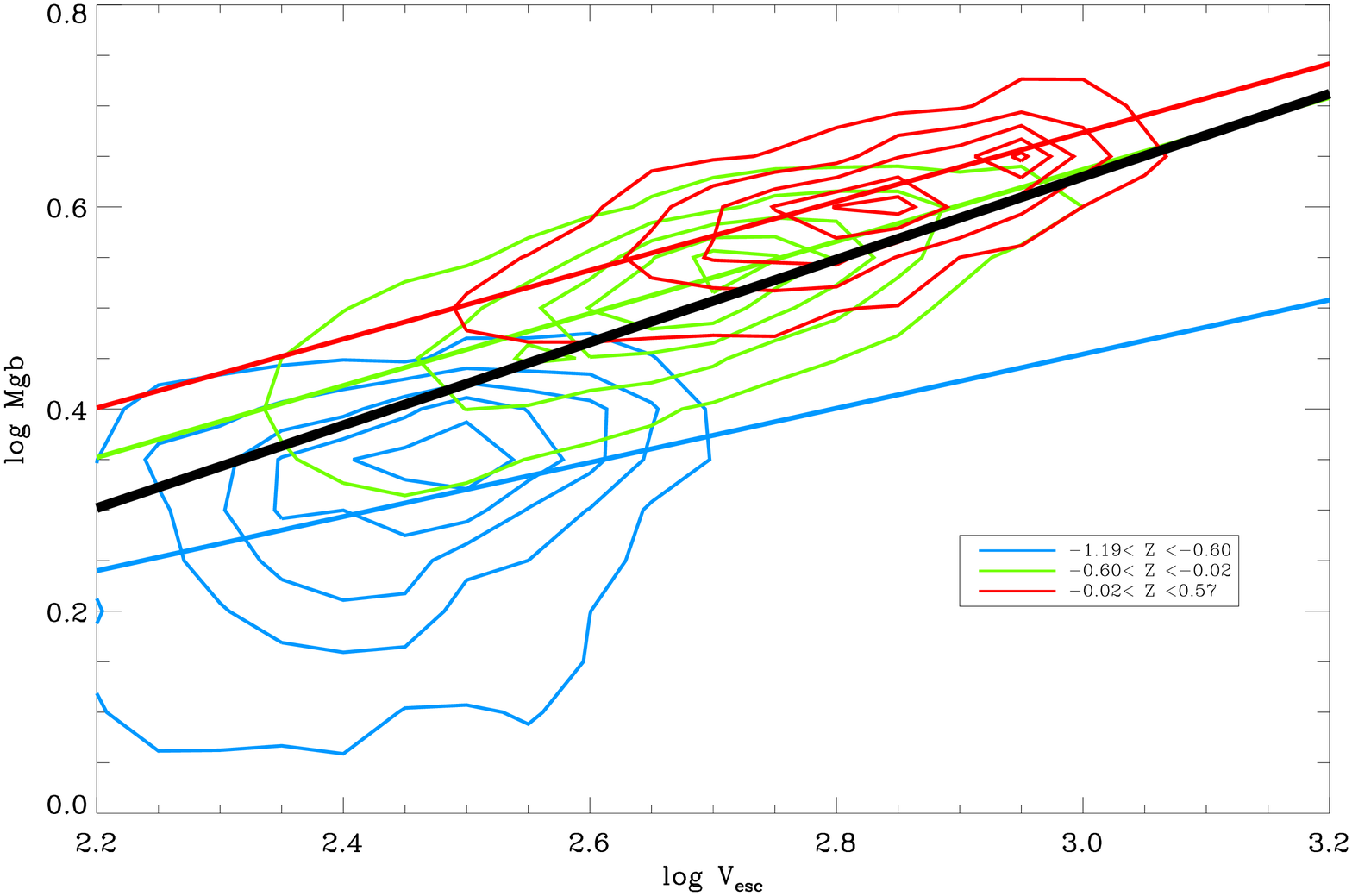}
\includegraphics[width=4.3in]{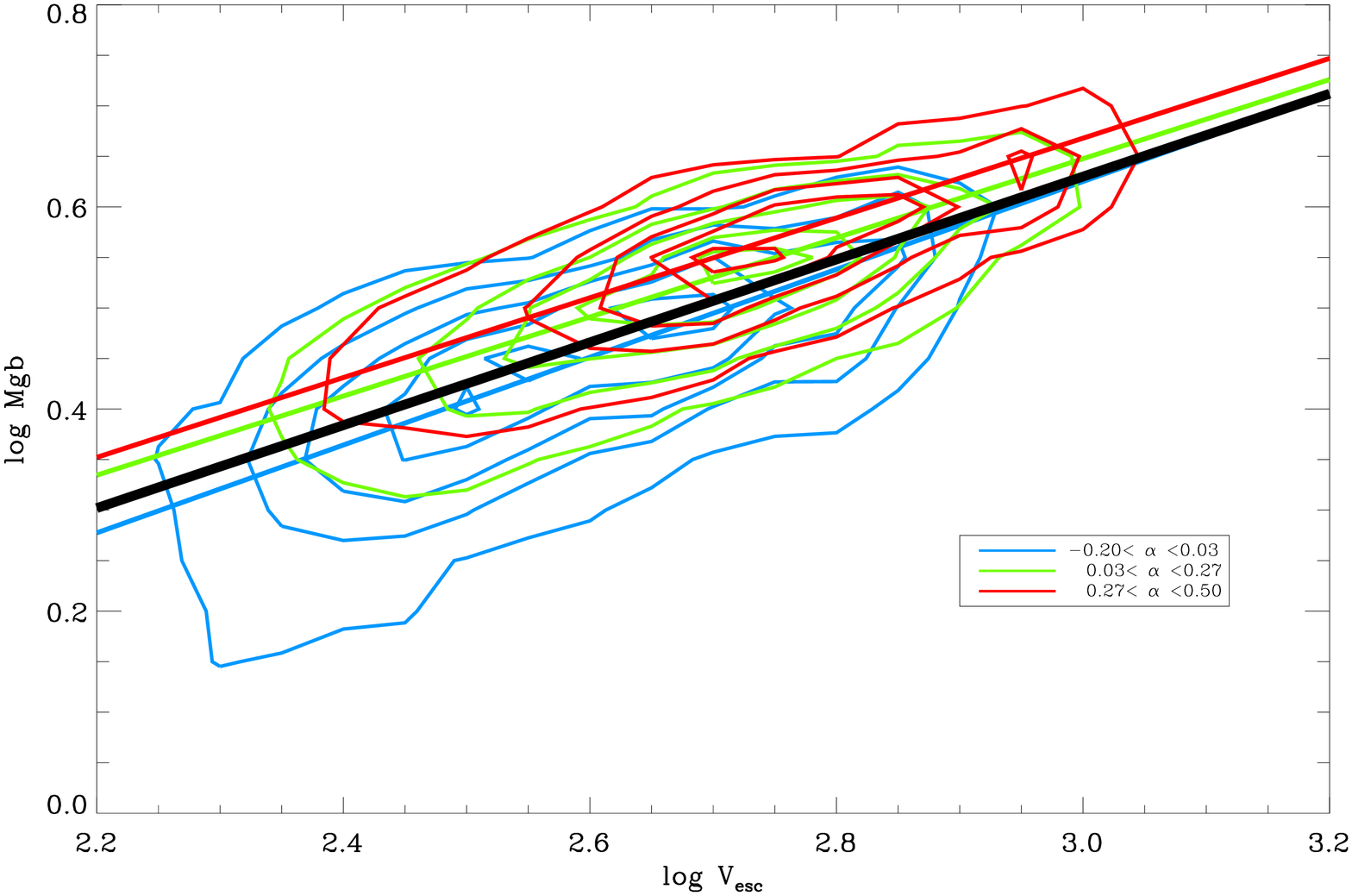}
\caption{The Mg$\,b$-V$_\mathrm{esc}$ relation for all individual bins, divided into three ranges of age (upper panel), metallicity (middle panel) and enhancement (lower panel). The coloured contours indicate the region occupied by the points with a given range of t, [Z/H] or [$\alpha$/Fe]. The contours enclose a set fraction of the bins with a given range of t, [Z/H] or [$\alpha$/Fe] -- for example there are far fewer total points in the lowest age bin than in intermediate age bins even though the same number of contours are displayed. The solid black line shows the global relation for all galaxies while the coloured lines are a fit to the SSP subsets. The range of SSP values indicated by each colour is given in the legend on each panel. The negative gradient galaxies stand out clearly at young ages and low metallicities. The relation shows the strongest dependence on [Z/H].}
\label{fig:mgb_vesc_ssp_bins}
\end{figure*}

\subsection{SSP parameters}
\label{sec:resid_ssp}
In Section \ref{Sec:SSP} we explored the connection between the global SSP parameters and V$_\mathrm{esc}$. In this section we go beyond globally averaged quantities and instead examine the dependence of the residuals of the Mg$\,b$-V$_\mathrm{esc}$ relation on the {\it local} SSP parameters. Using the SSP models of \citet{Schiavon:2007} we apply the SSP method of McDermid et al. (in preparation) to the line-strength indices of individual IFU bins and construct maps of the three SSP equivalent parameters age, metallicity and alpha-enhancement.

In Fig. \ref{fig:mgb_vesc_ssp_bins} we show the Mg$\,b$-V$_\mathrm{esc}$ relation for individual bins falling within a given range of each of the SSP parameters. The coloured contours show the location of all IFU bins falling within that range of t (upper panel), [Z/H] (middle panel) or [$\alpha$/Fe] (lower panel) in the Mg$\,b$-V$_\mathrm{esc}$ plane. The solid line of the same colour shows the best fit to the bins with the associated SSP values. The contours for each SSP parameter subset are normalised to the total number of bins falling within that SSP range -- with the contours enclosing 10, 30, 50, 70 and 90 per cent of bins with the given SSP value.

 For the relation at fixed t or [$\alpha$/Fe] we find that, except for the very youngest or most $\alpha$-deficient bins, all bins follow the Mg$\,b$-V$_\mathrm{esc}$ relation within the 1 $\sigma$ errors. Bins at fixed t or [$\alpha$/Fe] span the full range in V$_\mathrm{esc}$ of the sample, again except for the youngest bins. The youngest bins with t $< 1.2$ Gyrs (not explicitly shown in Fig. \ref{fig:mgb_vesc_ssp_bins}) all lie below the relation; they are also only found at low V$_\mathrm{esc}$ (consistent with our discussion on negative-gradients galaxies in the following Section \ref{sec:RSF}). Bins with t $< 2.4$ Gyrs also typically lie below the relation, though at higher V$_\mathrm{esc}$ this effect is less pronounced. All bins with t $> 2.4$ Gyrs follow the Mg$\,b$-V$_\mathrm{esc}$ relation, even at V$_\mathrm{esc} < 400 \mathrm{kms}^{-1}$. When binning by [$\alpha$/Fe], only the most [$\alpha$/Fe] deficient bins with [$\alpha$/Fe] $< -0.06$ show any significant deviations from the Mg$\,b$-V$_\mathrm{esc}$ relation (again, not explicitly shown in Fig. \ref{fig:mgb_vesc_ssp_bins}), and only apparent at V$_\mathrm{esc} < 400 \mathrm{kms}^{-1}$. There is a small but significant trend for the relation to be offset to higher values of Mg$\,b$ as [$\alpha$/Fe] increases: $\Delta$ Mg$\,b$ increases with increasing [$\alpha$/Fe]. There is no clear trend of $\Delta$ Mg$\,b$ with t.

The picture is very different when binning by [Z/H]. Bins of fixed [Z/H] only span a narrow range in V$_\mathrm{esc}$. The low-[Z/H] bins are concentrated at low V$_\mathrm{esc}$ and lie below the Mg$\,b$-V$_\mathrm{esc}$ relation, in the region occupied by the negative gradient galaxies. The high-[Z/H] bins are concentrated at high V$_\mathrm{esc}$ and lie above the main Mg$\,b$-V$_\mathrm{esc}$ relation. When [Z/H] is restricted to even smaller ranged than in the figure, each unique metallicity bin spans only a narrow range in V$_\mathrm{esc}$ and falls as a clump along the relation. The Mg$\,b$-V$_\mathrm{esc}$ relation disappears for fixed [Z/H]. This behaviour is not seen for t or [$\alpha$/Fe]; for even narrow ranges of t or [$\alpha$/Fe] the bins span a broad range in V$_\mathrm{esc}$ and the Mgb-V$_\mathrm{esc}$ relation is still observed. This is not surprising, given that the strength of the Mg$\,b$ absorption index is predominantly determined by [Z/H], and so Mg$\,b$ can vary little at fixed [Z/H]. The Mg$\,b$-V$_\mathrm{esc}$ relation is predominantly driven by the variation of [Z/H] with V$_\mathrm{esc}$, which is again unsurprising given the tight mass-metallicity relation of ETGs (McDermid et al, in preparation). 

\section{Discussion}
\label{Sec:Discussion}
In the previous two sections we have presented many pieces of observational evidence connected to the role of V$_\mathrm{esc}$ in local early-type galaxies. In this section we consider the above evidence in context of galaxy evolution theory and use it to discriminate between the different potential formation histories of the nearby early-type galaxy population. 

\subsection{The effect of recent star formation on the V$_\mathrm{esc}$ relation}
\label{sec:RSF}
\begin{figure}
\includegraphics[width=3.25in]{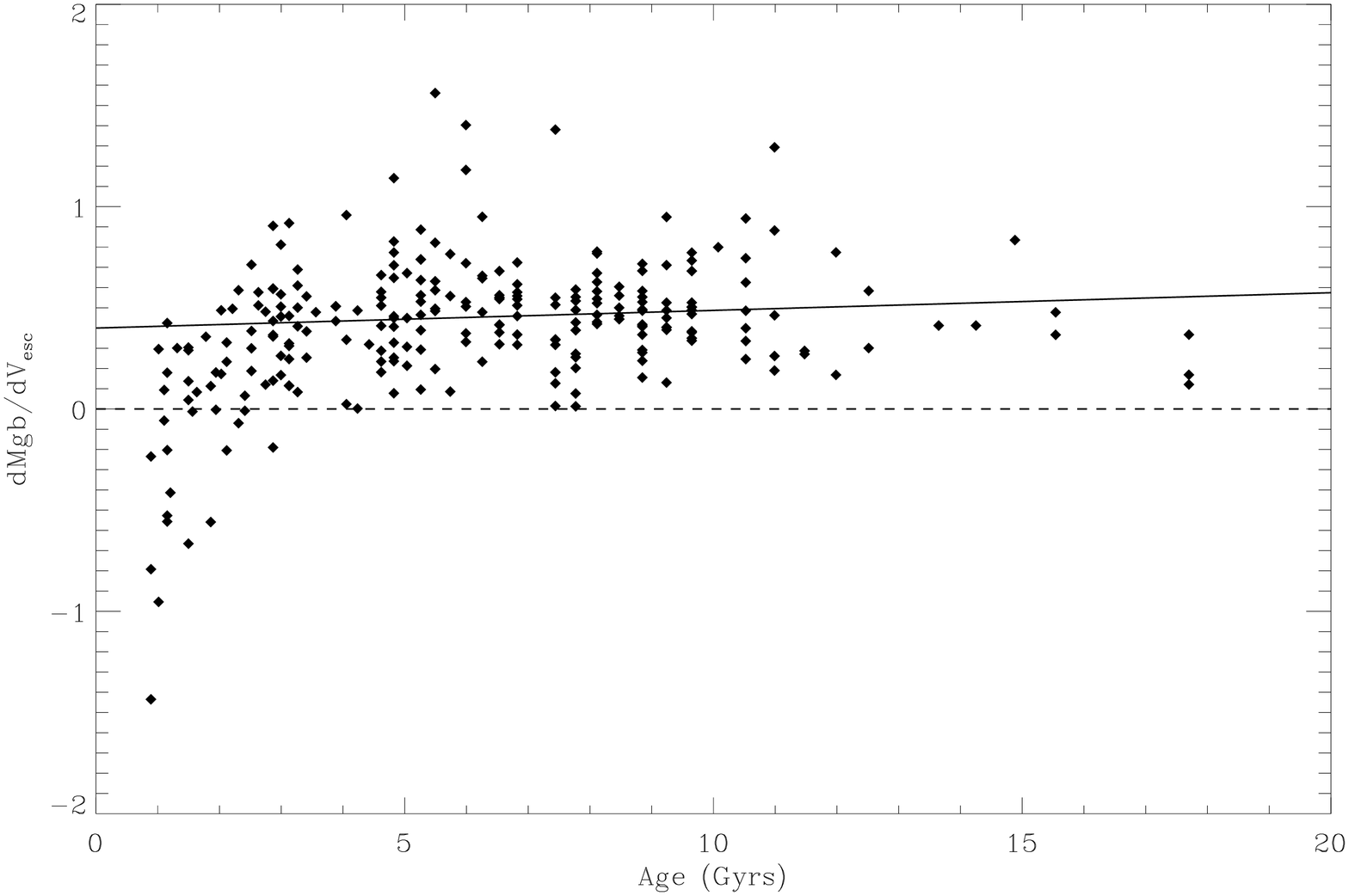}
\caption{Mg$\,b$-V$_\mathrm{esc}$ gradient vs. central age, measured within an aperture with radius R$_\mathrm{e}$/8. The solid line is a fit to the data. The dashed line divides galaxies with negative and positive gradients. Negative gradient galaxies all have young central ages.}
\label{fig:resid_agere8}
\end{figure}

The most striking aspect of the Mg$\,b$-V$_\mathrm{esc}$ relation is its tightness above V$_\mathrm{esc} \sim 400\mathrm{kms}^{-1}$ and the sharp increase in scatter below this value. This increase in scatter can largely be attributed to a population of galaxies showing negative gradients, with centrally depressed values of Mg$\,b$ (and to a lesser extent Fe5015) and enhanced values of H$\beta$. These negative gradient galaxies are found almost exclusively at low values of V$_\mathrm{esc}$, apart from a single exception. As noted above high values of H$\beta$ are usually indicative of a young stellar population, with our SSP analysis confirming that these galaxies have, on average, lower central SSP ages. All negative gradient galaxies have SSP equivalent ages less than 3 Gyrs (5 Gyrs) within an R$_e$/8 (R$_e$) aperture (Fig. \ref{fig:resid_agere8}), or, in observational terms, H$\beta > 2.5 (2.3)$. In the case of NGC4150, a prominent negative gradient galaxy, \citet{Crockett:2011} confirm the presence of a centrally concentrated $\sim 300$ Myr population using HST UV  and visible photometry. These galaxies also tend to be dusty and contain significant amounts of molecular gas, again indicative of recent star formation. 75 per cent of negative gradient galaxies are detected in CO, compared to 18 per cent of the normal galaxies. Similarly, 71 per cent of negative gradient galaxies contain dust identified in the {\it ugriz} photometry, compared to only 15 per cent of the normal galaxies.

\begin{figure}
\includegraphics[width=3.25in]{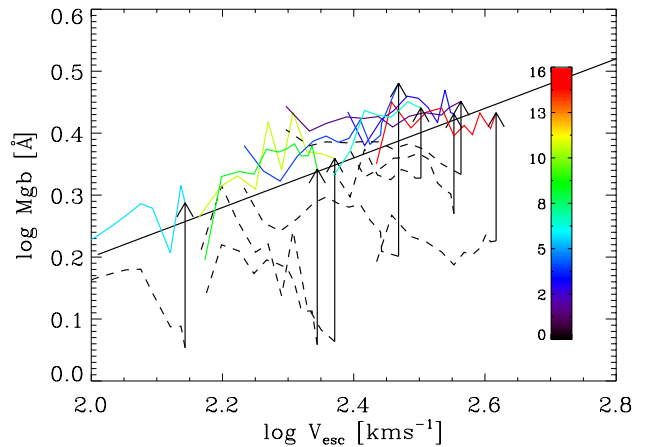}
\caption{The Mg$\,b$-V$_\mathrm{esc}$ relations for a selection of negative gradient galaxies (dashed lines) and their artificially aged counterparts (coloured lines). The `aged' galaxies are produced by keeping their SSP determined [Z/H] and [$\alpha$/Fe] fixed while allowing their SSP age to vary. The colours of the `aged' galaxies indicates the amount of ageing required to return them to the Mg$\,b$-V$_\mathrm{esc}$ relation. Purple/blue corresponds to $\sim$ 2 Gyrs of passive ageing, whereas red corresponds to $\sim$ 15 Gyrs.}
\label{Fig:passive}
\end{figure}

These young cores typically represent a small (by mass fraction) perturbation on top of an underlying older population. As the galaxy ages and the younger population fades, reducing its contribution to the total luminosity of the galaxy, we might expect these galaxies to return to the Mg$\,b$-V$_\mathrm{esc}$ relation as the older population comes to dominate the light. We can test this idea by artificially ageing these negative gradient galaxies using our SSP models. This analysis is shown in Fig. \ref{Fig:passive}. As can be seen the negative gradient galaxies typically return to the relation after a few Gyrs of evolution, assuming no further star formation takes place. The timescales for this are likely shorter than those predicted by the SSP models, as the assumptions made in the SSP modelling mean the SSP ages are strongly influenced by even a small fraction of very young stars. In an upcoming paper in the series we will explore more complex star formation histories for the sample, allowing us to accurately constrain the mass fractions and ages of these young central stellar populations.

These negative gradients are predominantly found in galaxies with central V$_\mathrm{esc} < 400 \mathrm{kms}^{-1}$. Not all galaxies below this V$_\mathrm{esc}$ threshold show negative gradients, the majority of galaxies in this region still follow the relation. However, these galaxies are much more common below this value. Galaxies with central V$_\mathrm{esc} = 400$ kms$^{-1}$ have JAM dynamical masses of $3\times10^{10}$ M$_\odot$. \citet{Kauffmann:2003} identified this mass scale as marking the transition from young, low surface-density galaxies to old, more concentrated objects. The same value M$_\mathrm{JAM} \approx3 \times 10^{10}$ M$_\odot$ was found in \citet[hereafter Paper XX]{Cappellari:2012c} to characterize the galaxy distribution on the $(M,R_{\rm e})$ and $(M,\sigma)$ planes. This mass defines a break in the ``zone of exclusion'' of the galaxy distribution. Below this mass ETGs are found to define a sequence of increasing $\sigma$, bulge fraction, $M/L$ and redder colours. Above this mass bulges dominate and the $M/L$ and colour are more homogeneous (see Paper XX for details). 

\subsection{Constraints on merging}
It is perhaps surprising that we do not find any difference, in terms of the Mg$\,b$-V$_\mathrm{esc}$ relation, between the fast and slow rotators in the ATLAS$^\mathrm{3D}$ sample, given the differing relative importance of gaseous and dissipationless processes in their formation (as discussed in Section \ref{sec:intro}). Taking an extreme example, if we were to merge two completely gas free, equal mass fast rotators that both initially lie on the observed Mg$\,b$-V$_\mathrm{esc}$ relation, would we expect the resulting merger remnant to lie on the Mg$\,b$-V$_\mathrm{esc}$ relation? A naive initial answer would suggest no; by merging the two galaxies we are significantly altering the mass distribution and hence the local V$_\mathrm{esc}$, while the stellar population remains unchanged as there is no gas to form new stars. If this simple analysis is valid then it is possible to constrain the number of major, dry mergers a typical early-type galaxy has experienced. However, the picture may be considerably more complicated than this simple analysis allows for, even in the case of a dry major merger. While the global potential of the galaxy must deepen as more mass is added, the merger process may redistribute the stars in such a way that the local V$_\mathrm{esc}$ is essentially unchanged. In support of this idea it has long been shown that the rank order of binding energies of the stars is preserved in a collisionless merger \citep{White:1978a,Barnes:1988}. 

We can estimate the effect of a major dry merger by calculating the change in V$_\mathrm{esc}$ from simple analytic expressions. The central V$_\mathrm{esc}$ will be affected by changes in the final mass and radius and due to the redistribution of the stellar material due to violent relaxation. Following a simple Virial estimate as in \citet{Naab:2009} (assuming Virialised initial systems formed dissipatively of stars and conservation of energy in the following merger) it can be shown that:
\begin{equation*}
\frac{r_f}{r_i} = \frac{\left(1 + \eta \right)^2}{\left( 1 + \eta \epsilon \right)}
\end{equation*}
where the subscripts $i$ and $f$ denote initial and final quantities, $r$ and $M$ are the radius and mass of the object, $\eta$ is the mass ratio of the merger, $1+\eta = M_f/M_i$ and $\epsilon$ is the ratio between the accreted and initial mean square speed of the stars, $\epsilon = \left< v_a^2\right>/ \left< v_i^2\right>$. If the two progenitors are identical ($\eta =1, \epsilon = 1$) the radius and mass of the remnant will double. This leads to a change in V$_\mathrm{esc}$ of:
\begin{equation*}
\frac{\mathrm{V}_\mathrm{esc,f}}{\mathrm{V}_\mathrm{esc,i}} \propto \left( \frac{r_i}{r_f}  \frac{M_f}{M_i} \right)^{1/2} = 1.
\end{equation*}
For a 1:1 major merger this analytic estimate predicts V$_\mathrm{esc}$ will remain unchanged, however this only applies to central values and under the assumptions outlined above. If energy is not conserved or relaxation is not complete the change in V$_\mathrm{esc}$ may differ. Indeed, \citet[their fig. 1]{Hopkins:2009} find in their simulations that $r_f/r_i \sim 1.8$, as opposed to the value of $2$ predicted by the above analytic estimate. This would yield $\frac{\mathrm{V}_\mathrm{esc,f}}{\mathrm{V}_\mathrm{esc,i}} \sim 1.05$. Equally importantly, our analytic estimate is too simplistic to predict how V$_\mathrm{esc}$ varies with radius after a merger. 

In order to better assess the effect of mergers on the Mg$\,b$-V$_\mathrm{esc}$ relation in real systems we make use of the N-body binary merger simulations of \citet[hereafter Paper VI]{Bois:2011}. By 'tagging' the stars in the merger progenitors with a given age and metallicity we were able to follow how the stellar population is redistributed in a dry major merger. Here we examine only the effect of major (1:1 mass ratio), dry ($< 3$ per cent gas) binary mergers on the Mg$\,b$-V$_\mathrm{esc}$ relation. In a future work in this series we will fully describe this approach including the extension to gas-rich and minor mergers. We consider two merger cases from Paper VI with early-type galaxies as progenitors\footnote{their so-called `remerger 2x11 dd' and 'remerger 2x11rr'. See Paper VI for details.}. All models are projected at an inclination of 60 degrees to the line of sight. We assign ages and metallicities to the progenitor galaxies such that they lie on the observed Mg$\,b$-V$_\mathrm{esc}$ relation. In the model galaxies we use the MILES SSP library \citep{Sanchez-Blazquez:2006b} to convert the projected metallicity and age of the stars into a map of the Mg$\,b$ line strength. The model galaxies are then 'observed' exactly as the real galaxies were and Mg$\,b$ and V$_\mathrm{esc}$ profiles are constructed as described in Section \ref{Sec:Index_Vesc}. 

\begin{figure}
\includegraphics[width=3.25in]{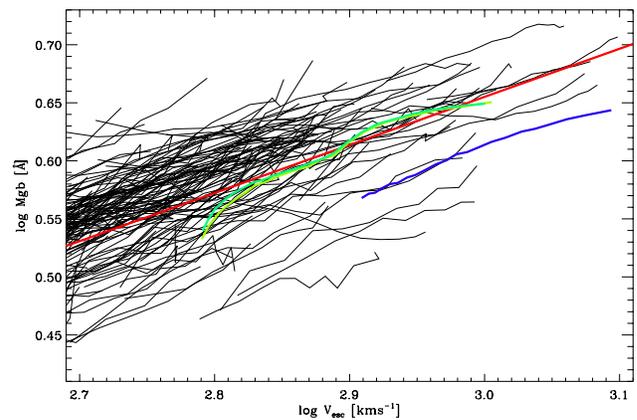}
\caption{A zoom in on the high-V$_\mathrm{esc}$ end of the Mg$\,b$-V$_\mathrm{esc}$ relation with the results of 1:1 mass-ratio, dry, binary merger simulations overplotted. The green lines show the Mg$\,b$-V$_\mathrm{esc}$ profile for the two (similar) progenitor galaxies and the blue line shows the profile for the merger remnant. The black lines indicate the observed galaxy profiles. The merger simulations are fully described in the text. The effect of the mergers is to increase V$_\mathrm{esc}$ without significantly changing the observed Mg$\,b$.}
\label{fig:mgb_vesc_sim}
\end{figure}

The results of this analysis are shown in Fig. \ref{fig:mgb_vesc_sim}. The progenitors are shown in green, with the corresponding remnant shown in blue. The black lines show the observed ATLAS$^\mathrm{3D}$ sample and the solid red line is the best fitting relation as described in Section \ref{Sec:Index_Vesc}. The dry merger simulations confirm our naive prediction that Mg$\,b$ remains essentially unchanged while the merger remnants are shifted to higher V$_\mathrm{esc}$ by an amount $\Delta \log$ V$_\mathrm{esc} = 0.08$. We also find that Mg$\,b$-V$_\mathrm{esc}$ gradients are largely preserved in the merger, in that the gradient of the remnant is the average of the progenitors' gradients. It is clear that dry major mergers move galaxies to the right of the relation, rather than along the relation. From this result we can draw two conclusions. Firstly, massive early-type galaxies are unlikely to have been formed purely by multiple dry major mergers of low-mass present-day early-type galaxies. An increase in their log V$_\mathrm{esc}$ by $\sim 0.4$ due to major mergers (moving them from the middle to the high end of the V$_\mathrm{esc}$ range of our sample), with no corresponding change in log Mg$\,b$ would give these hypothetical galaxies values of $\log$ Mg$\,b$ $> 3 \sigma$ below the relation. Secondly, the observed scatter in the Mg$\,b$-V$_\mathrm{esc}$ relation can be used to constrain the number of major dry mergers a typical early-type galaxy can have undergone. Given an observed scatter of $\sigma = 0.04$, and a typical $\Delta \log$ V$_\mathrm{esc}$ of 0.08 per major merger, we find that a typical early-type galaxy can only have experienced $\sim 1.5$ major dry mergers since joining the Mg$\,b$-V$_\mathrm{esc}$ relation.

\section{Conclusions}
\label{Sec:Conclusions}
In this work we have presented the Index-V$_\mathrm{esc}$ relations for the ATLAS$^\mathrm{3D}$ survey, a complete sample of nearby early-type galaxies. This survey is based on SAURON integral-field spectroscopy and is supported by {\it ugriz} photometry. Using these data we accurately modelled the mass distributions of 258 galaxies in the sample using a combination of MGE photometric models and JAM dynamical modelling. We again emphasise that our V$_\mathrm{esc}$ is derived from the surface brightness, so may differ from the true V$_\mathrm{esc}$ due to the influence of dark matter, and that the results we present here are predominantly empirical rather than causal. An important, and straightforward, test of a potential causal relationship would be to examine the stellar population - V$_\mathrm{esc}$ relations are large radii, where dark matter is expected to dominate the gravitational potential. From these results we are able to draw the following conclusions: 

\newcounter{Lcount3}
\begin{list}{\roman{Lcount3})}
{\usecounter{Lcount3}}
\item The Mg$\,b$ - V$_\mathrm{esc}$ and relation is extremely tight (with a scatter of 7 per cent) and displays a local and global connection, in that the relation is the same {\it within} individual galaxies as it is between different galaxies. The Colour - V$_\mathrm{esc}$ relations exhibit very similar behaviour. These correlations provide a compact way to describe gradients in stellar populations in ETGs. They are an ideal tool for comparison with numerical simulations, to test whether they reproduce real galaxies.
\item Below V$_\mathrm{esc} \sim 400\ \mathrm{kms}^{-1}$ the scatter in the Mg$\,b$ - V$_\mathrm{esc}$ relation increases significantly due to a population of negative gradient galaxies. These negative gradient galaxies typically have centrally concentrated young stellar populations and often contain molecular gas and dust. This strongly suggests that these negative gradients are due to recent star formation. The SSP models suggest that as these galaxies age the majority will return to the Mg$\,b$ - V$_\mathrm{esc}$ relation - negative gradients are largely a transient phenomena.
\item Excluding the negative gradient galaxies, the Mg$\,b$ - V$_\mathrm{esc}$ gradients do not correlate with any other galaxy property. The residuals from the V$_\mathrm{esc}$ relation correlate with: environment, dust and molecular gas content and $\alpha$ enhancement. 
\item In the four dimensional space of V$_\mathrm{esc}$, Age, [Z/H], [$\alpha$/Fe] the galaxies in our sample are approximately confined to a plane. This plane can be defined using just three variables: $\log \mathrm{V}_\mathrm{esc} = 0.62 \mathrm{[Z/H]} + 0.26 \log t + 2.53$. This SSP relation is tighter than any of the individual SSP - V$_\mathrm{esc}$ relations and also exhibits the same local and global behaviour as the Mg$\,b$ - V$_\mathrm{esc}$ relation.
\item At the high V$_\mathrm{esc}$ end of the Mg$\,b$ - V$_\mathrm{esc}$ relation a galaxy can typically have experienced only $\sim 1.5$ dry mergers without shifting the galaxy so far away from the relation that it is no longer consistent with the observed scatter. 

\end{list} 

\section{Acknowledgements}
We would like to thank Ignas Snellen, Ernst de Mooij and Mark den Brok for their help in acquiring INT photometry of several galaxies in the sample. This research has made use of the NASA/IPAC Extragalactic Database (NED) which is operated by the Jet Propulsion Laboratory, California Institute of Technology, under contract with the National Aeronautics and Space Administration. NS and TAD were supported by a STFC Postgraduate Studentship. NS additionally acknowledges support of Australian Research Council grant DP110103509. MC acknowledges support from a Royal Society University Research Fellowship. This work was supported by the rolling grants ÔAstrophysics at OxfordÕ PP/E001114/1 and ST/H002456/1 and visitors grants PPA/V/S/2002/00553, PP/E001564/1 and ST/H504862/1 from the UK Research Councils. RLD acknowledges travel and computer grants from Christ Church, Oxford and support from the Royal Society in the form of a Wolfson Merit Award 502011.K502/jd. RLD is also grateful for support from the Australian Astronomical Observatory Distinguished Visitors programme, the ARC Centre of Excellence for All Sky Astrophysics, and the University of Sydney during a sabbatical visit. SK acknowledges support from the Royal Society Joint Projects Grant JP0869822. RMcD is supported by the Gemini Observatory, which is operated by the Association of Universities for Research in Astronomy, Inc., on behalf of the international Gemini partnership of Argentina, Australia, Brazil, Canada, Chile, the United Kingdom, and the United States of America. TN and MBois acknowledge support from the DFG Cluster of Excellence `Origin and Structure of the Universe'. MS acknowledges support from a STFC Advanced Fellowship ST/F009186/1. PS is a NWO/Veni fellow. (TAD) The research leading to these results has received funding from the European Community's Seventh Framework Programme (/FP7/2007-2013/) under grant agreement No 229517. MBois has received, during this research, funding from the European Research Council under the Advanced Grant Program Num 267399-Momentum. The authors acknowledge financial support from ESO. 
\bibliographystyle{mn2e}
\bibliography{a3d_mnras}

\begin{thebibliography}{64}
\expandafter\ifx\csname natexlab\endcsname\relax\def\natexlab#1{#1}\fi

\bibitem[{{Abazajian} {et~al}\mbox{.}(2009){Abazajian}, {Adelman-McCarthy},
  {Ag{\"u}eros}, {Allam}, {Allende Prieto}, {An}, {Anderson}, {Anderson},
  {Annis}, {Bahcall}, \& et~al.}]{Abazajian:2009}
{Abazajian} K.~N. {et~al.}, 2009, ApJS, 182, 543

\bibitem[{{Adelman-McCarthy} {et~al}\mbox{.}(2007){Adelman-McCarthy},
  {Ag{\"u}eros}, {Allam}, {Anderson}, {Anderson}, {Annis}, {Bahcall},
  {Bailer-Jones}, \& et~al.}]{Adelman-McCarthy:2007}
{Adelman-McCarthy} J.~K. {et~al.}, 2007, ApJS, 172, 634

\bibitem[{{Aihara} {et~al}\mbox{.}(2011){Aihara}, {Allende Prieto}, {An},
  {Anderson}, {Aubourg}, {Balbinot}, {Beers}, {Berlind}, {Bickerton},
  {Bizyaev}, \& et~al.}]{Aihara:2011}
{Aihara} H. {et~al.}, 2011, ApJS, 193, 29

\bibitem[{{Barnes}(1988)}]{Barnes:1988}
{Barnes} J.~E., 1988, ApJ, 331, 699

\bibitem[{{Bender} {et~al}\mbox{.}(1993){Bender}, {Burstein}, \&
  {Faber}}]{Bender:1993}
{Bender} R., {Burstein} D., {Faber} S.~M., 1993, ApJ, 411, 153

\bibitem[{{Bernardi} {et~al}\mbox{.}(2010){Bernardi}, {Shankar}, {Hyde}, {Mei},
  {Marulli}, \& {Sheth}}]{Bernardi:2010}
{Bernardi} M., {Shankar} F., {Hyde} J.~B., {Mei} S., {Marulli} F., {Sheth}
  R.~K., 2010, MNRAS, 404, 2087

\bibitem[{{Bernardi} {et~al}\mbox{.}(2003){Bernardi}, {Sheth}, {Annis},
  {Burles}, {Finkbeiner}, {Lupton}, {Schlegel}, {SubbaRao}, {Bahcall},
  {Blakeslee}, \& et~al.}]{Bernardi:2003b}
{Bernardi} M. {et~al.}, 2003, AJ, 125, 1882

\bibitem[{{Binggeli} {et~al}\mbox{.}(1984){Binggeli}, {Sandage}, \&
  {Tarenghi}}]{Binggeli:1984}
{Binggeli} B., {Sandage} A., {Tarenghi} M., 1984, AJ, 89, 64

\bibitem[{{Blanton} {et~al}\mbox{.}(2011){Blanton}, {Kazin}, {Muna}, {Weaver},
  \& {Price-Whelan}}]{Blanton:2011}
{Blanton} M.~R., {Kazin} E., {Muna} D., {Weaver} B.~A., {Price-Whelan} A.,
  2011, AJ, 142, 31

\bibitem[{{Blanton} \& {Roweis}(2007)}]{Blanton:2007}
{Blanton} M.~R., {Roweis} S., 2007, AJ, 133, 734

\bibitem[{{Bois} {et~al}\mbox{.}(2011){Bois}, {Emsellem}, {Bournaud},
  {Alatalo}, {Blitz}, {Bureau}, {Cappellari}, {Davies}, {Davis}, {de Zeeuw}, \&
  et~al.}]{Bois:2011}
{Bois} M. {et~al.}, 2011, MNRAS, 416, 1654, Paper VI

\bibitem[{{Bower} {et~al}\mbox{.}(1998){Bower}, {Kodama}, \&
  {Terlevich}}]{Bower:1998}
{Bower} R.~G., {Kodama} T., {Terlevich} A., 1998, MNRAS, 299, 1193

\bibitem[{{Cappellari}(2002)}]{Cappellari:2002b}
{Cappellari} M., 2002, MNRAS, 333, 400

\bibitem[{{Cappellari}(2008)}]{Cappellari:2008}
{Cappellari} M., 2008, MNRAS, 390, 71

\bibitem[{{Cappellari} {et~al}\mbox{.}(2011{\natexlab{a}}){Cappellari},
  {Emsellem}, {Krajnovi{\'c}}, {McDermid}, {Scott}, {Verdoes Kleijn}, {Young},
  {Alatalo}, {Bacon}, {Blitz}, \& et~al.}]{Cappellari:2011a}
{Cappellari} M. {et~al.}, 2011{\natexlab{a}}, MNRAS, 413, 813, Paper I

\bibitem[{{Cappellari} {et~al}\mbox{.}(2011{\natexlab{b}}){Cappellari},
  {Emsellem}, {Krajnovi{\'c}}, {McDermid}, {Serra}, {Alatalo}, {Blitz}, {Bois},
  {Bournaud}, {Bureau}, \& et~al.}]{Cappellari:2011b}
{Cappellari} M. {et~al.}, 2011{\natexlab{b}}, MNRAS, 416, 1680, Paper VII

\bibitem[{{Cappellari} {et~al}\mbox{.}(2012{\natexlab{a}}){Cappellari},
  {McDermid}, {Alatalo}, {Blitz}, {Bois}, {Bournaud}, {Bureau}, {Crocker},
  {Davies}, {Davis}, \& et~al.}]{Cappellari:2012a}
{Cappellari} M. {et~al.}, 2012{\natexlab{a}}, Nat, 484, 485

\bibitem[{{Cappellari} {et~al}\mbox{.}(2012{\natexlab{b}}){Cappellari},
  {Scott}, {McDermid}, {Alatalo}, {Blitz}, {Bois}, {Bournaud}, {Bureau},
  {Crocker}, {Davies}, {Davis}, \& et~al.}]{Cappellari:2012b}
{Cappellari} M. {et~al.}, 2012{\natexlab{b}}, MNRAS, arXiv:1208.3522, Paper XIX

\bibitem[{{Cappellari} {et~al}\mbox{.}(2012{\natexlab{c}}){Cappellari},
  {Scott}, {McDermid}, {Alatalo}, {Blitz}, {Bois}, {Bournaud}, {Bureau},
  {Crocker}, {Davies}, {Davis}, \& et~al.}]{Cappellari:2012c}
{Cappellari} M. {et~al.}, 2012{\natexlab{c}}, MNRAS, arXiv:1208.3523, Paper XX

\bibitem[{{Cappellari} {et~al}\mbox{.}(2002){Cappellari}, {Verolme}, {van der
  Marel}, {Kleijn}, {Illingworth}, {Franx}, {Carollo}, \& {de
  Zeeuw}}]{Cappellari:2002}
{Cappellari} M., {Verolme} E.~K., {van der Marel} R.~P., {Kleijn} G.~A.~V.,
  {Illingworth} G.~D., {Franx} M., {Carollo} C.~M., {de Zeeuw} P.~T., 2002,
  ApJ, 578, 787

\bibitem[{{Carollo} \& {Danziger}(1994)}]{Carollo:1994}
{Carollo} C.~M., {Danziger} I.~J., 1994, MNRAS, 270, 523

\bibitem[{{Carollo} {et~al}\mbox{.}(1997){Carollo}, {Franx}, {Illingworth}, \&
  {Forbes}}]{Carollo:1997}
{Carollo} C.~M., {Franx} M., {Illingworth} G.~D., {Forbes} D.~A., 1997, ApJ,
  481, 710

\bibitem[{{Chen} {et~al}\mbox{.}(2010){Chen}, {C{\^o}t{\'e}}, {West}, {Peng},
  \& {Ferrarese}}]{Chen:2010}
{Chen} C.-W., {C{\^o}t{\'e}} P., {West} A.~A., {Peng} E.~W., {Ferrarese} L.,
  2010, ApJS, 191, 1

\bibitem[{{Colless} {et~al}\mbox{.}(1999){Colless}, {Burstein}, {Davies},
  {McMahan}, {Saglia}, \& {Wegner}}]{Colless:1999}
{Colless} M., {Burstein} D., {Davies} R.~L., {McMahan} R.~K., {Saglia} R.~P.,
  {Wegner} G., 1999, MNRAS, 303, 813

\bibitem[{{Conselice}(2006)}]{Conselice:2006}
{Conselice} C.~J., 2006, MNRAS, 373, 1389

\bibitem[{{Crockett} {et~al}\mbox{.}(2011){Crockett}, {Kaviraj}, {Silk},
  {Whitmore}, {O'Connell}, {Mutchler}, {Balick}, {Bond}, {Calzetti}, \&
  {Carollo}}]{Crockett:2011}
{Crockett} R.~M. {et~al.}, 2011, ApJ, 727, 115

\bibitem[{{Davies} {et~al}\mbox{.}(1993){Davies}, {Sadler}, \&
  {Peletier}}]{Davies:1993}
{Davies} R.~L., {Sadler} E.~M., {Peletier} R.~F., 1993, MNRAS, 262, 650

\bibitem[{{Davis} {et~al}\mbox{.}(2011){Davis}, {Alatalo}, {Sarzi}, {Bureau},
  {Young}, {Blitz}, {Serra}, {Crocker}, {Krajnovi{\'c}}, {McDermid}, \&
  et~al.}]{Davis:2011}
{Davis} T.~A. {et~al.}, 2011, MNRAS, 417, 882, Paper X

\bibitem[{{de Zeeuw} {et~al}\mbox{.}(2002){de Zeeuw}, {Bureau}, {Emsellem},
  {Bacon}, {Carollo}, {Copin}, {Davies}, {Kuntschner}, {Miller}, {Monnet},
  {Peletier}, \& {Verolme}}]{deZeeuw:2002}
{de Zeeuw} P.~T. {et~al.}, 2002, MNRAS, 329, 513

\bibitem[{{Emsellem} {et~al}\mbox{.}(1996){Emsellem}, {Bacon}, {Monnet}, \&
  {Poulain}}]{Emsellem:1996}
{Emsellem} E., {Bacon} R., {Monnet} G., {Poulain} P., 1996, AAP, 312, 777

\bibitem[{{Emsellem} {et~al}\mbox{.}(2011){Emsellem}, {Cappellari},
  {Krajnovi{\'c}}, {Alatalo}, {Blitz}, {Bois}, {Bournaud}, {Bureau}, {Davies},
  {Davis}, \& et~al.}]{Emsellem:2011}
{Emsellem} E. {et~al.}, 2011, MNRAS, 414, 888, Paper III

\bibitem[{{Emsellem} {et~al}\mbox{.}(1994){Emsellem}, {Monnet}, \&
  {Bacon}}]{Emsellem:1994}
{Emsellem} E., {Monnet} G., {Bacon} R., 1994, AAP, 285, 723

\bibitem[{{Faber}(1973)}]{Faber:1973}
{Faber} S.~M., 1973, ApJ, 179, 731

\bibitem[{{Ferrarese} {et~al}\mbox{.}(2006){Ferrarese}, {C{\^o}t{\'e}},
  {Jord{\'a}n}, {Peng}, {Blakeslee}, {Piatek}, {Mei}, {Merritt},
  {Milosavljevi{\'c}}, {Tonry}, \& {West}}]{Ferrarese:2006}
{Ferrarese} L. {et~al.}, 2006, ApJS, 164, 334

\bibitem[{{Francis} \& {Wills}(1999)}]{Francis:1999}
{Francis} P.~J., {Wills} B.~J., 1999, in Astronomical Society of the Pacific
  Conference Series, Vol. 162, Quasars and Cosmology, {G.~Ferland \&
  J.~Baldwin}, ed., p. 363

\bibitem[{{Franx} \& {Illingworth}(1990)}]{Franx:1990}
{Franx} M., {Illingworth} G., 1990, ApJL, 359, L41

\bibitem[{{Hopkins} {et~al}\mbox{.}(2009){Hopkins}, {Lauer}, {Cox},
  {Hernquist}, \& {Kormendy}}]{Hopkins:2009}
{Hopkins} P.~F., {Lauer} T.~R., {Cox} T.~J., {Hernquist} L., {Kormendy} J.,
  2009, ApJS, 181, 486

\bibitem[{{Janz} \& {Lisker}(2008)}]{Janz:2008}
{Janz} J., {Lisker} T., 2008, ApJ, 689, L25

\bibitem[{{Jarrett} {et~al}\mbox{.}(2000){Jarrett}, {Chester}, {Cutri},
  {Schneider}, {Skrutskie}, \& {Huchra}}]{Jarrett:2000}
{Jarrett} T.~H., {Chester} T., {Cutri} R., {Schneider} S., {Skrutskie} M.,
  {Huchra} J.~P., 2000, AJ, 119, 2498

\bibitem[{{Jorgensen}(1997)}]{Jorgensen:1997}
{Jorgensen} I., 1997, MNRAS, 288, 161

\bibitem[{{Kauffmann} {et~al}\mbox{.}(2003){Kauffmann}, {Heckman}, {White},
  {Charlot}, {Tremonti}, {Peng}, {Seibert}, {Brinkmann}, {Nichol}, {SubbaRao},
  \& {York}}]{Kauffmann:2003}
{Kauffmann} G. {et~al.}, 2003, MNRAS, 341, 54

\bibitem[{{Khochfar} {et~al}\mbox{.}(2011){Khochfar}, {Emsellem}, {Serra},
  {Bois}, {Alatalo}, {Bacon}, {Blitz}, {Bournaud}, {Bureau}, {Cappellari}, \&
  et~al.}]{Khochfar:2011}
{Khochfar} S. {et~al.}, 2011, MNRAS, 417, 845, Paper VIII

\bibitem[{{Kormendy} {et~al}\mbox{.}(2009){Kormendy}, {Fisher}, {Cornell}, \&
  {Bender}}]{Kormendy:2009}
{Kormendy} J., {Fisher} D.~B., {Cornell} M.~E., {Bender} R., 2009, ApJS, 182,
  216

\bibitem[{{Krajnovi{\'c}} {et~al}\mbox{.}(2008){Krajnovi{\'c}}, {Bacon},
  {Cappellari}, {Davies}, {de Zeeuw}, {Emsellem}, {Falc{\'o}n-Barroso},
  {Kuntschner}, {McDermid}, {Peletier}, {Sarzi}, {van den Bosch}, \& {van de
  Ven}}]{Krajnovic:2008}
{Krajnovi{\'c}} D. {et~al.}, 2008, MNRAS, 390, 93

\bibitem[{{Krajnovi{\'c}} {et~al}\mbox{.}(2011){Krajnovi{\'c}}, {Emsellem},
  {Cappellari}, {Alatalo}, {Blitz}, {Bois}, {Bournaud}, {Bureau}, {Davies},
  {Davis}, \& et~al.}]{Krajnovic:2011}
{Krajnovi{\'c}} D. {et~al.}, 2011, MNRAS, 414, 2923, Paper II

\bibitem[{{Kuntschner} {et~al}\mbox{.}(2006){Kuntschner}, {Emsellem}, {Bacon},
  {Bureau}, {Cappellari}, {Davies}, {de Zeeuw}, {Falc{\'o}n-Barroso},
  {Krajnovi{\'c}}, {McDermid}, {Peletier}, \& {Sarzi}}]{Kuntschner:2006}
{Kuntschner} H. {et~al.}, 2006, MNRAS, 369, 497

\bibitem[{{Kuntschner} {et~al}\mbox{.}(2010){Kuntschner}, {Emsellem}, {Bacon},
  {Cappellari}, {Davies}, {de Zeeuw}, {Falc{\'o}n-Barroso}, {Krajnovi{\'c}},
  {McDermid}, {Peletier}, {Sarzi}, {Shapiro}, {van den Bosch}, \& {van de
  Ven}}]{Kuntschner:2010}
{Kuntschner} H. {et~al.}, 2010, MNRAS, 408, 97

\bibitem[{{Lablanche} {et~al}\mbox{.}(2012){Lablanche}, {Cappellari},
  {Emsellem}, {Bournaud}, {Michel-Dansac}, {Alatalo}, {Blitz}, {Bois},
  {Bureau}, {Davies}, \& et~al.}]{Lablanche:2012}
{Lablanche} P.-Y. {et~al.}, 2012, MNRAS, 424, 1495, Paper XII

\bibitem[{{McFarland} {et~al}\mbox{.}(2011){McFarland}, {Verdoes-Kleijn},
  {Sikkema}, {Helmich}, {Boxhoorn}, \& {Valentijn}}]{McFarland:2011}
{McFarland} J.~P., {Verdoes-Kleijn} G., {Sikkema} G., {Helmich} E.~M.,
  {Boxhoorn} D.~R., {Valentijn} E.~A., 2011, arXiv:1110.2509

\bibitem[{{Monet} {et~al}\mbox{.}(2003){Monet}, {Levine}, {Canzian}, {Ables},
  {Bird}, {Dahn}, {Guetter}, {Harris}, \& et~al.}]{Monet:2003}
{Monet} D.~G. {et~al.}, 2003, AJ, 125, 984

\bibitem[{{Naab} {et~al}\mbox{.}(2009){Naab}, {Johansson}, \&
  {Ostriker}}]{Naab:2009}
{Naab} T., {Johansson} P.~H., {Ostriker} J.~P., 2009, ApJL, 699, L178

\bibitem[{{S{\'a}nchez-Bl{\'a}zquez}
  {et~al}\mbox{.}(2006){S{\'a}nchez-Bl{\'a}zquez}, {Peletier},
  {Jim{\'e}nez-Vicente}, {Cardiel}, {Cenarro}, {Falc{\'o}n-Barroso}, {Gorgas},
  {Selam}, \& {Vazdekis}}]{Sanchez-Blazquez:2006b}
{S{\'a}nchez-Bl{\'a}zquez} P. {et~al.}, 2006, MNRAS, 371, 703

\bibitem[{{Schiavon}(2007)}]{Schiavon:2007}
{Schiavon} R.~P., 2007, ApJS, 171, 146

\bibitem[{{Schlegel} {et~al}\mbox{.}(1998){Schlegel}, {Finkbeiner}, \&
  {Davis}}]{Schlegel:1998}
{Schlegel} D.~J., {Finkbeiner} D.~P., {Davis} M., 1998, ApJ, 500, 525

\bibitem[{{Scott} {et~al}\mbox{.}(2009){Scott}, {Cappellari}, {Davies},
  {Bacon}, {de Zeeuw}, {Emsellem}, {Falc{\'o}n-Barroso}, {Krajnovi{\'c}},
  {Kuntschner}, {McDermid}, \& et~al.}]{Scott:2009}
{Scott} N. {et~al.}, 2009, MNRAS, 398, 1835, S09

\bibitem[{{Serra} {et~al}\mbox{.}(2012){Serra}, {Oosterloo}, {Morganti},
  {Alatalo}, {Blitz}, {Bois}, {Bournaud}, {Bureau}, {Cappellari}, {Crocker}, \&
  et~al.}]{Serra:2012}
{Serra} P. {et~al.}, 2012, MNRAS, 422, 1835, Paper XIII

\bibitem[{{Strateva} {et~al}\mbox{.}(2001){Strateva}, {Ivezi{\'c}}, {Knapp},
  {Narayanan}, {Strauss}, {Gunn}, {Lupton}, {Schlegel}, {Bahcall}, {Brinkmann},
  \& et~al.}]{Strateva:2001}
{Strateva} I. {et~al.}, 2001, AJ, 122, 1861

\bibitem[{{Thomas} {et~al}\mbox{.}(2005){Thomas}, {Maraston}, {Bender}, \&
  {Mendes de Oliveira}}]{Thomas:2005}
{Thomas} D., {Maraston} C., {Bender} R., {Mendes de Oliveira} C., 2005, ApJ,
  621, 673

\bibitem[{{Trager} {et~al}\mbox{.}(2000){Trager}, {Faber}, {Worthey}, \&
  {Gonz{\'a}lez}}]{Trager:2000b}
{Trager} S.~C., {Faber} S.~M., {Worthey} G., {Gonz{\'a}lez} J.~J., 2000, AJ,
  120, 165

\bibitem[{{Valentijn} {et~al}\mbox{.}(2007){Valentijn}, {McFarland}, {Snigula},
  {Begeman}, {Boxhoorn}, {Rengelink}, {Helmich}, {Heraudeau}, {Kleijn},
  {Vermeij}, \& et~al.}]{Valentijn:2007}
{Valentijn} E.~A. {et~al.}, 2007, in Astronomical Society of the Pacific
  Conference Series, Vol. 376, Astronomical Data Analysis Software and Systems
  XVI, {R.~A.~Shaw, F.~Hill, \& D.~J.~Bell}, ed., p. 491

\bibitem[{{van den Bergh}(2007)}]{van_den_Bergh:2007}
{van den Bergh} S., 2007, AJ, 134, 1508

\bibitem[{{Verdoes Kleijn} {et~al}\mbox{.}(2007){Verdoes Kleijn}, {Vermeij},
  {Valentijn}, \& {Kuijken}}]{Verdoes_Kleijn:2007}
{Verdoes Kleijn} G., {Vermeij} R., {Valentijn} E., {Kuijken} K., 2007, in
  Astronomical Society of the Pacific Conference Series, Vol. 364, The Future
  of Photometric, Spectrophotometric and Polarimetric Standardization,
  {Sterken} C., ed., p. 103

\bibitem[{{Visvanathan} \& {Sandage}(1977)}]{Visvanathan:1977}
{Visvanathan} N., {Sandage} A., 1977, ApJ, 216, 214

\bibitem[{{White}(1978)}]{White:1978a}
{White} S.~D.~M., 1978, MNRAS, 184, 185

\end{thebibliography}

\appendix
\section{MGE models for all 258 galaxies}
Contour plots showing observations (black) and MGE models (red) for all 258 galaxies with {\it r}-band imaging available. We show only the central 40" x 40", approximately the region covered by the SAURON observations. The MGE models are able to reproduce a range of morphologies from round galaxies to edge-on, dusty, disky objects. Bars, which our MGE modelling procedure is designed not to fit, stand out clearly. Contours are shown at 0.5 magnitude intervals down to the standard deviation of the sky background level.
\begin{figure*}
\includegraphics[width=6.9in]{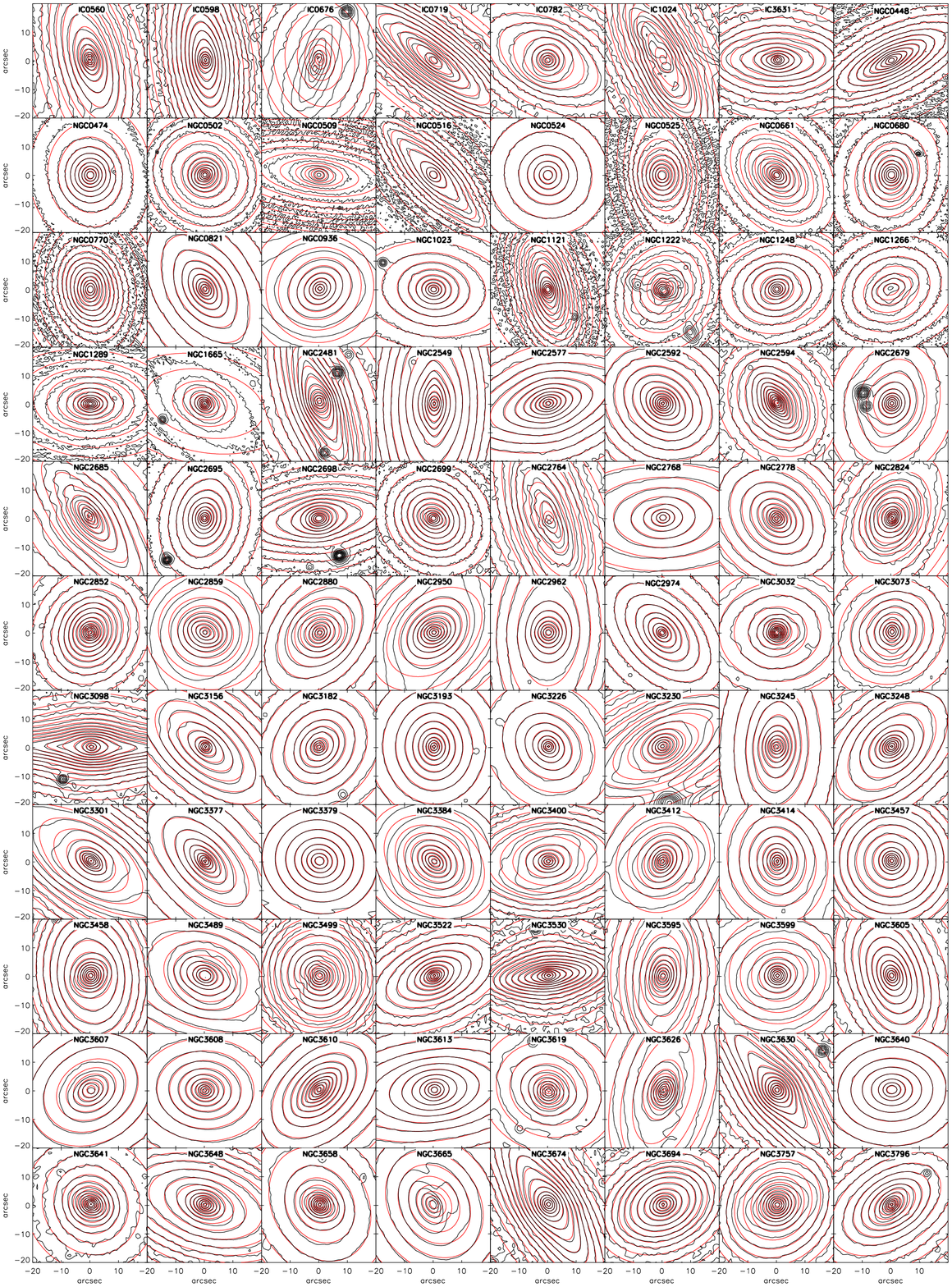}
\caption{MGE models (red) and observations (black) for all 258 galaxies with {\it r}-band imaging available, arranged by name.}
\label{fig:mge_models}
\end{figure*}

\begin{figure*}
\includegraphics[width=6.9in]{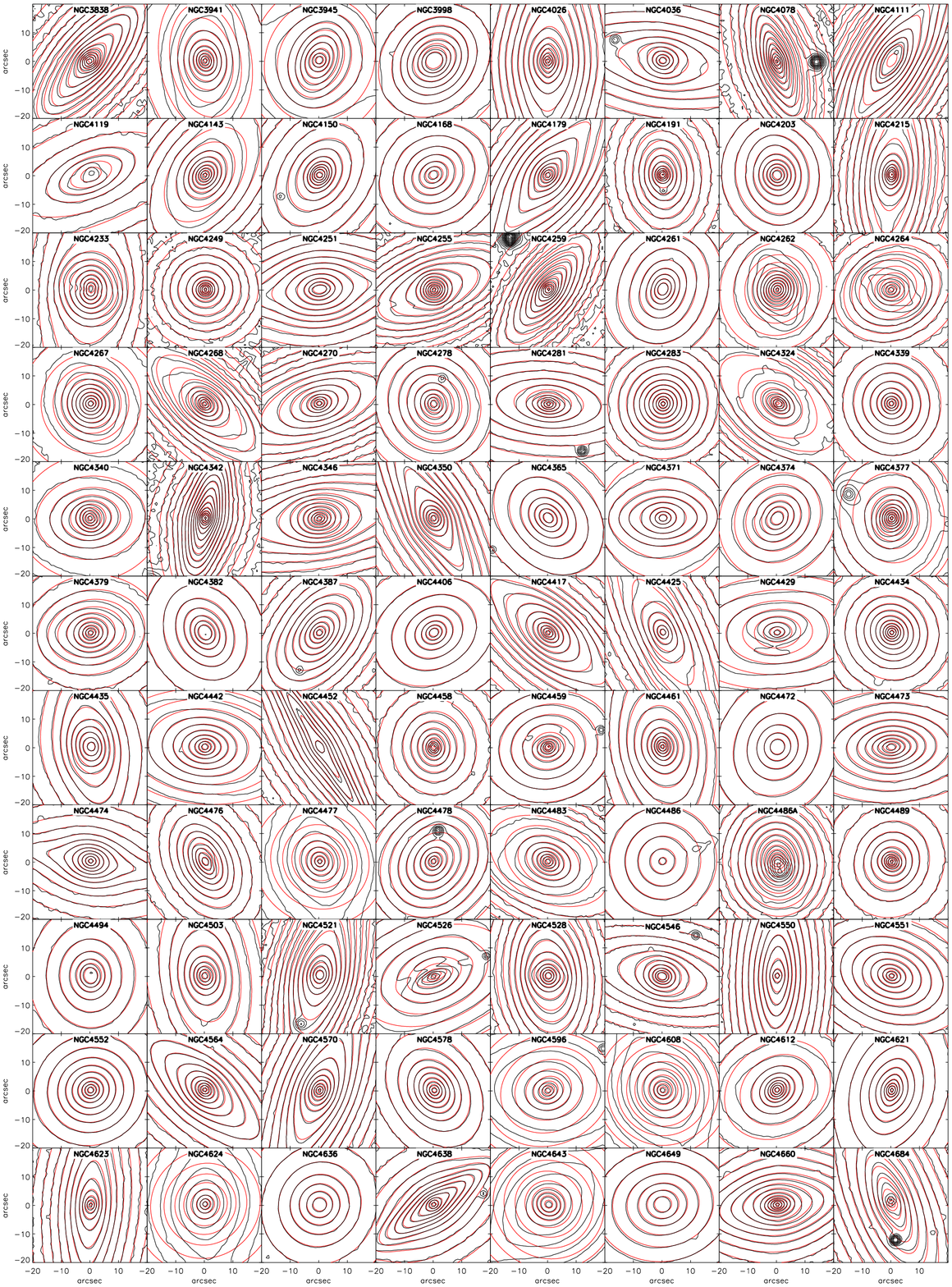}
Fig. \ref{fig:mge_models} continued.
\end{figure*}

\begin{figure*}
\includegraphics[width=6.9in]{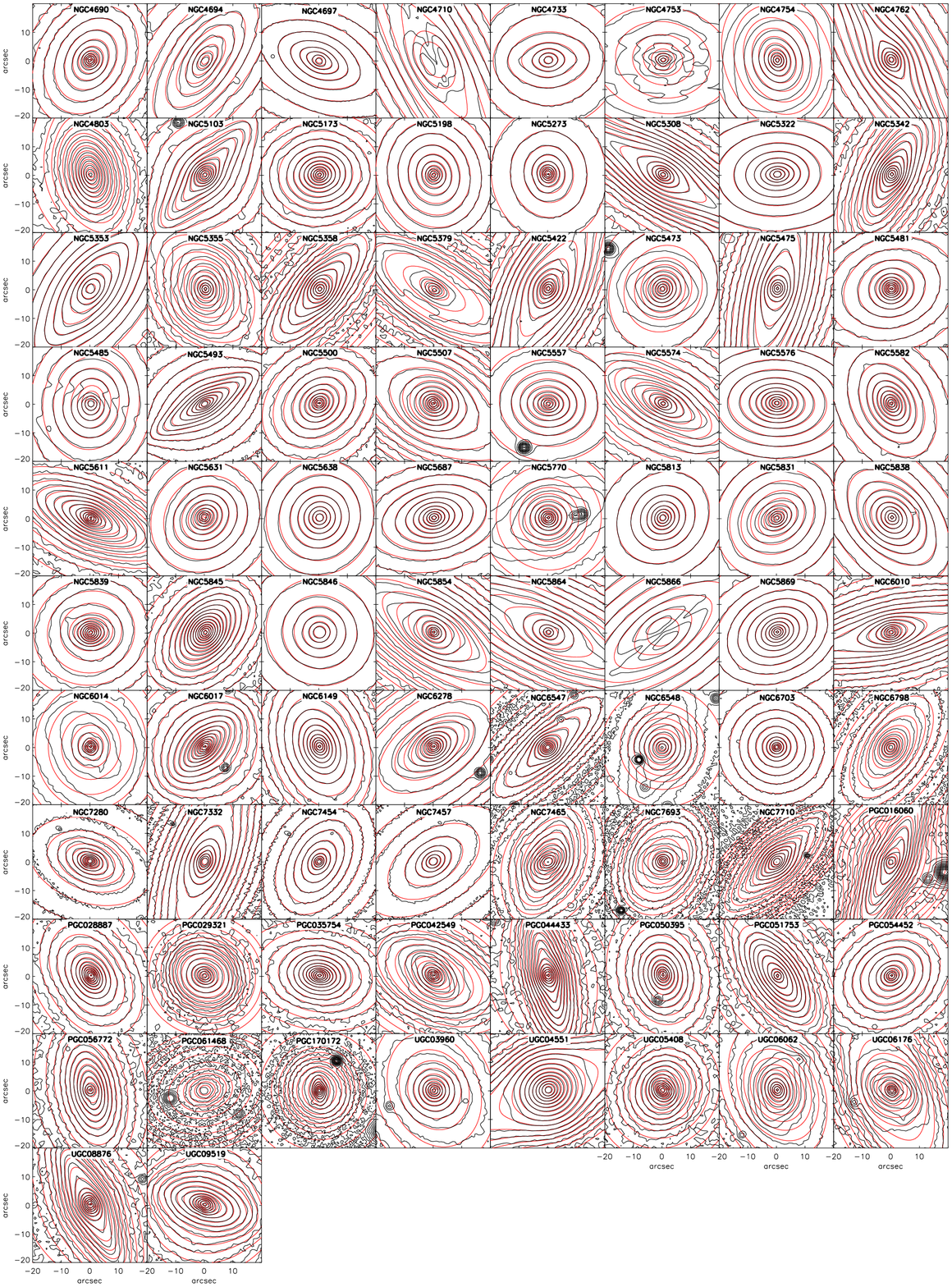}
Fig. \ref{fig:mge_models} continued.
\end{figure*}

\section{Comparison between MGE and literature surface brightness profiles}
Comparison of surface brightness profiles derived from our MGE models with those of \citet{Kormendy:2009}, for all 23 galaxies in common between the two samples. The agreement between the two sets of profiles is excellent, with typical residuals of 0.04 magnitudes.
\begin{figure*}
\includegraphics[width=3.125in]{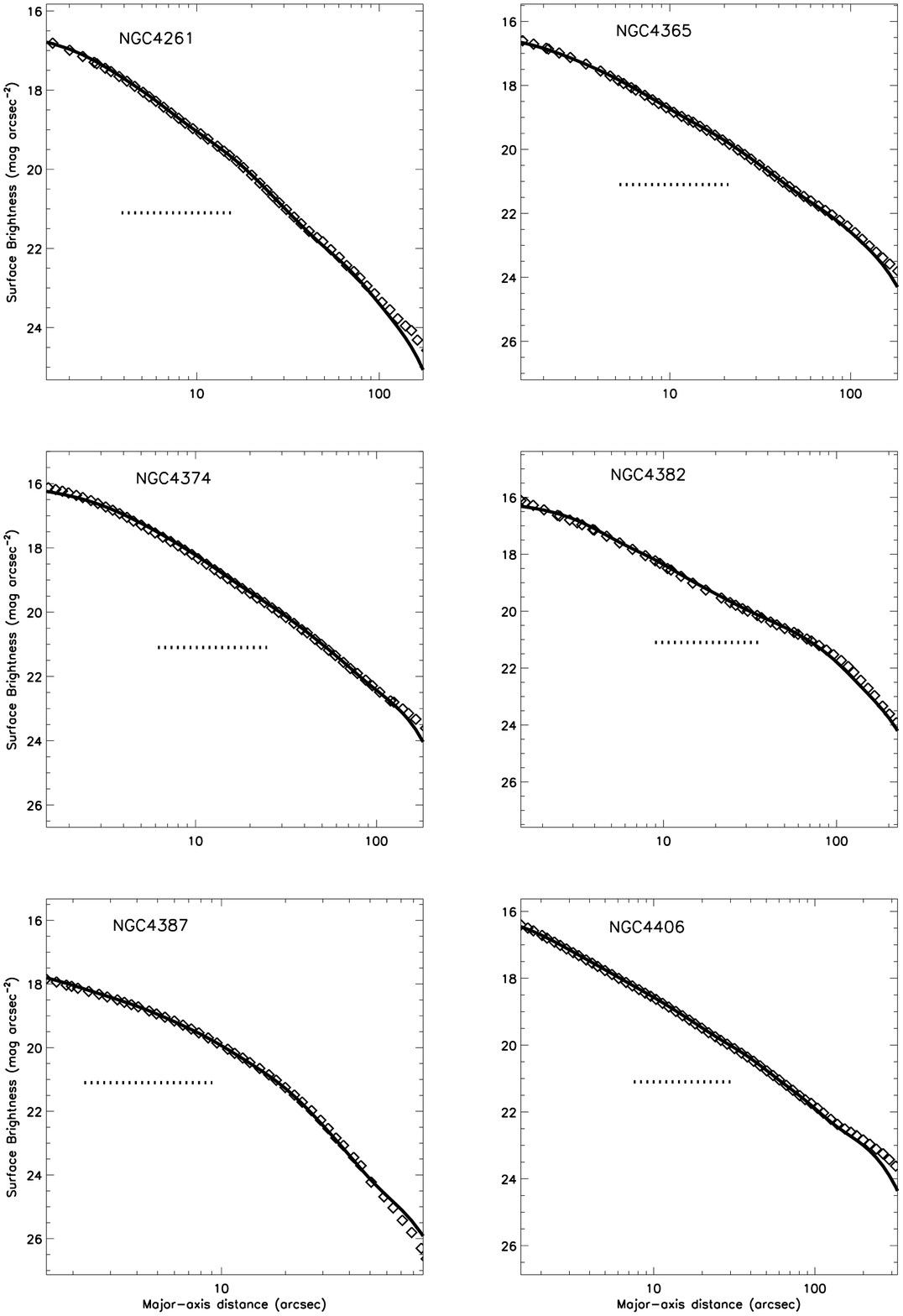}
\includegraphics[width=3.125in]{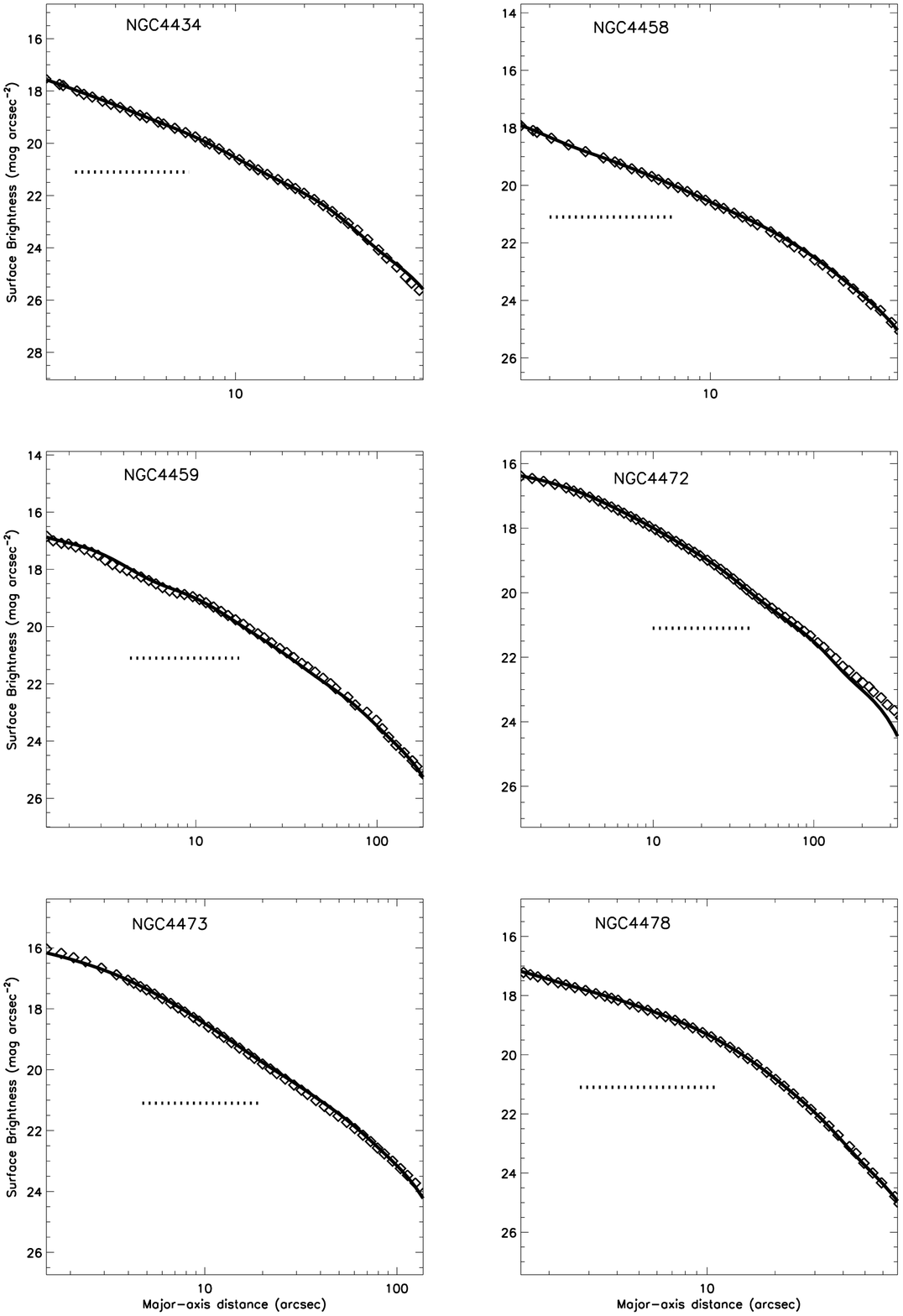}
\includegraphics[width=3.125in]{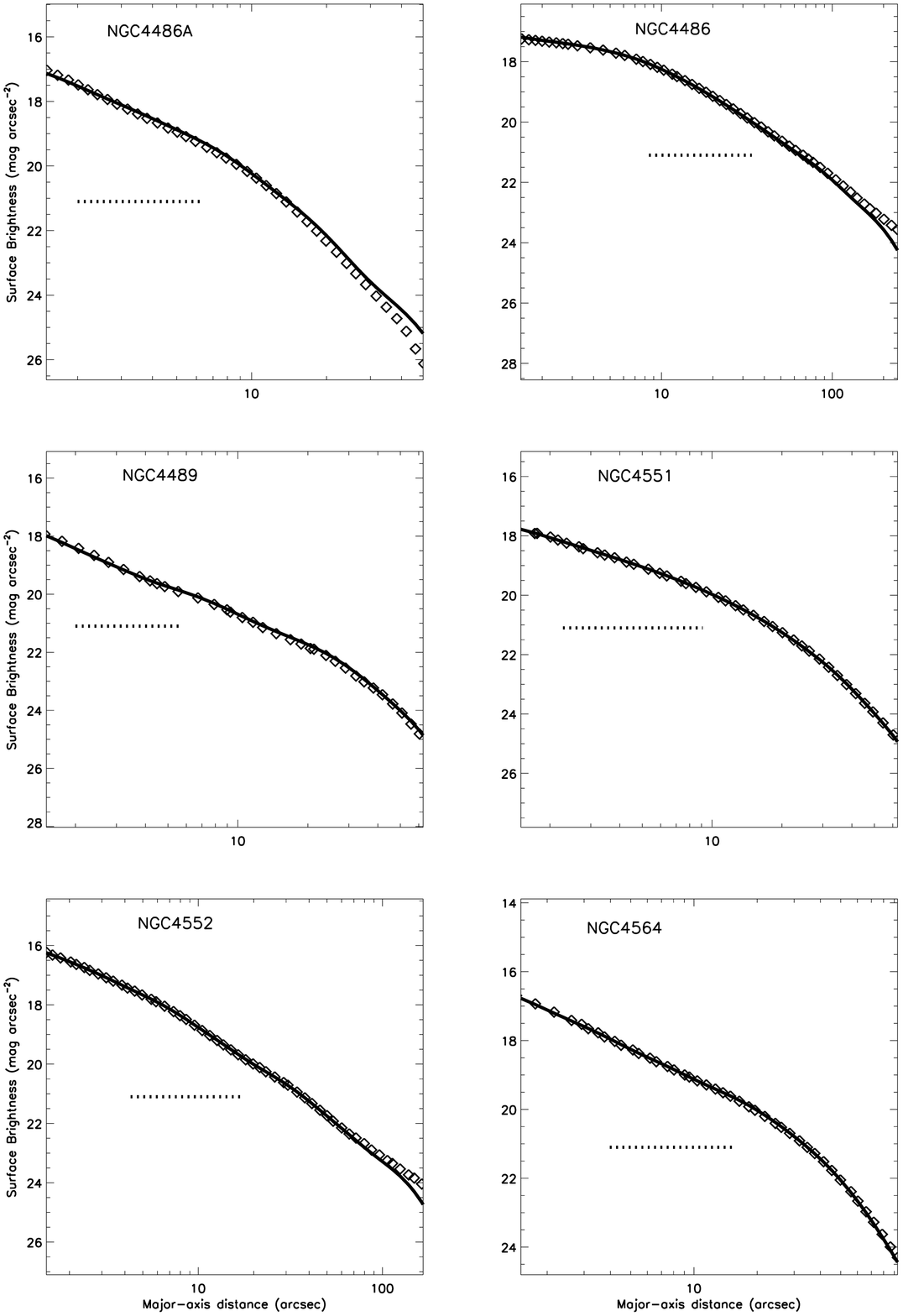}
\includegraphics[width=3.125in]{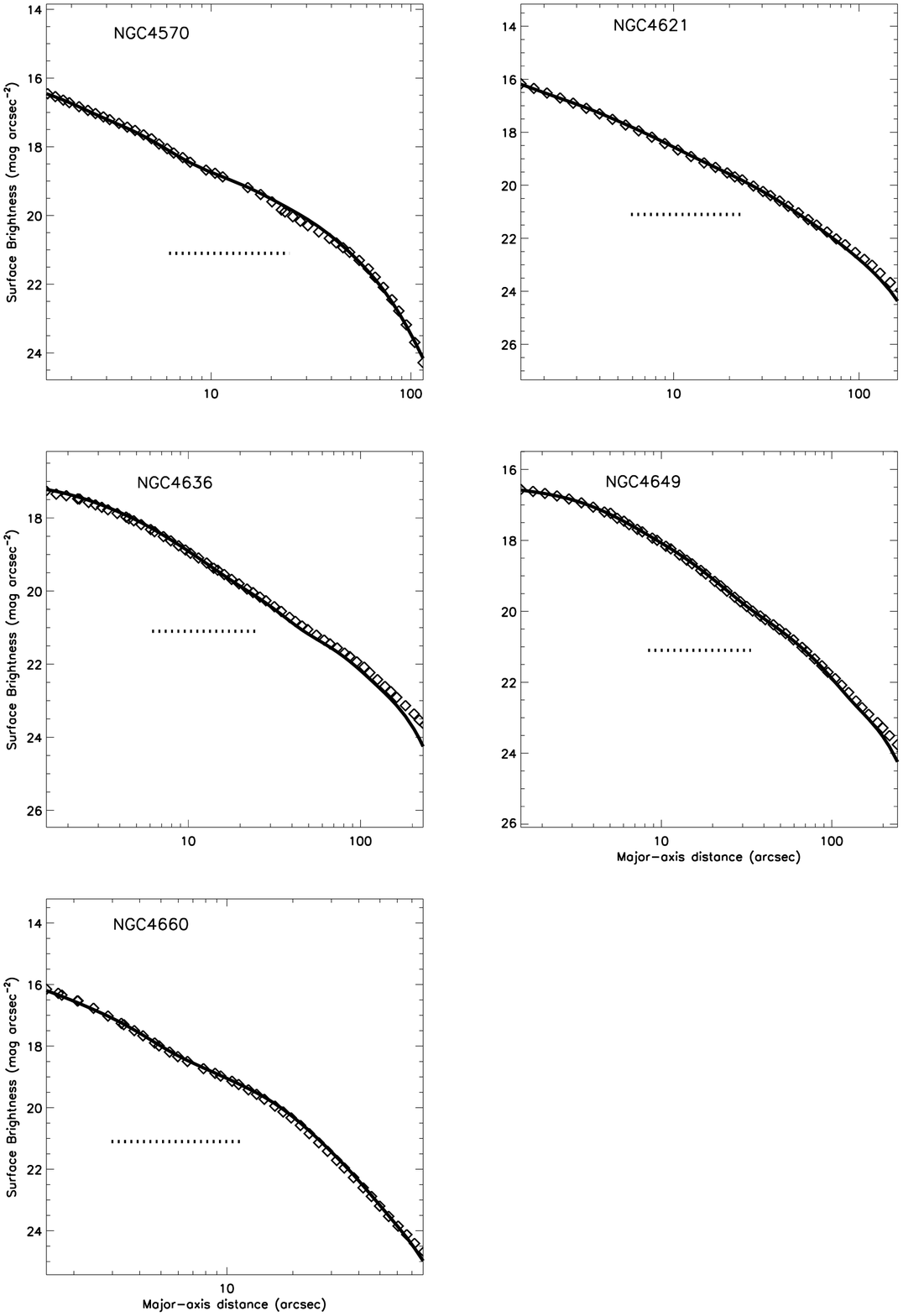}
\caption{Comparison of surface brightness profiles from MGE models (solid line) and observations from \citet{Kormendy:2009} for all 23 galaxies in common between the two samples. The comparison is shown only over the region for which the MGE is valid, 1.5" $< r < 2 \mathrm{max}(\sigma_\mathrm{MGE})$. The sky background level in the SDSS {\it r}-band images is indicated by the horizontal dotted line.}
\end{figure*}

\label{lastpage}

\end{document}